\UseRawInputEncoding

\documentclass[11pt]{article}
\usepackage[margin=0.8in]{geometry}
\usepackage{amsmath,amssymb,bbm,amsthm}
\usepackage{thm-restate,color,xcolor,xspace}
\usepackage{hyperref,cleveref}
\usepackage{algorithm,algorithmicx}
\usepackage[noend]{algpseudocode}
\usepackage{graphicx}
\usepackage{stmaryrd,mathtools,nicefrac}
\usepackage{tikz}
\usepackage[export]{adjustbox}
\newcommand{\f}{\frac}
\newcommand{\cd}{\cdot}
\newcommand{\bn}{\binom}
\newcommand{\sr}{\sqrt}
\newcommand{\cds}{\cdots}
\newcommand{\lds}{\ldots}

\newcommand{\s}{\subseteq}

\newcommand{\BE}{\begin{enumerate}}
\newcommand{\EE}{\end{enumerate}}
\newcommand{\im}{\item}

\newcommand{\Prod}{\displaystyle\prod\limits}

\newcommand{\logn}{\log n}

\newcommand{\inv}{^{-1}}

\newcommand{\R}{\mathbb R}
\newcommand{\Z}{\mathbb Z}

\newcommand{\e}{\epsilon}
\newcommand{\de}{\delta}
\newcommand{\De}{\Delta}

\newcommand{\G}{\Gamma}

\newcommand{\al}{\alpha}
\newcommand{\be}{\beta}
\newcommand{\om}{\omega}
\newcommand{\Om}{\Omega}
\newcommand{\el}{\ell}

\newcommand{\Th}{\Theta}
\newcommand{\m}{\mathcal}

\newcommand{\lf}{\lfloor}
\newcommand{\rf}{\rfloor}
\newcommand{\lc}{\lceil}
\newcommand{\rc}{\rceil}

\newcommand{\E}{\mathbb E}

\newcommand{\poly}{\textup{poly}}
\newcommand{\pl}{\textup{polylog}}
\newcommand{\polylog}{\textup{polylog}}
\newcommand{\norm}[1]{\left\lVert#1\right\rVert}
\newcommand{\1}[1]{\left\lVert#1\right\rVert_1}

\newcommand{\lp}{\left(}
\newcommand{\rp}{\right)}
\newcommand{\lb}{\left[}
\newcommand{\rb}{\right]}
\newcommand{\lmt}{\left[\begin{matrix}}
\newcommand{\rmt}{\end{matrix}\right]}

\newtheorem{theorem}{Theorem}[section]
\newtheorem{lemma}[theorem]{Lemma}
\newtheorem{definition}[theorem]{Definition}
\newtheorem{corollary}[theorem]{Corollary}
\newtheorem{observation}[theorem]{Observation}
\newtheorem{claim}[theorem]{Claim}
\newtheorem{subclaim}{Subclaim}
\newtheorem{fact}[theorem]{Fact}
\newtheorem{assumption}[theorem]{Assumption}
\newtheorem{remark}[theorem]{Remark}
\newcommand{\BT}{\begin{theorem}}
\newcommand{\ET}{\end{theorem}}
\newcommand{\BL}{\begin{lemma}}
\newcommand{\EL}{\end{lemma}}
\newcommand{\BD}{\begin{definition}}
\newcommand{\ED}{\end{definition}}
\newcommand{\BC}{\begin{corollary}}
\newcommand{\EC}{\end{corollary}}
\newcommand{\BO}{\begin{observation}}
\newcommand{\EO}{\end{observation}}
\newcommand{\BCL}{\begin{claim}}
\newcommand{\ECL}{\end{claim}}
\newcommand{\BSCL}{\begin{subclaim}}
\newcommand{\ESCL}{\end{subclaim}}
\newcommand{\BF}{\begin{fact}}
\newcommand{\EF}{\end{fact}}
\newcommand{\BA}{\begin{assumption}}
\newcommand{\EA}{\end{assumption}}
\newcommand{\BR}{\begin{remark}}
\newcommand{\ER}{\end{remark}}
\newcommand{\BP}{\begin{proof}}
\newcommand{\EP}{\end{proof}}
\newcommand{\BPS}{\begin{proof}[Proof (Sketch)]}
\newcommand{\EPS}{\end{proof}}
\Crefname{observation}{Observation}{Observations}
\Crefname{claim}{Claim}{Claims}
\Crefname{subclaim}{Subclaim}{Subclaims}
\Crefname{fact}{Fact}{Facts}
\Crefname{assumption}{Assumption}{Assumptions}
\newenvironment{subproof}[1][\proofname]{%
  \begin{proof}[#1]%
}{%
  \end{proof}%
}
\newcommand{\BSP}{\begin{subproof}}
\newcommand{\ESP}{\end{subproof}}

\newcommand{\alert}{\textcolor{red}}
\newcommand{\para}{\paragraph}

\newcommand{\tO}{\widetilde{O}}

\newcommand{\thml}[1]{\label{thm:#1}}
\newcommand{\thm}[1]{\Cref{thm:#1}}
\newcommand{\leml}[1]{\label{lem:#1}}
\newcommand{\lem}[1]{\Cref{lem:#1}}
\newcommand{\defnl}[1]{\label{def:#1}}
\newcommand{\defn}[1]{\Cref{def:#1}}
\newcommand{\clml}[1]{\label{clm:#1}}
\newcommand{\clm}[1]{\Cref{clm:#1}}
\newcommand{\corl}[1]{\label{cor:#1}}
\newcommand{\cor}[1]{\Cref{cor:#1}}
\newcommand{\obsl}[1]{\label{obs:#1}}
\newcommand{\obs}[1]{\Cref{obs:#1}}
\newcommand{\eqnl}[1]{\label{eq:#1}}
\newcommand{\eqn}[1]{(\ref{eq:#1})}
\newcommand{\linel}[1]{\label{line:#1}}
\renewcommand{\line}[1]{line~\ref{line:#1}}
\newcommand{\figl}[1]{\label{fig:#1}}
\newcommand{\fig}[1]{Figure~\ref{fig:#1}}

\renewcommand{\b}{\textbf}

\newcommand{\Wembed}{W_{\textup{embed}}}

\newcommand{\WSSSP}{W_{\textup{SSSP}}}
\newcommand{\WTS}{W_{\textup{TS}}}

\newcommand{\Tembed}{T_{\textup{embed}}}

\newcommand{\TSSSP}{T_{\textup{SSSP}}}
\newcommand{\TTS}{T_{\textup{TS}}}

\newcommand{\ka}{\kappa}
\newcommand{\sm}{\setminus}
\newcommand{\up}{\uparrow}
\newcommand{\n}{\nicefrac}

\newcommand{\opt}{\mathsf{opt}}
\newcommand{\cost}{\mathsf{cost}}

\newcommand{\wG}{\widehat G}

\newcommand{\mf}{maximum flow\xspace}
\newcommand{\ts}{transshipment\xspace}

\newcommand{\secl}[1]{\label{sec:#1}}
\renewcommand{\sec}[1]{\Cref{sec:#1}}

\newtheorem{invariant}[theorem]{Invariant}
\Crefname{invariant}{Invariant}{Invariants}

\makeatletter
\newcounter{algocounter}
\@ifpackageloaded{hyperref}%
  {\newcommand{\mylabel}[2]
    {\refstepcounter{algocounter}\protected@write\@auxout{}{\string\newlabel{#1}{{\textcolor{black}{\textup{#2}}}{\thepage}%
      {\@currentlabelname}{\@currentHref}{}}}}}%
\makeatother

\allowdisplaybreaks

\begin{document}

\title{Faster Parallel Algorithm for Approximate Shortest Path}
\author{Jason Li\thanks{Carnegie Mellon University, Pittsburgh, PA}}
\date{\today}
\maketitle
\abstract{
We present the first $m\,\text{polylog}(n)$ work, $\text{polylog}(n)$ time algorithm in the PRAM model that computes $(1+\epsilon)$-approximate single-source shortest paths on weighted, undirected graphs. This improves upon the breakthrough result of Cohen~[JACM'00] that achieves $O(m^{1+\epsilon_0})$ work and $\text{polylog}(n)$ time. While most previous approaches, including Cohen's, leveraged the power of hopsets, our algorithm builds upon the recent developments in \emph{continuous optimization}, studying the shortest path problem from the lens of the closely-related \emph{minimum transshipment} problem. To obtain our algorithm, we demonstrate a series of near-linear work, polylogarithmic-time reductions between the problems of approximate shortest path, approximate transshipment, and $\ell_1$-embeddings, and establish a recursive algorithm that cycles through the three problems and reduces the graph size on each cycle. As a consequence, we also obtain faster parallel algorithms for approximate transshipment and $\ell_1$-embeddings with polylogarithmic distortion. The minimum transshipment algorithm in particular improves upon the previous best $m^{1+o(1)}$ work sequential algorithm of Sherman~[SODA'17].

To improve readability, the paper is almost entirely self-contained, save for several staple theorems in algorithms and combinatorics.
}

\newpage

\section{Introduction}
The single-source shortest path problem is one of the most fundamental combinatorial optimization problems, and is also among the most notorious in parallel computation models. While the sequential model has simple near-linear time algorithm dating back to Dijkstra, much remains unknown for even the PRAM model despite decades of extensive research.

One of the most well-known settings studied so far in the PRAM\ model is the case of $(1+\e)$-approximate single-source shortest paths in undirected graphs. Early work on this problem produced algorithms in sublinear time~\cite{klein1992parallel,klein1997randomized}, until the breakthrough result of Cohen~\cite{Cohen}, who presented an algorithm in $O(m^{1+\e_0})$ work (for any constant $\e_0>0$) and $\polylog(n)$ time through the use of \emph{hopsets}: additional edges added to the graph so that short paths in the graph span few edges. Since then, it was a long-standing open problem whether Cohen's algorithm could be improved to run in $m\,\pl(n)$ work while keeping the time $\pl(n)$.


Recently, this question was partially answered by Abboud,~Bodwin~and~Pettie~\cite{hopset}, surprisingly in the negative: they showed that there exist families of graphs for which any hopsets on these graphs must have size $\Om(m^{1+\e_0})$, thereby lower bounding the work by $\Om(m^{1+\e_0})$ for any purely hopset-based algorithm like Cohen's. While their lower bound does not rule out other approaches to this problem, no other directions of attack have come close to matching Cohen's method of hopsets.

In this paper, we tackle this problem from a new perspective: \emph{continuous optimization}, especially the methods pioneered by Sherman~\cite{Sherman13} for the maximum flow problem. By reducing to studying the closely-related and more continuous \emph{minimum transshipment} problem, we provide the first $(1+\e)$-approximate SSSP algorithm for weighted, undirected graphs in $m\,\pl(n)$ work and $\pl(n)$ time in the PRAM model, bypassing the hopset lower bound and resolving the aforementioned open problem. This serves as evidence that continuous optimization, with its rich theory in graph algorithm and inherent parallelism, is a promising research direction in parallel graph algorithms and can bypass known barriers to other common approaches.

\BT[Parallel SSSP]\thml{t1}
There exists a parallel algorithm that, given an undirected graph with nonnegative weights, computes a $(1+\e)$-approximate single-source shortest path tree in $m\,\pl(n)\,\e^{-2}$ work and $\pl(n)\,\e^{-2}$ time in the PRAM model.
\ET




Our SSSP algorithm is recursive, cycling through three problems in a round-robin fashion: SSSP, \ts, and the problem of computing an $\el_1$-embedding of a graph with $\pl(n)$ distortion in $O(\logn)$ dimensions. That is, each problem calls the next problem on the cyclic list possibly many times, and possibly on a smaller graph instance.  Hence, we obtain parallel algorithms with similar running times for the other two problems as well.

\BT[Parallel \ts]\thml{t2}
There exists a parallel algorithm that, given an undirected graph with nonnegative weights and 
 polynomial aspect ratio, computes a $(1+\e)$-approximation to minimum transshipment in $m\,\pl(n)\,\e^{-2}$ work and $\pl(n)\,\e^{-2}$ time in the PRAM model.
\ET

\BT[Parallel $\el_1$-embedding]\thml{t3}
There exists a parallel algorithm that, given an undirected graph with nonnegative weights and polynomial aspect ratio, computes an $\ell_1$-embedding with $\pl(n)$ distortion in $O(\logn)$ dimensions in $m\,\pl(n)$ work and $\pl(n)$ time in the PRAM\ model.
\ET

\thm{t2} also establishes the first $m\,\pl(n)$ time \emph{sequential} algorithm for $(1+\e)$-approximate \ts, improving upon the $m^{1+o(1)}$-time algorithm of Sherman~\cite{Sherman17}. For readers primarily interested in the sequential setting, we further optimize our parameters to the following. Note that the best algorithm for the closely-related \mf problem \cite{Peng} requires $O(m\log^{41}n)$ time in comparison. Our algorithm is also technically considerably simpler than the \mf algorithm, and may serve as a gentler introduction to readers new to continuous optimization methods in graph algorithms.
\begin{restatable}[Sequential \ts]{theorem}{Main}\label{thm:main}
There is an algorithm that, given an undirected graph with nonnegative weights and 
 polynomial aspect ratio, computes a $(1+\e)$-approximation to minimum transshipment in time $O( (m\log^{10}n + n\log^{15}n) \, \e^{-2} \, (\log\logn)^{O(1)})$.
\end{restatable}

\subsection{Our Techniques}

Our recursive algorithm is inspired by a similar recursive algorithm by Peng~\cite{Peng} for \mf. It is instructive to compare our result to that of Peng~\cite{Peng}, the first $\tO(m)$ time\footnote{Throughout the paper, we use the standard $\tO(\cd)$ notation to hide polylogarithmic factors in the running time.} algorithm for $(1-\e)$-approximate maximum flow.\footnote{Note that \mf and minimum \ts are closely related: for graph incidence matrix $A$ and a diagonal matrix $C$ capturing the edge capacities/costs, and for a given demand vector $b$, the \mf problem is equivalent to $\min \norm f_\infty$ subject to $Af=b$, and the minimum \ts problem is exactly $\min\1f$ subject to $Af=b$.} Peng~\cite{Peng} uses an oblivious routing scheme for \mf that achieves $\polylog(n)$-approximation, but requires $\polylog(n)$ calls to $(1-\e)$-maximum flow~\cite{RST}. This oblivious routing scheme produced a chicken-and-egg situation for \mf and oblivious routing, since each one required calls to the other. Peng's main contribution is breaking this cycle, by allowing the oblivious routing to call maximum flow on sufficiently smaller-sized graphs to produce an efficient recursive algorithm. Here, we adopt a similar recursive approach, cycling through the problems of shortest path, minimum \ts, oblivious routing, and $\el_1$-embedding.

\para{Step 1: reduce to transshipment.}
The first step of the algorithm is to reduce the approximate SSSP problem to the approximate \emph{minimum transshipment} problem, which was previously done in \cite{becker} for various other computational models. Making it work in the PRAM model requires a little more care, and for completeness, we provide a self-contained reduction in \Cref{sec:esssp,sec:sample-tree}. 

Note that if we were in the exact case, then the reduction would be immediate: there is a straightforward reduction from exact SSSP\ to exact \ts: set $-(n-1)$ demand on the source vertex and $+1$ demand on the rest, and from the transshipment flow we can recover the exact SSSP relatively easily. However, in the approximate case, an approximate \ts solution in the same reduction only satisfies distances on ``average''. \cite{becker} handles this issue through $O(\logn)$ calls to approximate \ts with carefully and adaptively constructed demands on each call; we use $O(\log^2n)$ calls instead with a more sophisticated reduction.

\para{Step 2: $\ell_1$-oblivious routing.}
Sherman's framework reduces the problem of approximate \ts to that of \emph{$\el_1$-oblivious routing}, which we define later. For readers familiar with traditional oblivious routing (for \mf) which we will henceforth call $\el_\infty$-oblivious routing, the $\el_1$-oblivious routing problem is the same except the cost is measured by the total (weighted) \emph{sum} of congestions on edges rather than the maximum congestion. and our main technical contribution of the paper. 

Just like $\el_\infty$-oblivious routing is harder than \mf, $\el_1$-oblivious routing is harder than \ts. However, the benefit to considering $\el_1$- or $\el_\infty$-oblivious routing is that only a $\pl(n)$-approximate solution is needed to obtain a $(1+\e)$-approximate \mf or \ts solution; that is, Sherman's framework can be thought of as \emph{boosting} the error from $\pl(n)$ to $(1+\e)$ (at the expense of solving a harder problem).

In \cite{Sherman17}, Sherman uses his framework to solve $(1+\e)$-approximate \ts in $m^{1+o(1)}$ sequential time by providing an $m^{o(1)}$-approximate $\el_1$-oblivious routing scheme that runs in $m^{1+o(1)}$ time. Our main technical contribution in this entire paper is providing an improved scheme that is $\pl(n)$-approximate and runs in $m\,\pl(n)$ time. Like Sherman, our algorithm requires an initial \emph{$\el_1$-embedding} of the vertices of the graph to establish some geometric structure on the vertices.

The $\el_1$-oblivious routing algorithm is mostly self-contained and has no relation to the parallel sections of the paper. We therefore isolate it in its own section, \sec{seq}, for the convenience of readers primarily interested in transshipment in the sequential setting.

\para{Step 3: $\el_1$-embedding and ultra-sparsification.}
If we could compute an $\el_1$-embedding with $\pl(n)$ distortion, then we would be done. (We can obtain $O(\logn)$ dimensions for free with Johnson-Lindenstrauss dimension reduction.) Unfortunately, while this task is simple to compute sequentially, no work-efficient parallel algorithm is known. This is because the popular algorithms that compute $\el_1$-embeddings sequentially all require distance computations as subroutines, and no work-efficient parallel algorithm for SSSP is known. (If one were known, then there would be no need for our result in the first place!)

Recall that we sought out to solve SSSP, and currently, our $\el_1$-embedding problem \emph{requires} an SSSP routine on its own. This is where Peng's key insight comes to play: while recursing naively on the same graph will not work (since it would loop endlessly), if we can recurse on \emph{sufficiently smaller} graphs, then the recursion analysis would produce an algorithm with the desired running time. This is indeed Peng's approach for \mf: he makes one \mf instance call $\el_\infty$-oblivious routing, which in turn calls maximum flow a number of times, but ensures that the total size of the recursive calls is at most half the size of the original graph. The recursion then works out to roughly $T(m) = \sum_iT(m_i)+\tO(m)$ where $\sum_im_i\le m/2$, which solves to $T(m)=\tO(m)$.

How does Peng achieve the reduction in size? Instead of computing $\el_\infty$-oblivious routing in the original graph $G$, he first \emph{(edge-)sparsifies} $G$ into a graph $H$ on $n$ vertices and $(n-1)+O(\f m{\pl(n)})$ edges by computing an \emph{ultra-sparsifier} of the graph \cite{UltraSpanner}. This is a graph that is so sparse that it is almost ``tree-like'' (at least when $m=\tO(n)$). Of course, this alone might not achieve the desired size reduction, for example if $m\approx n$. Therefore, he next \emph{vertex-sparsifies} $H$ into a graph $H'$ with $O(\f m{\pl(n)})$ vertices and $O(\f m{\pl(n)})$ edges using a $j$-tree construction of Madry~\cite{Madry10}. He now calls $\el_\infty$-oblivious routing on $H'$ (instead of $G$), which again calls maximum flow, but this time on graphs of small enough size (w.r.t.\ the original graph) to make the recursion work out. Moreover, by the properties of the ultra-sparsifier and the vertex-sparsifier, a $\pl(n)$-approximate $\el_\infty$-oblivious routing scheme for $H'$ is also a $\pl(n)$-approximate $\el_\infty$-oblivious routing scheme for $G$ (that is, the approximation suffers an extra $\pl(n)$ factor). The specific $\pl(n)$ factor does not matter at the end, since in Sherman's framework, any $\pl(n)$ factor is sufficient to boost the error to $(1+\e)$ for \mf at an additional \emph{additive} cost of $\tO(m)$.

Our approach is similar, but adapted from $\el_\infty$/\mf to $\el_1$/\ts. The $\el_1$-analogy of an ultrasparsifier has been studied previously by Elkin~and~Neiman~\cite{Elkin}, who coined the term \emph{ultra-sparse spanner}; in this paper, we will use \emph{ultra-spanner} instead to emphasize its connection to ultra-sparsifiers. Instead of running $\el_1$-embedding on $G$, we compute an ultra-spanner $H$, and then vertex-sparsify it in the same manner as Peng; again, the resulting graph $H'$ has $O(\f m{\pl(n)})$ vertices and edges. We then run $\el_1$-embedding on $H'$, making calls to (approximate) SSSP on graphs of much smaller size. It turns out that approximate SSSP works for the $\el_1$-embedding algorithm that we use, provided that the distances satisfy a certain triangle inequality condition that our SSSP algorithm obtains for free.

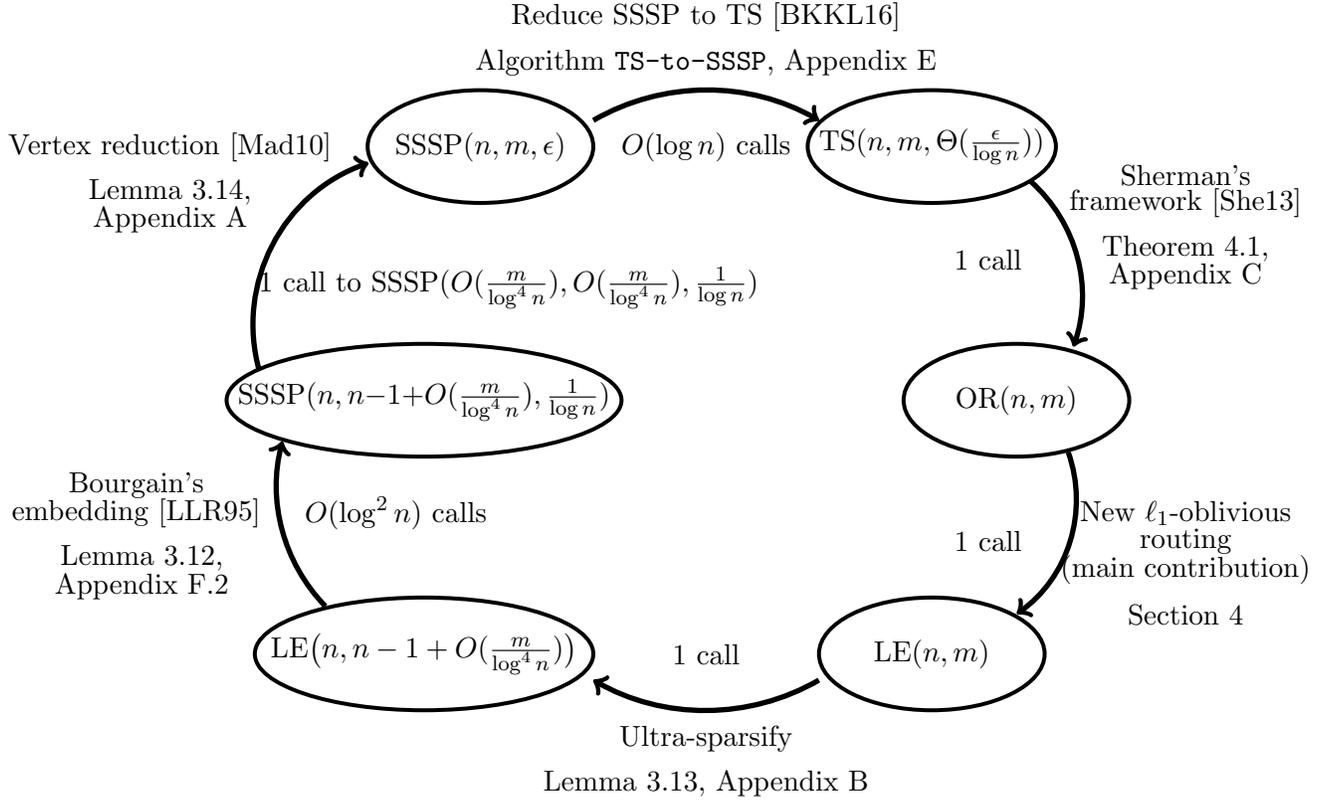
\begin{figure}\centering
\begin{tikzpicture}[scale=.75]
\tikzstyle{every node} = [scale=1]

\draw[line width=1.5]  (-4,4) ellipse (2 and 1);
\draw[line width=1.5]  (4,4) ellipse (2.2 and 1);
\draw[line width=1.5] (-5,-5) ellipse (3 and 1);
\draw[line width=1.5]  (4,-5) ellipse (2 and 1);
\node at (-4,4) {SSSP$(n,m,\epsilon)$};
\node at (4,4) {TS($n,m,\Theta(\frac{\epsilon}{\log n}))$};
\node[scale=1] at (-5,-5) {LE$\big(n,n-1+O(\frac m{\log^4n})\big)$};
\node at (4,-5) {LE$(n,m)$};
\node at (-9.5,3.2) {\lem{ultra-S},};
\node at (-9.5,2.7) {\sec{ultra-S}};
\node at (0,6.3) {Reduce SSSP to TS~\cite{becker}};
\node at (0,5.5) {Algorithm~\ref{sssp}, \sec{sample-tree}};
\node at (5,2) {$1$ call};
\node at (5,-3) {$1$ call};
\node at (0,-5) {$1$ call};
\node at (0,-7.3) {\thm{ultra}, \sec{ultra}};
\draw[->, line width=2] (-2,4.4641) arc (120:60:4);
\draw [->, line width=2](2,-5.4641) arc (-60:-120:4);
\node at (0,4) {$O(\log n)$ calls};
\node at (0,-6.5) {Ultra-sparsify};
\node at (-3.5,1.5) {$1$ call to SSSP$(O(\frac m{\log^4n}),O(\frac m{\log^4n}),\frac1{\log n})$};
\draw [line width=1.5] (5.5,-0.5) ellipse (2 and 1);
\draw [line width=1.5] (-5,-0.5) ellipse (3.5 and 1);
\node at (5.5,-0.5) {OR$(n,m)$};
\node at (-5,-0.5) {SSSP$(n,n{-}1{+}O(\frac m{\log^4n}),\frac1{\log n})$};
\node at (8.5,3.5) {Sherman's};
\node at (8.5,3) {framework \cite{Sherman13}};
\node at (8.5,2.2) {\thm{sherman},};
\node at (8.5,1.7) {\sec{mwu}};
\node at (8.5,-3.5) {(main contribution)};
\node at (8.5,-4.3) {\sec{seq}};
\node at (8.5,-2.5) {New $\ell_1$-oblivious};
\node at (8.5,-3) {routing};
\node at (-10.1,-2) {Bourgain's};
\node at (-10.1,-2.5) {embedding \cite{linial}};
\node at (-5.5,-2.5) {$O(\log^2n)$ calls};
\node at (-10,-3.3) {\lem{embed},};
\node at (-10,-3.8) {\sec{om27}};
\node at (-9.5,4) {Vertex reduction \cite{Madry10}};
\draw [->,line width=2](5.7569,3.3783) arc (48.1795:-19.3119:2.7203);
\draw [->,line width=2](6.4185,-1.4232) arc (19.5206:-54.7719:2.4906);
\draw [->,line width=2](-6.7483,-4.1547) arc (-135.8461:-194.9314:3.0635);
\draw [->,line width=2](-7.922,0.028) arc (-164.4981:-251.0418:3.0339);
\end{tikzpicture}
\caption{Our recursive approach, inspired by \cite{Peng}'s for \mf. SSSP$(n,m,\e)$ is the work required to compute $(1+\e$)-approximate SSSP (on a graph with $n$ vertices and $m$ edges) that satisfies a certain triangle inequality condition that we omit here. TS$(n,m,\e)$ is the work required to compute $(1+\e)$-approximate \ts. OR$(n,m)$ is the work required to compute a $\pl(n)$-approximate $\el_1$-oblivious routing (matrix), and LE$(n,m)$ is the work required to compute an $\el_1$-embedding in $O(\logn)$ dimensions with at most $\pl(n)$ distortion.
}
\label{fig:recursion}
\end{figure}

\subsection{Concurrent Work: \cite{andoni}}

At around the same time, Andoni, Stein, and Zhong \cite{andoni} obtained similar results, solving approximate shortest paths in the PRAM\ model in $\tO(m)$ work and $\pl(n)$ time. Their algorithm only computes an $(1+\e)$-approximate $s$---$t$ path, while ours works for full SSSP. Both algorithms are based on transshipment and Sherman's framework, but they deviate in the general algorithmic structure. Our routing algorithm recursively calls itself and reduces the sizes of the recursive calls via ultra-spanners, while they develop the concepts of subemulators and compressed preconditioners to construct the oblivious routing (or preconditioner) directly.

While both results were arXived on the same day, our result had a nontrivial omission that was patched a month and a half later. The author acknowledges that the work of Andoni et al.\ was completed first, and ours is a chronologically later but slightly stronger result. See the Acknowledgments for more details.

\subsection{Related Work}

There has been a recent trend in applying continuous optimization techniques to graph optimization problems in the sequential setting, bypassing long-standing running time barriers to the purely combinatorial approaches. The first such algorithm is due to Daitch and Spielman~\cite{Daitch}, who combined the Laplacian/SDD solvers of Spielman and Teng~\cite{ST} with recent developments in interior point methods~\cite{Renegar,Ye} to provide a faster algorithm for minimum-cost flow with small capacities. For general minimum cost flow, Lee~and~Sidford~\cite{lee2014path} used interior point methods to obtain the current best $\tO(m\sr n)$ time.

Christiano et al.~\cite{christiano2011electrical} combine Laplacian solvers with electrical flows to obtain an $\tO(m+n^{4/3}\poly(1/\e))$ time algorithm for $(1-\e)$-approximate maximum flow. Sherman~\cite{Sherman13} obtained the first $m^{1+o(1)}\poly(1/\e)$-time algorithms for $(1-\e)$-approximate maximum flow, using an algorithm that combines oblivious routing~\cite{Racke} with steepest descent methods. This running time was also obtained independently by Kelner et al.~\cite{KLOS}, and then improved to $\tO(m\e^{-2})$ by Peng~\cite{Peng}, and then $\tO(m\e^{-1})$ by Sherman~\cite{ShermanArea}. Sherman~\cite{Sherman17} also investigated the minimum transshipment problem and obtained an $m^{1+o(1)}$ time $(1+\e)$-approximation, again through oblivious routing. Until now, an $\tO(m)$ time algorithm for this problem remained open. Cohen~et~al.~\cite{cohen2017negative} obtained the fastest algorithm for negative-length shortest paths in $\tO(m^{10/7}\log W)$ time for graphs whose weights have absolute value at most $W$. Other notable results for flow-related problems include~\cite{madry2013navigating,becker2013combinatorial}.

Continuous optimization techniques have also seen recent success under the parallel computation setting, due to its inherent parallelism. One significant result in this area is the parallel Laplacian/SDD solvers of~\cite{PS}, which run in $\tO(m)$ work and $\polylog(n)$ (parallel) time and has applications to approximate flow problems in parallel. In addition, the algorithms of Sherman~\cite{Sherman13,Sherman17} can be adapted to the parallel setting, producing $(1\pm\e)$-approximate maximum flow and minimum transshipment in $m^{1+o(1)}$ work and $m^{o(1)}$ time.

For the parallel single source shortest path problem itself, there were a few early results~\cite{klein1992parallel,klein1997randomized} before the breakthrough algorithm of Cohen~\cite{Cohen}, which computes $(1+\e)$-approximate solutions to undirected, weighted graphs in $O(m^{1+\epsilon_0})$ work and $\text{polylog}(n)$ time. The running time was improved slightly in~\cite{Elkin} with a better dependence of $\pl(n)$. The algorithm of \cite{klein1997randomized} also works on \emph{directed} graphs as well, as does the more recent~\cite{forster2018faster}. 




\subsection{Self-containment}
Last but not least, it is worth mentioning that the results in this paper are almost entirely \emph{self-contained}. Throughout the paper, we make an effort to re-prove known theorems for the sake of self-containment and to improve readability. Most notably, we simplify Sherman's continuous optimization framework for \ts through \emph{multiplicative weights update (MWU)}, with only slightly worse guarantees, in \sec{mwu}. (In contrast, Sherman's framework is centered around the technically heavy $\el_1$-steepest descent of Nesterov~\cite{nesterov2005smooth}.) We believe that with our self-contained treatment, the \ts problem serves as a smoother introduction to continuous optimization techniques in graph algorithms compared to its sister problem \mf. Indeed, the companion $(1+\e)$-approximate maximum flow result of Peng~\cite{Peng} builds upon a multitude of previous work \cite{RST,Sherman13,KMP,abraham2012using}, many of which are technically dense and lead to a steep learning curve. Thus, we hope that the relative simplicity and self-containment of this paper will have broad appeal to interested readers outside the area.

\subsection{Organization}

In \sec{par}, we introduce the high-level components of our recursive  parallel algorithm (see \Cref{fig:recursion}), leaving the details to later sections and the appendix.

\sec{seq} is focused exclusively on the sequential transshipment result (\thm{main}). The algorithm is almost completely self-contained, save for Sherman's framework and an initial $\el_1$-embedding step (which can be computed quickly sequentially~\cite{linial}). It has nothing deferred to the appendix in an attempt to make it a standalone section for readers primarily interested in \thm{main}.



\section{Preliminaries}

All graphs in the paper are undirected and (positively) weighted, with the exception of \sec{esssp}, where directed graphs and edges of zero weights are defined explicitly. Given a graph $G$, we define $V(G)$ and $E(G)$ as the vertices and edges of the graph. For two vertices $u,v\in V(G)$, we define $d_G(u,v)$ as the (weighted) distance between $u$ and $v$ in $G$; if the graph $G$ is clear from context, we sometimes use $d(u,v)$ instead.

\subsection{PRAM Model}


Our PRAM model is based off of the one in~\cite{fineman2018nearly}, also called the \emph{work-span} model. An algorithm in the PRAM model proceeds identically to a sequential algorithm except for the addition of the parallel foreach loop. In a parallel foreach, each iteration of the loop must run independently of the other tasks, and the parallel algorithm may execute all iterations in parallel instead of sequentially. The \emph{work} of a PRAM algorithm is the same as the sequential running time if each parallel foreach was executed sequentially instead. To determine the \emph{time} of the algorithm, for every parallel foreach, we calculate the maximum sequential running time over all iterations of the loop, and sum this quantity over all parallel foreach loops. We then add onto the total the sequential running time outside the parallel foreach loops to determine the total time. There are different variants of the PRAM model, such as the binary-forking model and the unlimited forking model, that may introduce additional overhead in foreach loops. However, these all differ by at most polylogarithmic factors in their work and span, which we always hide behind $\tO(\cd)$ notation, so we do not concern ourselves with the specific model.



\subsection{Transshipment Preliminaries}
The definitions below are central for our sequential \ts algorithm (\thm{main}, \sec{seq}) and are also relevant for the parallel algorithms.
\BD[Transshipment]
The \emph{minimum transshipment} problem inputs a (positively) weighted, undirected graph $G=(V,E)$, and defines the following auxiliary matrices:
\BE
\im Incidence matrix $A\in\R^{V\times E}$: for each edge $e=(u,v)$, the column of $A$ indexed by $e$ equals either $\mathbbm1_u-\mathbbm1_v$ or $\mathbbm1_v-\mathbbm1_u$.
\im Cost matrix $C\in\R^{E\times E}$: a diagonal matrix with entry $C_{e,e}$ equal to the weight of edge $e$.
\EE

In a \emph{transshipment instance}, we are also given a \emph{demand vector} $b\in\R^V$ satisfying $\mathbbm1^Tb=0$.
\ED

Consider now the LP formulation for minimum \ts: $\min\1{Cf}:Af=b$, and its dual, $\max b^T\phi:\norm{C\inv A^T\phi}_\infty\le1$. Let us define the solutions to the primal and dual formulations as \emph{flows} and \emph{potentials}:

\BD[Flow]
Given a transshipment instance, a $\emph{flow vector}$ (or \emph{flow}) is a vector $f\in\R^E$ satisfying the \emph{primal constraints} $Af=b$, and it has \emph{cost} $\1{Cf}$. The flow minimizing $\1{Cf}$ is called the \emph{optimal flow} of the transshipment instance. For any $\al\ge1$, an \emph{$\al$-approximate flow} is a flow whose value $\1{Cf}$ is at most $\al$ times the minimum possible (over all flows).
\ED
\BD[Potential]\defnl{pot}
Given a transshipment instance, a \emph{set of potentials} (or \emph{potential}) is a vector $\phi\in\R^V$ satisfying the \emph{dual constraints} $\norm{C\inv A^T\phi}_\infty\le1$. The potential maximizing $b^T\phi$ is called the \emph{optimal potential} of the transshipment instance.
\ED
For convenience, we will treat potentials as functions on $V$; that is, we will use the notation $\phi(v)$ instead of $\phi_v$.
\BD[Flow-potential pair]
For any flow $f\in\R^E$ and potential $\phi\in\R^V$, the pair $(f,\phi)$ is called a \emph{flow-potential pair}. For $\al\ge1$, $(f,\phi)$ is an \emph{$\al$-approximate flow-potential pair} if $\1{Cf}\le\al\,b^T\phi$.
\ED

\BF
If $(f,\phi)$ is an $\al$-approximate flow-potential pair, then $f$ is an $\al$-approximate flow.
\EF
\BP
Let $f^*$ be the optimal flow. The two LPs $\min \1{Cf}: Af=b$ and $\max b^T\phi:\norm{C\inv A^T\phi}_\infty\le1$ are duals of each other, so by (weak) LP duality, the potential $\phi$ satisfies $b^T\phi\le\1{Cf^*}$. Since $(f,\phi)$ is an $\al$-approximate flow-potential pair, we have $\1{Cf}\le\al\, b^T\phi\le\al\1{Cf^*}$.
\EP

\BD[$\opt$]
Given a transshipment problem and demand vector $b$, define $\opt(b)$ as the cost of the optimal flow of that instance, that is:
\[ \opt(b) := \min_{f:Af=b} \1{Cf} .\]
When the underlying graph $G$ is ambiguous, we use the notation $\opt_G(b)$ instead.
\ED

\subsection{Parallel Shortest Path Preliminaries}
The definitions below are confined to the parallel algorithms in the paper, so a reader primarily interested in the sequential \ts algorithm (\thm{main}, \sec{seq}) may skip these.

We first introduce a notion of approximate SSSP distances which we call \emph{approximate SSSP potentials}.

\BD[Approximate $s$-SSSP potential]\defnl{SSSP-dual}
Given a graph $G=(V,E)$ and a source $s$, a vector $\phi\in\R^V$ is an \b{$\al$-approximate $s$-SSSP potential} if:
 \BE
 \im For all $v\in V$, $\phi(v)-\phi(s) \ge \f1\al\cd d(s,v)$
 \im For each edge $(u,v)$, $|\phi(u)-\phi(v)| \le w(u,v)$.
 \EE
When the source $s$ is either irrelevant or clear from context, we may use \emph{$\al$-approximate SSSP potential} (without the $s$) instead.
\ED
Observe that the approximate SSSP potential problem is slightly more stringent than simply approximate shortest path distances: the second condition of \defn{SSSP-dual} requires that distances satisfy a sort of approximate \emph{subtractive} triangle inequality. To illustrate why this condition is more restrictive, imagine a graph on three vertices $s,u,v$, with $d(s,u)=d(s,v)=100$ and $d(u,v)=1$, and let $\al:=10/9$. Then, the distance estimates $\tilde d(s)=0$ and $\tilde d(u)=90$ and $\tilde d(v)=100$ are $\al$-approximate SSSP distances with source $s$, but the vector $\phi$ with $\phi(s)=0$ and $\phi(u)=90$ and $\phi(v)=100$ is not a $(1+\e)$-approximate SSSP potential because it violates the second condition of \defn{SSSP-dual} for edge $(u,v)$: we have $|\phi(u)-\phi(v)|=10 > w(u,v)=1$.

\BO
An $\al$-approximate $s$-SSSP potential is also an $\al$-approximate potential for the transshipment instance with demands $\sum_v(\mathbbm1_v-\mathbbm1_s)$ (but the converse is not true).
\EO

\BO\obsl{dual-upper}
Given a graph $G=(V,E)$ and a source $s$, any $\al$-approximate $s$-SSSP potential $\phi$ satisfies $|\phi(u)-\phi(v)|\le d(u,v)$ for all $u,v\in V$.
\EO
\BP
Let $u=v_0,v_1,\lds,v_\el=v$ be the shortest path from $s$ to $v$. By property~(2), we have
\[ |\phi(u)-\phi(v)| \le \bigg| \sum_{i=1}^{\el} \phi(v_i)-\phi(v_{i-1}) \bigg| \le \sum_{i=1}^\el\left|\phi(v_i)-\phi(v_{i-1})\right|\le \sum_{i=1}^\el d(v_i,v_{i-1})=d(u,v) .\]
\EP

\BO\label{obs:shift}
If $\phi$ is an $\al$-approximate $s$-SSSP potential, then $\phi+c\cd\mathbbm1$ is also one for any scalar $c\in\R$.
Therefore, we can always assume w.l.o.g.\ that $\phi(s)=0$. In that case, by property~(1), we also have $\phi(v)\ge0$ for all $v\in V$.
\EO

\BO\label{obs:max}
Given two vectors $\phi_1$ and $\phi_2$ that satisfy property~(2), the vectors $\phi_{\min},\,\phi_{\max}\in\R^V$ defined as $\phi_{\min}(v):=\min\{\phi_1(v),\phi_2(v)\}$ and $\phi_{\max}(v):=\max\{\phi_1(v),\phi_2(v)\}$ for all $v\in V$ also satisfy property~(2).
\EO

We now generalize the notion of SSSP potential to the case when the ``source'' is a subset $S\s V$, not a single vertex. Essentially, the definition is equivalent to contracting all vertices in $S$ into a single source $s$, taking an $s$-SSSP potential, and setting the potential of each vertex in $S$ to the potential of $s$.

\BD[Approximate $S$-SSSP potential]\defnl{SSSP-dual-S}
Given a graph $G=(V,E)$ and a vertex subset $S\s V$, a vector $\phi\in\R^V$ is an \b{$\al$-approximate $S$-SSSP potential} if:
 \BE
 \im[0.] For all $s\in S$, $\phi(s)$ takes the same value
 \im For all $v\in V$ and $s\in S$, $\phi(v)-\phi(s) \ge \f1\al\cd d(s,v)$
 \im For each edge $(u,v)$, $|\phi(u)-\phi(v)| \le w(u,v)$.
 \EE
When the set $S$ is either irrelevant or clear from context, we may use \emph{$\al$-approximate SSSP potential} (without the $S$) instead.
\ED

\BO\label{obs:S-SSSP}
Given a graph $G=(V,E)$ and a vertex subset $S\s V$, let $G'$ be the graph with all vertices in $S$ contracted into a single vertex $s'$. Then, if $\phi$ is an $\al$-approximate $S$-SSSP potential, then the vector $\phi'$ defined as $\phi'(v)=v$ for $v\in V\sm S$ and $\phi'(s')=\phi(s)$ for some $s\in S$ is an $\al$-approximate $s$-SSSP potential in $G'$.
\EO

Also, we will need the notion of a \emph{spanner} throughout the paper:
\BD[Spanner]
Given a graph $G=(V,E)$ and a parameter $\al\ge1$, a subgraph $H\s G$ is an \emph{$\al$-spanner} of $G$ if for all $u,v\in V$, we have $d_G(u,v)\le d_H(u,v)\le\al\, d_G(u,v)$.
\ED

\subsection{Polynomial Aspect Ratio}
Throughout the paper, we assume that the initial input graph for the approximate SSSP problem has \emph{polynomially bounded aspect ratio}, defined below:
\BD[Aspect ratio]
The \emph{aspect ratio} of a graph $G=(V,E)$ is the quantity $\f{\max_{u,v\in V}d_G(u,v)}{\min_{u,v\in V}d_G(u,v)}$.
\ED

 This assumption can be safely assumed: there is a reduction by Klein and Subramanian~\cite{klein1992parallel} (also used by Cohen~\cite{Cohen}) that transforms the $(1+\e)$-approximate SSSP problem on a graph with arbitrary, nonnegative weights to solving $(1+\e/2)$-approximate SSSP on a collection of graphs of total size $O(m\log n)$, each with polynomially bounded aspect ratio, and requiring an additional $O(m\logn)$ work and $O(\logn)$ time. Since polynomially bounded aspect ratio is a common assumption in graph optimization problems, we will not present this reduction for sake of self-containment.

Since our SSSP algorithm is recursive, and the SSSP problem that we solve is actually the (slightly more general) SSSP potential problem, we do not apply the reduction of Klein and Subramanian again in each recursive call. Rather, we take some care to show that the aspect ratio does not blow up over recursive calls.

For the $\el_1$-embedding and \ts problems, we will handle the aspect ratio issue differently. For the $\el_1$-embedding problem, we will explicitly require that the input graph has aspect ratio at most $n^C$ for some fixed constant $C$ (which can be made arbitrarily large). In particular, this assumption translates over in our theorem statement for parallel $\el_1$-embedding (\thm{t3}). For the transshipment problem, we will not assume that the graph has polynomial aspect ratio, but every time we recursively call \ts, we will ensure that the \emph{demand vector} has small, integral entries in the recursive instance. 
Assuming this guarantee on the demand vector, we reduce the \ts problem to the case when the graph also has polynomial aspect ratio like in the SSSP case, but here, the reduction is simple enough that we include it in the paper for completeness (\lem{aspect-ratio}).

\section{The Recursive Algorithm}\secl{par}
Our algorithm will \emph{recursively} cycle through three problems: approximate SSSP potentials, approximate \ts, and $\el_1$-embedding. For the $\el_1$-embedding and SSSP potential problems, we will always assume that the input graph has aspect ratio at most $n^C$ for some arbitrarily large but fixed constant $C>0$ (that remains unchanged throughout the recursion). The transshipment problem will require no bound on aspect ratio: we provide a simple transformation on the graph to ensure that the aspect ratio is polynomial. Let us now define the work required to solve the three problems below:

\BE
\im $\Wembed(m)$ and $\Tembed(m)$ are the work and time to $\el_1$-embed a connected graph with $m$ edges and aspect ratio at most $n^{5}$  into $O(\logn)$ dimensions with distortion $O(\log^{10.5}n)$, where the $O(\cd)$ hides an arbitrarily large but fixed constant.
\im $\WSSSP(m,\e)$ and $\TSSSP(m,\e)$ are the work and time to compute an $(1+\e)$-approximate SSSP potential of a connected graph with $m$ edges and  aspect ratio at most $\tO(n^{5})$, where the $\tO(\cd)$ hides a factor $c\log^cn$ for an arbitrarily large but fixed constant $c>0$.
\im $\WTS(m,\e)$ and $\TTS(m,\e)$ are the work and time to compute a $(1+\e)$-approximate transshipment instance of a connected graph with $m$ edges, where the demand vector $b$ is integral and satisfies $|b_v|\le n-1$ for all vertices $v$.
\EE



The following is the main result of \sec{ultra-main}:

\begin{restatable}{theorem}{ultraSSSP}\thml{ultra-SSSP}
Let $G=(V,E)$ be a connected graph with $n$ vertices and $m$ edges with aspect ratio $M$, let $\be\ge1$ be a parameter, and let $\m A$ be an algorithm that inputs (i) a connected graph on at most $m/\be$ vertices and edges with aspect ratio $\tO(\be^2M)$ and (ii) a source vertex $s$, and outputs a $(1+1/\logn)$-approximate $s$-SSSP potential. Then, there is an algorithm that computes an $\el_1$-embedding of $G$ into $O(\logn)$ dimensions with distortion $O(\be^2\log^{6.5}n)$ and calls $\m A$ at most $O(\log^2n)$ times in parallel, plus $\tO(m)$ additional work and $\pl(n)$ additional time.
\end{restatable}

\BC\clml{embed}
$\Wembed(m) \le O(\log^2n)\cd\WSSSP(\de m/\log^4n,1/\logn) + \tO(m)$ for any fixed, arbitrarily small constant $\de>0$, and $\Tembed(m)\le\TSSSP(\de m/\log^4n,1/\logn)+\pl(n)$.
\EC
\BP
Apply \thm{ultra-SSSP} with $\be:=\f1\de\log^2n$, obtaining distortion $O(\be^2\log^{6.5}n)=O(\log^{10.5}n)$.
\EP

The following is a corollary of our sequential transshipment result in \sec{seq} which constitutes our main technical contribution of the paper:
\begin{restatable}{corollary}{LOne}\corl{l1}
Given an undirected graph with nonnegative weights and polynomial aspect ratio, and given an $\el_1$-embedding of the graph with $\pl(n)$ distortion in $O(\logn)$ dimensions, there is a parallel algorithm to compute a $(1+\e)$-approximate minimum \ts instance in $\tO(m\e^{-2})$ work and $\pl(n)\e^{-2}$ time.
\end{restatable}

The following is Sherman's framework for the minimum \ts problem, for which we provide a self-contained treatment through the \emph{multiplicative weights} method in \sec{mwu}. This is where the error boosting takes place: given a lossy $\pl(n)$-approximate $\el_1$-oblivious routing algorithm encoded by the matrix $R$, we can boost the error all the way to $(1+\e)$ for \ts. The only overhead in Sherman's framework is an \emph{additive} $\tO(m)$ work and $\pl(n)$ time (where these polylogarithmic factors depend on the approximation of the $\el_1$-oblivious routing), which is ultimately what makes the recursion work out.

\begin{restatable}{theorem}{ShermanPara}
\thml{sherman-para}
Given a transshipment problem, suppose we have already computed a matrix $R$ satisfying:
  \BE
  \im For all demand vectors $b\in\R^n$,
\begin{gather}\opt(b)\le\1{Rb}\le\kappa\cd\opt(b)\eqnl{kappa}\end{gather}

  \im Matrix-vector products with $R$ and $R^T$ can be computed in $M$ work and $\pl(n)$ time\footnote{$M$ can potentially be much lower than the number of nonzero entries in the matrix $R$ if it can be efficiently compressed.}
  \EE
Then, for any transshipment instance with demand vector $b$, we can compute a flow vector $ f$ and a vector of potentials $\tilde\phi$ in $\tO(\kappa^2(m+n+M)\,\e^{-2})$ time that satisfies:
  \BE
  \im $\big\lVert Cf\big\rVert_1\le(1+\e)b^T\tilde\phi \le (1+\e)\,\opt(b)$
  \im $\opt(Af-b) \le \be\,\opt(b)$
  \EE
\end{restatable}

Lastly, there is one minor mismatch: \cor{l1} assumes that the graph has polynomial aspect ratio, while the problem for $\WTS(\cd)$ does not assume such a thing, but rather assumes that the demand vector has entries restricted to $\{-(n-1),-(n-2),\lds,n-2,n-1\}$. It turns out that given this restriction on the demand vector, the polynomial aspect ratio of the graph can be obtained for free. We defer this proof to \sec{om21}.

\begin{restatable}{lemma}{AspectRatio}\leml{aspect-ratio}
Given a transshipment instance with graph $G=(V,E)$ with $n$ vertices and $m$ edges and an integer demand vector $b$ satisfying $|b_v|\le M$ for all $v\in V$, we can transform $G$ into another graph $\wG$ on $n$ vertices and at most $m$ edges such that $\wG$ has aspect ratio at most $n^{4}M$, and $\opt_G(b) \le \opt_{\wG}(b) \le (1+1/n^2)\,\opt_G(b)$. The transformation takes $\tO(m)$ work and $\pl(n)$ time.
\end{restatable}

\BC\thml{TS}
$\WTS(m,\e) \le \Wembed(m)+ \tO(m/\e^2)$. That is, outside of an $\el_1$-embedding into $O(\logn)$ dimensions with distortion $O(\log^{10.5}n)$, the additional work to compute $(1+\e)$-approximate transshipment is $\tO(m/\e^2)$, and the additional time is $\tO(1/\e^2)$.
\EC
\BP
By assumption, the demand vector $b_v$ is integral and satisfies $|b_v|\le n-1$ for all vertices $v$. Apply \lem{aspect-ratio} with $M:=n-1$ so that the aspect ratio of the modified graph $\wG$ is at most $n^5$, which is polynomial, and the optimal solution changes by factor at most $(1+1/n^2)$.
Compute an $\el_1$-embedding of $\wG$ into $\pl(n)$ dimensions (in $\Wembed(m)$ work), and then apply \cor{l1} with approximation factor $(1+\e/2)$. The final approximation factor is $(1+1/n^2)(1+\e/2)$, which is at most $(1+\e)$ for $\e\ge\Om(1/n^2)$. (If $\e=O(1/n^2)$, then an algorithm running in time $\tO(1/\e^2)\ge\tO(n^4)$ is trivial.)
\EP

We now present the reduction from approximate SSSP to approximate \ts, partially inspired by a similar routine in \cite{becker}; for completeness, we give a self-contained proof of the reduction in \Cref{sec:esssp,sec:sample-tree} in the form of this theorem:

\begin{restatable}{theorem}{SampleTree}\thml{sample-tree}
Let $G=(V,E)$ be a graph with $n$ vertices and $m$ edges, and let $\e>0$ be a parameter. 
Given graph $G$, a source $s\in V$, and an $\el_1$-embedding of it into $O(\logn)$ dimensions with distortion $\pl(n)$, we can compute a $(1+\e)$-approximate SSSP tree and potential in additional $\tO(m/\e^2)$ work and $\tO(1/\e^2)$ time.
\end{restatable}

\BC\clml{SSSP}
$\WSSSP(m,\e)\le\Wembed(m) + \tO(m/\e^2)$ and $\TSSSP(m,\e)\le \Tembed(m) + \tO(1/\e^2)$.
\EC
\BP
This is essentially \thm{sample-tree} in recursive form.
\EP

\BC\corl{SSSP-r}
$\WSSSP(m,\e) \le O(\log^2n)\cd\WSSSP(\de m/\log^2n,1/\logn)+ \tO(m/\e^2)$ and $\TSSSP(m,\e) \le\TSSSP(\de m/\log^2n,1/\logn)+ \tO(1/\e^2)$ for any fixed, arbitrarily small constant $\de>0$. 
\EC
\BP
Follows directly from \Cref{clm:embed,clm:SSSP}.
\EP



\BC\corl{SSSP-final}
$\WSSSP(m,\e)\le\tO(m/\e^2)$ and $\TSSSP(m,\e)\le\tO(1/\e^2)$.
\EC
\BP
Observe that in the recursion of \cor{SSSP-r}, by setting $\de>0$\ small enough, the total graph size $O(\log^2n)\cd\de m/\log^2n \le m/2$ drops by at least half on each recursion level. The time bound follows immediately, and the total work is dominated by the work at the root of
the recursion tree, which is $\tO(m/\e^2)$. 
\EP

Finally, \thm{t3} follows from \Cref{clm:embed,cor:SSSP-final}, and \thm{t1} and \thm{t2} follow from the addition of \thm{sample-tree} and \thm{TS}, respectively.

\subsection{$\el_1$-Embedding from Approximate SSSP Potential}

In this section, we briefly overview our $\el_1$-embedding algorithm, which is necessary for \thm{ultra-SSSP} and hence, the reduction from $\el_1$-embedding to smaller instances of approximate SSSP potentials.
Our $\el_1$-embedding algorithm is very similar to Bourgain's embedding as presented in~\cite{linial}, except utilizing approximate SSSP instead of exact, as well as slightly simplified at the expense of several logarithmic factors. Due to its similarily, we defer its proof to \sec{om}.
\BT\thml{embed}
Let $G=(V,E)$ be a graph with $n$ vertices and $m$ edges, and let $\m A$ be an algorithm that inputs any vertex set $S\s V$ and  outputs a $(1+1/\logn)$-approximate $S$-SSSP potential of $G$. Then, there is an algorithm that computes an $\el_1$-embedding of $G$ into $O(\logn)$ dimensions with distortion $O(\log^{4.5}n)$ and calls $\m A$ at most $O(\log^2n)$ times, plus $\tO(m)$ additional work and $\pl(n)$ additional time.
\ET

We will focus our attention on a slightly different variant which we show implies \thm{embed}:

\begin{restatable}{lemma}{Embed}\leml{embed}
Let $G=(V,E)$ be a graph with $n$ vertices and $m$ edges, and let $\m A$ be an algorithm that inputs any vertex set $S\s V$ and  outputs a $(1+1/\logn)$-approximate $S$-SSSP potential of $G$. Then, there is an algorithm that computes an $\el_1$-embedding of $G$ into \textcolor{blue}{$O(\log^2n)$} dimensions with distortion \textcolor{blue}{$O(\log^3n)$} and calls $\m A$ at most $O(\log^2n)$ times, plus $\tO(m)$ additional work and $\pl(n)$ additional time.
\end{restatable}

\lem{embed} is proved in \sec{om27}.
We now show that \lem{embed} implies \thm{embed}. Since the $\el_1$ and $\el_2$ metrics are at most a multiplicative $\sr k$ factor apart in dimension $k$, the embedding of \lem{embed} has distortion $O(\log^3n)\cd\sr{O(\log^2n)}=O(\log^4n)$ in the $\el_2$ metric. Next, apply Johnson-Lindenstrauss dimensionality reduction~\cite{JL} on this set of vectors, reducing the dimension to $O(\logn)$ with a constant factor increase in the distortion. We now move back to the $\el_1$ metric, incurring another $O(\sr{\logn})$ factor in the distortion, for a total of $O(\log^{4.5}n)$ distortion.

\subsection{Sparsification and Recursion to Smaller Instances}\secl{ultra-main}
In this section, we briefly overview the main ideas behind our sparsification process in order to reduce the $\el_1$-embedding problem to approximate SSSP instances of sufficiently smaller size:
\ultraSSSP*

One key tool we will use is the concept of \emph{ultra-sparse spanners}, introduced by Elkin~and~Neiman~\cite{Elkin}. Here, we will rename them to \emph{ultra-spanners} to further emphasize their connection to \emph{ultra-sparsifiers} in \cite{KMP,Peng}. These are spanners that are so sparse that they are almost ``tree-like'' when the graph is sparse enough: a graph with $(n-1)+t$ edges for some small $t$ (say, $t=m/\pl(n)$). We will utilize the following ultra-spanner construction, which is adapted from the one of \cite{MPV}; while theirs is not ultra-sparse, we modify it to be, at the expense of an additional $k$ factor in the stretch. The ultra-spanner algorithm is deferred to \sec{ultra}.
\begin{restatable}{lemma}{ultra}\thml{ultra}
Given a weighted graph $G$ with polynomial aspect ratio and a parameter $k\ge\Om(1)$, there is an algorithm to compute a $k^2$-spanner of $G$ with $(n-1)+O(\f{m\log n}k)$ edges in $\tO(m)$ work and $\pl(n)$ time.
\end{restatable}

Why are ultra-spanners useful for us? Their key property, stated in the lemma below, is that
we can compute an $\al$-approximate SSSP potential on an ultra-spanner by recursively calling $\al$-approximate SSSP potentials on a graph with potentially \emph{much fewer vertices}. To develop some intuition on why this is possible, observe first that if a connected graph has $(n-1)$ edges, then it is a tree, and SSSP is very easy to solve on trees. If the graph has $(n-1)+t$ edges instead for some small value of $t$, then the graph is almost ``tree-like'' outside of at most $2t$ vertices: take an arbitrary spanning tree, and let these vertices be the endpoints of the $t$ edges not on the spanning tree. We want to say that the graph is ``easy'' outside a graph on $2t$ vertices, so that we can solve a SSSP problem on the ``hard'' part of size $O(t)$ and then extend the solution to the rest of the graph in an efficient manner. This is indeed our approach, and it models closely off the concept of a $j$-tree by Madry~\cite{Madry10}, which is also used in Peng's recursive \mf algorithm~\cite{Peng}.

This recursion idea can be considered a \emph{vertex-sparsification} step, following the edge-sparsification that the ultra-spanner achieves. We package the vertex-sparsification in the lemma below; while this lemma works for all $t$, the reader should imagine that $t = m/\pl(n)$, since that is the regime where the lemma will be applied. Due to its length and technical involvement, the proof is deferred to \sec{ultra-S}.

\begin{restatable}{lemma}{ultraS}\leml{ultra-S}
Let $G=(V,E)$ be a connected graph with aspect ratio $M$ with $n$ vertices and $(n-1)+t$ edges, and let $\al>0$ be a parameter. Let $\m A$ be an algorithm that inputs a connected graph on at most $70t$ vertices and edges and aspect ratio $\tO(M)$ and outputs an $\al$-approximate $s$-SSSP potential of that graph. Then, for any subset $S\s V$, we can compute an $\al$-approximate $S$-SSSP potential of $G$ through a single call to $\m A$, plus $\tO(m)$ additional work and $\pl(n)$ additional time.
\end{restatable}

We now prove \thm{ultra-SSSP} assuming \lem{ultra-S}:
\BP[Proof (\thm{ultra-SSSP})]
Invoke \thm{ultra} with $k:=C\be\logn$ for a large enough constant $C>0$, producing a spanner $H$ with $(n-1)+O(\f{m\log n}k)$
edges and stretch at most $k^2=O(\be^2\log^2n)$. Since $H$ is a spanner, we have $\min_{u,v\in V}d_H(u,v)\ge\min_{u,v\in V}d_G(u,v)$ and $\max_{u,v\in V}d_H(u,v)\le k^2\max_{u,v\in V}d_G(u,v)$, so $H$ has aspect ratio $\tO(\be^2M)$. Since $G$ is connected, we have $O(m\log n/k) \le m/(70\be)$ for $C$ large enough, so $H$ has at most $(n-1)+m/(70\be)$ edges. Then, apply \lem{ultra-S} on $H$ with $t:=m/(70\be)$, $\al:=1+1/\logn$, and the algorithm $\m A$, producing an algorithm $\m A_H$ that inputs any vertex set $S\s V$ and outputs an $(1+1/\logn)$-approximate $S$-SSSP potential on $H$ through a single call to $\m A$, plus $\tO(m)$ additional work and $\pl(n)$ additional time.

Next, apply \thm{embed} on the spanner $H$ with algorithm $\m A_H$, embedding $H$ into $O(\logn)$ dimensions with distortion $O(\log^{4.5}n)$ through $O(\log^2n)$ calls to $\m A_H$, which in turn makes $O(\log^2n)$ calls to $\m A$; the additional work and time remain $\tO(m)$ and $\pl(n)$, respectively. 

Finally, since $H$ is a spanner for $G$ with stretch $O(\be^2\log^2
n)$, the $\el_1$-embedding of $H$ with stretch $O(\log^{4.5}n)$ is automatically an $\el_1$-embedding of $G$ with distortion $O(\be^2\log^2n)\cd O(\log^{4.5}n)=O(\be^2\log^{6.5}n)$.
\EP

\section{$\el_1$-Oblivious Routing and Sequential Transshipment}
\label{sec:seq}

This section is dedicated to the sequential \ts result (\thm{main}, restated below) and constitutes our main technical contribution of the paper.
\Main*
Throughout the section, we make no references to parallel algorithms, keeping all our algorithms entirely sequential in an effort to focus solely on \thm{main}. Nevertheless, to a reader with parallel algorithms in mind, it should be clear that all algorithms in this section can be parallelized to require $\pl(n)$ parallel time. To streamline the transition to parallel algorithms in the rest of the paper, we package a parallel version of the main routine in this section in an easy-to-use statement, \cor{l1}.



\subsection{Improved $\el_1$-Oblivious Routing: Our Techniques}

The key technical ingredient in our \ts algorithm is an improved $\el_1$-oblivious routing, scheme. Our algorithm begins similarly to Sherman's~\cite{Sherman17}: compute an $\el_1$-embedding into low dimensions at a small loss in approximation. Sherman chooses dimension $O(\sr{\logn})$ and loses a $2^{O(\sr{\logn})}$ factor in the distortion, and then constructs an oblivious routing in the embedded space in time exponential in the dimension. Our oblivious routing is polynomial in the dimension, so we can afford to choose dimension $O(\logn)$, giving us $\polylog(n)$ distortion. The benefit in the $\el_1$-embedding is that we now have a nice geometric property of the vertices, which are now points in $O(\logn)$-dimensional space under the $\el_1$ metric. 

At this point, let us provide some intuition for the oblivious routing problem in $\el_1$ space. Suppose for simplicity that the dimension is $1$ (i.e., we are on the real line) and that all vertices have integer coordinates. That is, every vertex $v\in V$ is now an integer on the real line, i.e., $V\s\Z$. We will now (informally) define the problem of oblivious routing on the line:\footnote{Our formal definition of oblivious routing is in matrix notation, and is considerably less intuitive. Therefore, we hope to present enough of our intuition in this section.}
\BE
\im Our input is a set of points $V\s\Z$. There is also a function $b:V\to\R$ of demands with $\sum_{v\in V}b(v)=0$ that is \emph{unknown} to us.
\im On each step, we can choose any two points $x,y\in\Z$ and a scalar $c\in\R$, and ``shift'' $c$ times the demand at $x$ to location $y$. That is, we simultaneously update $b(x)\gets b(x)-c\cd b(x)$ and $b(y)\gets b(y)+c\cd b(x)$. We pay $c\cd b(x)\cd|x-y|$ total cost for this step. Again, we do not know how much we pay. Let an \emph{iteration} be defined as one or more such steps executed in parallel.
\im After a number of iterations, we declare that we are done. At this point, we \emph{must} be certain that the demand is $0$ everywhere: $b(x)=0$ for all $x\in\Z$. 
\im Once we are done, we learn the set of initial demands, sum up our total cost, and compare it to the optimal strategy we could have taken \emph{if we had known the demands beforehand}. We would like our cost to be comparable with this retrospective optimum. In particular, we would like to pay at most $\polylog(n)$ times this optimum.
\EE

\begin{figure}\centering
\begin{tikzpicture}[scale=.7]

\foreach \y in {0,2,4,6,8,-2} {

\foreach \x in {0,...,16}
\draw (\x, \y+0.1) -- (\x, \y-0.1) ;

\foreach \x in {0,...,8}
\draw[line width=.6] (2*\x, \y+0.15) -- (2*\x, \y-0.15) ;

\foreach \x in {0,...,4}
\draw[line width=1] (4*\x, \y+0.2) -- (4*\x, \y-0.2) ;

\draw[line width=1] (0,\y) -- (16,\y);
}

\node[below] at (0,8) {$0$};
\node[below] at (16,8) {$16$};
\node[below] at (0,6) {$0$};
\node[below] at (16,6) {$16$};
\node[below] at (0,4) {$0$};

\tikzstyle{every node} = [scale=2.27,fill,circle]

\node[green] (v1) at (5,8) {};
\node[red] (v0) at (14,8) {};
\node[green!50] (v2) at (4,6) {};
\node[green!50] (v3) at (6,6) {};
\node[red] (v6) at (14,6) {};
\node[green!75] (v4) at (4,4) {};
\node[green!25] (v5) at (8,4) {};
\node[red!50] (v7) at (12,4) {};
\node[red!50] (v8) at (16,4) {};
\node[green!37.5] (v9) at (0,2) {};
\node[green!37.5] (v10) at (8,2) {};
\node[red!75] (v11) at (16,2) {};
\node[green!56.25] (v12) at (0,0) {};
\node[red!37.5] (v13) at (16,0) {};
\node[lightgray] (v99) at (0,-2) {};

\tikzstyle{every node} = [];

\draw [->,line width=1.5] (v1) -- (v2) node[pos=.5,left] {$+\n12$};
\draw [->,line width=1.5] (v1) -- (v3) node[pos=.5,right] {$+\n12$};;
\draw [->,line width=1.5] (v3) -- (v4) node[pos=.5,left] {$+\n14$};
\draw [->,line width=1.5] (v3) -- (v5) node[pos=.5,right] {$+\n14$};;
\draw [->,line width=1.5] (v6) -- (v7) node[pos=.5,left] {$-\n12$};
\draw [->,line width=1.5] (v6) -- (v8) node[pos=.5,right] {$-\n12$};;
\draw [->,line width=1.5] (v4) -- (v9) node[pos=.6,above] {$+\n38$};
\draw [->,line width=1.5] (v4) -- (v10) node[pos=.6,above] {$+\n38$};
\draw [->,line width=1.5] (v7) -- (v10) node[pos=.6,above] {$-\n14$};
\draw [->,line width=1.5] (v7) -- (v11) node[pos=.6,above] {$-\n14$};
\draw [->,line width=1.5] (v10) -- (v12) node[pos=.6,above] {$+\n3{16}$};
\draw [->,line width=1.5] (v10) -- (v13) node[pos=.6,above] {$+\n3{16}$};
\draw [->,line width=1.5] (v13) -- (v99) node[pos=.6,above] {$-\n9{16}$};

\node at(v1){$+1$};
\node at(v0){$-1$};
\node at(v2){$+\n12$};
\node at(v3){$+\n12$};
\node at(v6){$-1$};
\node at(v4){$+\n34$};
\node at(v5){$+\n14$};
\node at(v7){$-\n12$};
\node at(v8){$-\n12$};
\node at(v9){$+\n38$};
\node at(v10){$+\n38$};
\node at(v11){$-\n34$};
\node at(v12){$+\n9{16}$};
\node at(v13){$-\n9{16}$};
\node at(v99){$0$};

\node at (5,8.8) {$5$};
\node at (14,8.8) {$14$};

\node[scale=1.5] at (-1.2,8) {$b_0{:}$};
\node[scale=1.5] at (-1.2,6) {$b_1{:}$};
\node[scale=1.5] at (-1.2,4) {$b_2{:}$};
\node[scale=1.5] at (-1.2,2) {$b_3{:}$};
\node[scale=1.5] at (-1.2,0) {$b_4{:}$};
\node[scale=1.5] at (-1.2,-2) {$b_5{:}$};
\end{tikzpicture}
\caption{Oblivious routing in $1$-dimensional ($\el_1$-) space. Here, there are only two locations with nonzero demand at the beginning: $+1$ demand at point $5$ and $-1$ demand at point $14$. The optimal routing for each $b_t$ has cost $18$, and the routing costs of iterations $t=1,2,3,4,5$ are $1,\,3,\,5,\,3$, and $9$, respectively.}\label{f}
\end{figure}
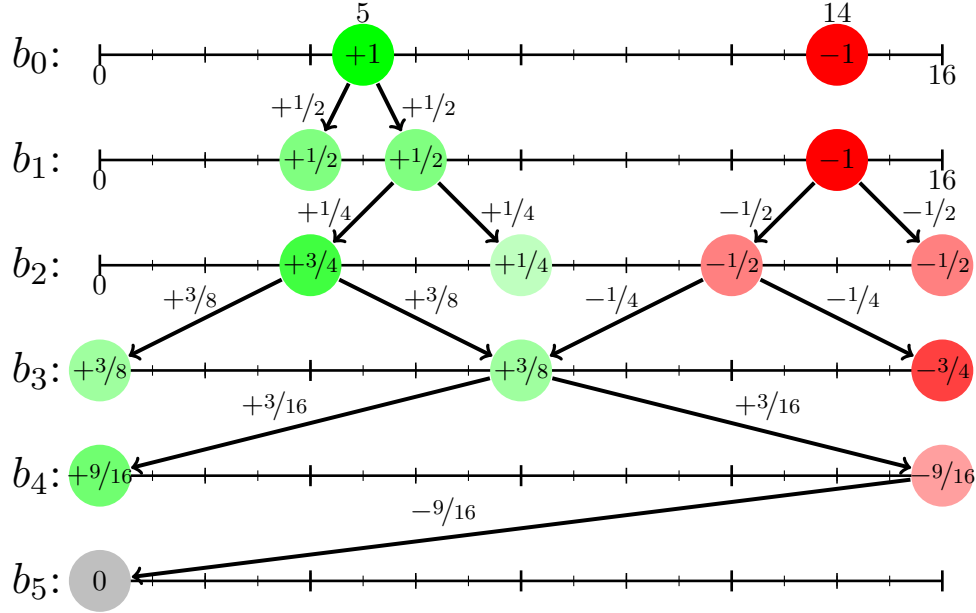

We maintain functions $b_0,b_1,b_2\lds:\Z\to\R$ that track how much demand remains at each (integer) point after each iteration. Given a demand vector (function) $b:V\to\R$, every vertex $v\in V$ has an initial demand $b_0(v):=b(v)$, and these demands sum to $0$.  Consider the following oblivious routing algorithm: for each iteration $t=1,2,\lds$, every point $x\in\Z$ with $x\equiv2^{t-1}\bmod2^{t}$ sends $b_t(x)/2$ flow to point $x-2^{t-1}$ and $b_t(x)/2$ flow to point $x+2^{t-1}$;  let $b_{t+1}(x)$ be the new set of demands (see Figure~\ref{f}).

This is actually Sherman's oblivious routing in $1$-dimensional space. He proves the following two properties of the routing:
\BE
\im After each iteration $t$, the optimal routing for the remaining points can never increase. (In Figure~\ref{f}, the optimal routing of each $b_t$ is exactly $9$.)
\im The routing cost at each iteration $t$ is at most the optimal cost of routing $b_t$. (In Figure~\ref{f}, the routing costs of iterations $t=1,2,3,4$ are $1,\,3,\,5$, and $3$, respectively.)
\EE

Let us assume that $V\s[0,1,2,\lds,n^c]$ for some constanct $c$, that is, all points in $V$ are nonnegative, polynomial-sized integers. Then, after $\lc\log_2(n^c)\rc=O(\logn)$ iterations, all points are either on $0$ or $2^{\lc\log_2(n^c)\rc}$. Thus, moving all demand from $0$ to $2^{\lc\log_2(n^c)\rc}$ finishes the oblivious routing. From the two properties above, this oblivious routing can be shown to be $O(\logn)$-competitive.

We believe this simple scheme provides a good intuition of what an oblivious routing algorithm requires. In particular, it must be \emph{unbiased}, in that demand from a given vertex must be spread evenly to the left and right. This is because we do not know where the demands lie, so our best bet is to spread equal amounts of demand left and right.

Sherman's oblivious routing extends this idea to higher dimensions. The actual routing is more complicated to describe, but as an example, on iteration $t=1$, a point $x=(1,1,1,\lds,1)$ will need to send $b(x)/2^k$ flow to each of the $2^k$ points in $\{0,2\}^k$. In other words, the running time can be exponential in the dimension.

This is where our oblivious routing algorithm deviates from Sherman's. To avoid the issue of sending flow to too many other points, we make use of random sampling: on each iteration, every point sends its flow to $\polylog(n)$ randomly chosen points close-by. These random points need to be correlated sufficiently well so that we can control the total number of points. (In particular, we do not want the number of points to increase by factor $O(\logn)$ each iteration, which would happen on a naive attempt.)

To solve this issue, we use the concept of \emph{randomly shifted grids} popular in low-dimensional computational geometric algorithms~\cite{har2011geometric}: overlay a randomly shifted grid of a specified size $W$ in the $\R^k$-dimensional space. Every point sends a fraction of its demand to (say) the midpoint of the grid cell containing it.\footnote{For notational simplicity, our algorithm will actually send to the ``lower-left'' corner of each grid, but for this section, midpoint is more intuitive to think about.} The benefit in grid shifting is that many nearby points can coalesce to the same midpoints of a grid, controlling the growth of the number of points. We compute $s=\polylog(n)$ such grids, with each point sending $1/s$ fraction to the midpoint of each grid; this is to control the variance, so that we can apply concentration bounds to show that we are still approximately unbiased from each point.

\subsection{Sherman's Framework}\label{sec:sherman}


Below, we state a paraphrased version of Sherman's framework \cite{Sherman17}. For the simplest reference, see Corollary~1 and Lemma~4 of~\cite{khesin2019preconditioning}. We also provide a proof via \emph{multiplicative weights} in \sec{mwu}, whose running time suffers an additional factor of $\log(n/\e)$ due to a binary search overhead. For \thm{main}, we will use the theorem below, while for the parallel algorithms, the weaker \thm{sherman-para} suffices.

\BT[Sherman, paraphrased]\label{thm:sherman}
Given a transshipment problem, suppose we have already computed a matrix $R$ satisfying:
  \BE
  \im For all demand vectors $b\in\R^n$, $\opt(b)\le\1{Rb}\le\kappa\cd\opt(b)$
  \im Matrix-vector products with $R$ and $R^T$ can be computed in $M$ time
  \EE
Then, for any transshipment instance with demand vector $b$, we can compute a flow vector $\tilde f$ and a vector of potentials $\tilde\phi$ in $O(\kappa^2(m+n+M)\log(m)(\e^{-2}+\log(1/\be)))$ sequential time that satisfies:
  \BE
  \im $\big\lVert C\tilde f\big\rVert_1\le(1+\e)b^T\tilde\phi \le (1+\e)\,\opt(b)$
  \im $\opt(A\tilde f-b) \le \be\,\opt(b)$
  \EE
\ET

The matrix $R$ encodes the oblivious routing algorithm. Also, intuitively, the more efficient the oblivious routing, the sparser the matrix $R$, although this relation is not as well-defined. Nevertheless, there is an equivalence between oblivious routing schemes and matrices $R$ that satisfy requirement~(1) of \Cref{thm:sherman}. But since Sherman's framework uses steepest descent methods that involve matrix algebra, a matrix $R$ with efficient matrix-vector multiplications is most convenient for the framework.

Our main technical result is computing such a matrix $R$ efficiently:
\BT[Computing $R$]\label{thm:routing-main}
Given a transshipment problem, we can compute a matrix $R$  with $O(n\log^5n(\log\logn)^{O(1)})$ nonzero entries, such that for any demand vector $b$,
\[ \opt(b)\le\1{Rb}\le O(\log^{4.5}n) \cd\opt(b) .\]
The algorithm succeeds w.h.p., and runs in $O(m\log^2n+n\log^{10}n(\log\logn)^{O(1)})$ sequential time.
\ET

With this fast routing algorithm in hand, our main theorem, restated below, follows immediately. Our proof uses low-stretch spanning trees~\cite{abraham2012using}, so for a self-contained rendition, we remark after the proof that low-stretch spanning trees can be removed at the expense of another $\logn$ factor.

\Main*
\BP
Apply \Cref{thm:sherman} with the parameters $\kappa:=O(\log^{4.5}n)$ and $M:=O(n\log^5n(\log\logn)^{O(1)})$ guaranteed by \Cref{thm:routing-main}, along with $\beta:=\Th(\e/(\logn\log\logn))$. This takes time
\[ O(\log^9n\cd(m+n\log^5n)\cd\logn\cd \e^{-2} \cd (\log\logn)^{O(1)} ) = O( (m\log^{10}n + n\log^{15}n) \cd \e^{-2} \cd (\log\logn)^{O(1)}) , \]
and outputs a flow $\tilde f$ with $\1{Cf}\le(1+\e)\,\opt(b)$
and $\opt(A\tilde f-b) \le \be\,\opt(b)$. 

To route the remaining demand $A\tilde f-b$, for $O(\logn)$ independent iterations, compute a low-stretch spanning tree in $O(n\logn\log\logn)$ time with expected stretch $O(\logn\log\logn)$~\cite{abraham2012using} and solve (exact) transshipment in linear time on the tree. In each iteration, the expected cost is at most $O(\logn\log\logn)\cd\be\,\opt(b)=\e\,\opt(b)$ for an appropriate choice of $\be$, so w.h.p., one iteration has cost at most twice the expectation. Let $f'$ be this flow, which satisfies $\1{Cf'}\le2\e\,\opt(b)$ and $Af'=A\tilde f-b$. The composed flow $\tilde f-f'$ is our final flow, which satisfies $\1{C(f-f')}\le\1{Cf}+\1{Cf'}\le(1+3\e)\,\opt(b)$ and $A(\tilde f-f')=b$. Finally, to obtain a $(1+\e)$-approximation, we can simply reset $\e\gets\e/3$.
\EP
\begin{remark}\label{rmk:sp-tree}
To eliminate the use of low-stretch spanning trees, we can set $\beta:=\e/n$ instead, picking up another $\logn$ factor in \Cref{thm:sherman}. Then, we can route the remaining demand along a minimum spanning tree, which is an $(n-1)$-approximation of optimum, or at most $(n-1)\,\beta\opt(b)\le \e\,\opt(b)$.
\end{remark}

\subsection{Polynomial Aspect Ratio}
Throughout this section, we assume that the input graph has polynomial aspect ratio, since that is assumed in \thm{main}. 

\subsection{Reduction to $\ell_1$ Metric}\label{sec:red}

The reduction to the $\ell_1$ metric is standard, via Bourgain's embedding:

\BD
For $p\ge1$, an \emph{$\ell_p$-embedding} of a graph $G=(V,E)$ with \emph{distortion} $\al$ and \emph{dimension} $k$ is a collection of vectors $\{x_v\in\R^k:v\in V\}$ such that
\[ \forall u,v\in V:\ d_G(u,v) \le \norm{x_u-x_v}_p \le \al \, d_G(u,v) . \]
\ED

\BT[Fast Bourgain's embedding]\label{thm:embed-seq}
Given a graph with $m$ edges, there is a randomized $O(m \log^2 n)$ time algorithm that computes an $\el_1$-embedding of the graph with distortion $O(\log^{1.5}n)$ and dimension $O(\log n)$.
\ET
\BPS

Apply the fast embedding algorithm of \cite{linial} in $\el_2$, which runs in $O(m\log^2n)$ randomized time and w.h.p., computes an $\el_2$-embedding of the graph with distortion $O(\logn)$ and dimension $O(\log^2n)$. Next, apply Johnson-Lindenstrauss dimensionality reduction~\cite{JL} on this set of vectors, reducing the dimension to $O(\logn)$ with a constant factor increase in the distortion. Lastly, since the $\ell_1$ and $\ell_2$ metrics are at most a multiplicative $\sr k$ factor apart in dimension $k$, this same set of vectors in $O(\logn)$ dimensions has distortion $O(\log^{1.5}n)$ in the $\el_1$ metric.
\EPS

Finally, since our input graph is assumed to have polynomial aspect ratio, so do the embedded points under the $\el_1$ metric. In particular, suppose that before applying \Cref{thm:embed-seq} we scaled the graph $G$ so that the smallest edge had length $1$. Then, the embedding satisfies the following:



\BA[Polynomial aspect ratio]\label{as:input}
For some constant $c>0$, the vectors $\{x_v:v\in V\}$ satisfy $1 \le d(u,v) \le n^{c}$ for all $u,v\in V$.
\EA

\subsection{Oblivious Routing on $\ell_1$ Metric}\label{sec:obl}

In this section, we work under the $\el_1$ metric in $O(\logn)$ dimensions (the setting established by \Cref{thm:embed-seq}) with the additional \Cref{as:input}. Our main technical result is:

\begin{restatable}{theorem}{RLone}\label{thm:R-L1}
We can compute a matrix $R$  with $O(n\log^5n(\log\logn)^{O(1)})$ nonzero entries, such that for any demand vector $b$,
\[ \opt(b)\le\1{Rb}\le O(\log^{3}n) \cd\opt(b) .\]
The algorithm succeeds w.h.p., and runs in $O(n\log^{10}(\log\logn)^{O(1)})$ sequential time.
\end{restatable}

Together with the $O(\log^{1.5}n)$ additional distortion from \Cref{thm:embed-seq}, this proves \Cref{thm:routing-main}.

Before we begin with the algorithm, we first make a reduction from ``w.h.p., for all $b$'' to the weaker statement ``for each $b$, w.h.p.''. The former requires that w.h.p., the statement holds for \emph{every} demand vector $b$, while the latter requires that for any \emph{given} demand vector $b$, the statement holds w.h.p. (Since there are uncountably many such $b$, the latter does not imply the former in general.) This simplifies our argument, since we only need to focus on a given demand vector $b$, which will often be fixed throughout a section. Before we state and prove the reduction, for each $v\in V$, let us define $\chi_v:V\to\R$ as the function that is $1$ at $v$ and $0$ elsewhere.
\BL
Suppose a randomized algorithm outputs a matrix $R$ such that for any given demand vector $b$, we have $\opt(b)\le\1{Rb}$ with probability $1$, and $\1{Rb}\le \kappa\cd\opt(b)$ w.h.p. Then, this same matrix $R$ satisfies the following stronger property: w.h.p., for any demand vector $b$, we have $\opt(b)\le\1{Rb}\le\kappa\cd\opt(b)$.
\EL
\BP
W.h.p., the matrix $R$ satisfies $\1{R(\chi_u-\chi_v)}\le\kappa\cd\opt(\chi_u-\chi_v)$ for each of the $O(n^2)$ demand vectors $\chi_u-\chi_v$ ($u,v\in V$). We claim that in this case, $R$ actually satisfies $\opt(b)\le\1{Rb}\le\kappa\cd\opt(b)$ for all demand vectors $b$.

Fix any demand vector $b$, and suppose that the flow achieving $\opt(b)$ routes $f(u,v)\ge0$ flow from $u$ to $v$ for each $u,v\in\R^k$, so that $b=\sum_{u,v}f(u,v)\cd(\chi_u-\chi_v)$ and $\opt(b)=\sum_{u,v}f(u,v)\cd\1{u-v}$.  Then, we still have $\opt(b)\le\1{Rb}$ by assumption, and for the other direction, we have
\[ \1{Rb} = \1{R\cd\sum_{u,v}f(u,v)(\chi_u-\chi_v)} \le \sum_{u,v}f(u,v)\1{R(\chi_u-\chi_v)} \le \kappa\cd\sum_{u,v}f(u,v)\,\opt(\chi_u-\chi_v) = \kappa\cd\opt(b) ,\]
as desired.
\EP

We also introduce a specific formulation of a \emph{routing} that helps in the analysis of our algorithm:

\BD[Routing]
Given a metric space $(V,d)$, a \emph{routing} is a function $R:V\times V\to\R$ such that
\[ \forall u,v\in V:\ R(u,v)=-R(v,u) .\]
A routing $R$ \emph{satisfies} demand vector $b\in\R^V$ if
\[ \forall v\in V:\ \sum_{u\in V}R(u,v) = b_v .\]
A routing $R$ has \emph{cost}
\[ \cost(R) := \sum_{u,v\in V}|R(u,v)|\cd\1{u-v} ,\]
and is \emph{optimal} for demand vector $b$ if it minimizes $\cost(R)$ over all routings $R'$ satisfying $b$.
\ED
For example, if $b=\chi_u-\chi_v$ for some $u,v\in V$, then one feasible routing (in fact, the optimal one) is $R(u,v)=1$, $R(v,u)=-1$, and $R(x,y)=0$ for all other pairs $(x,y)$, which has cost $2\1{u-v}$.

Note that $\cost(R)$ is actually \emph{twice} the value of the actual transshipment cost in the $\ell_1$ metric. However, since this notion of routing is only relevant in our analysis, and we are suffering a $\polylog(n)$ approximation anyway, we keep it this way for future simplicity.

\BO
Given a metric space $(V,d)$ and demand vector $b\in \R^V$, $\opt(b)$ is equal to the minimum value of $\f12\cost(R)$ over all routings $R$ that satisfy demand vector $b$.
\EO



We first introduce our algorithm in pseudocode below, along with the following notations. For real numbers $x$ and $W>0$, define $\lf x\rf_W:=\lf x/W\rf\cd W$ as the greatest (integer) multiple of $W$ less than or equal to $W$ (so that if $W=1$, then $\lf x\rf_W=\lf x\rf$), and similarly, define $\lc x\rc_W:=\lc x/W\rc\cd W$ as the smallest (integer) multiple of $W$ greater than or equal to $W$. 

The lines marked \emph{imaginary} are actually not executed by the algorithm. They are present to define the ``imaginary'' routings $R^*_t$, which exist only for our analysis. We could have defined the $R^*_t$ separately from the algorithm, but we decided that including them alongside the algorithm is more concise and (more importantly) illustrative.

Lastly, we remark that the algorithm does not require the input $b\in\R^V$ to be a demand vector. This observation is important when building the matrix $R$, where we will evaluate the algorithm on only the vectors $\chi_v$ for $v\in V$, which are not demand vectors.
\begin{algorithm}[H]
\caption{Routing$(V,b)$}
Input: \\
(1) $V$, a set of $n$ vectors in $\R^{k}$ satisfying \Cref{as:input}, where $k=O(\logn)$ \\
(2) $b\in\R^V$, a (not necessarily demand) vector
\begin{algorithmic}[1]
\State Initialize $w\gets\lc \log n\rc$, $s\gets \lc\log^4n \log\log n\rc$, $T\gets\lc\log_w(n^c)\rc$
\State Initialize $V_0\gets V$
\State Initialize function $b_0:\R^V\to\R$ satisfying $b_0(v)=b_v$ for all $v\in V$ and $b(x)=0$ for all $x\notin V$
\State Initialize function $R:\R^V\times\R^V\to\R$ as the zero function (i.e., $R(x,y)=0$ for all $x,y\in\R^V$)
\State Initialize $R^*_0:\R^V\times\R^V\to\R$ as the optimal routing satisfying $b_0$ \Comment{Imaginary}
\For {iteration $t=0,1,2,\lds,T$}
  \State $W \gets w^t$, a positive integer
  \State Initialize $V_{t+1}\gets\emptyset$
  \State Initialize $b_{t+1}:\R^V\to\R$ as the zero function 
  \State Initialize $R^*_{t+1}:\R^V\times\R^V\to\R$ as the zero function \Comment{Imaginary}
  \For {independent trial $j=1,2,\lds,s$}
    \State Choose independent, uniformly random real numbers $r_1,\lds,r_k\in[0,W)$
    \State Define $h_j:\R^k\to\R^k$ as $(h_j(x))_i = \lf x_i + r \rf_W$ for all $i\in[k]$ \label{line:13}
    \For {$x\in V_t: b_t(x)\ne0$} \label{line:14}
      \State $y\gets h_j(x)$
      \State $V_{t+1} \gets V_{t+1} \cup \{y\}$ \Comment{$V_{t+1}$ is the set of points with flow after iteration $t$}
      \State $R(x,y) \gets R(x,y) + b_t(x)/s$ \Comment{Send $b_t(x)/s$ flow from $x$ to $y$} \label{line:17}
      \State $R(y,x) \gets R(y,x) - b_t(x)/s$ \label{line:18}
      \State $b_{t+1}(y) \gets b_{t+1}(y) + b_t(x)/s$ \label{line:19}
    \EndFor
    \For {$(x,y)\in\R^k\times\R^k:R^*_t(x,y)\ne0$} \Comment{Imaginary} \label{line:20}
      \State $R^*_{t+1}(h_j(x),h_j(y)) \gets R^*_{t+1}(h_j(x),h_j(y)) + R^*_t(x,y)/s$ \Comment{Imaginary: move $1/s$ fraction flow
} \label{line:21}
    \EndFor
  \EndFor
\EndFor
\State Let $y\in V_{T+1}$ be arbitrary
\For {$x\in V_{T+1}:b_{T+1}(x)\ne0$}
  \State $R(x,y)\gets R(x,y)+b_{T+1}(x)$\label{line:24} \Comment{Route all demand in $V_{T+1}$ to arbitrary vertex $y$ in $V_{T+1}$}
  \State $R(y,x)\gets R(y,x)-b_{T+1}(x)$\label{line:25}
\EndFor
\end{algorithmic}
\label{alg:routing}
\end{algorithm}

\subsubsection{Proof Outline}

The purpose of the ``imaginary'' routing $R^*_t$ is to upper bound our actual cost. For $R^*_t$ to be a reasonable upper bound, it should not increase too much over the iterations. These properties are captured in the two lemmas below, proved in \Cref{sec:apx}:

\begin{restatable}{lemma}{RealCost}\label{lem:real-cost}
The total cost of routing on each iteration $t$ (lines~\ref{line:17}~and~\ref{line:18}) is at most $kw \cd \cost(R^*_{t})$.
\end{restatable}

\begin{restatable}{lemma}{Blowup}\label{lem:cost-blowup}
With probability $1-n^{-\om(1)}$,  $\cost(R^*_{t+1}) \le (1+\f1{\log n})\,\cost(R^*_t)$ for each iteration $t$.
\end{restatable}

The last routing on lines~\ref{line:24}~and~\ref{line:25} is handled in \Cref{sec:last}:

\begin{restatable}{lemma}{LastRouting}\label{lem:last}
The total cost of routing on lines~\ref{line:24}~and~\ref{line:25} is at most $O(kw)\cd \cost(R^*_{T+1})$.
\end{restatable}

The three lemmas above imply the following corollary:
\BC\label{cor:cost}
With probability $1-n^{-\om(1)}$, the total cost of routing in the algorithm is at most $O(kwT) \cd \opt(b)$.
\EC
\BP
By applying \Cref{lem:cost-blowup} inductively over all $t$, with probability $1-n^{-\om(1)}$,
\[ \cost(R^*_{t+1})\le\lp1+\f1{\logn}\rp^t\cost(R^*_0)=\lp1+\f1{\logn}\rp^t\opt(b)\le\lp1+\f1{\logn}\rp^{O(\logn)}\opt(b)=O(1)\cd\opt(b).\]
By \Cref{lem:real-cost}, the cost of routing on iteration $t$ is at most $kw\cd\cost(R^*_t)\le O(kw)\cd\opt(b)$. Summing over all $t$, we obtain a total cost of $O(kwT)\cd\opt(b)$ over iterations $0$ through $T$. Finally, by \Cref{lem:last}, the cost of routing on lines~\ref{line:24}~and~\ref{line:25} is at most $O(kw)\cd\cost(R^*_{T+1})\le O(kw)\cd\opt(b)$ as well.
\EP

At the same time, the routing should be ``sparse'', to allow for a near-linear time algorithm. Our sparsity is captured by the following lemma, proved in \Cref{sec:sparse}:

\begin{restatable}{lemma}{ComputeBasis}\label{lem:compute-basis}
For each $\chi_v$, if we run Algorithm~\ref{alg:routing} on demands $\chi_v$, every function $b_t$ has $O(s)$ nonzero values in expectation. Moreover, each function $b_t$ can be computed in $O(s^2)$ expected time.
\end{restatable}

This sparsity guarantee ensures that the matrix $R$ that we compute is also sparse, specified in the lemma below, proved in \Cref{sec:compute-R}:

\begin{restatable}{lemma}{computeR}\label{lem:compute-R}
We can compute a matrix $R$  such that $\1{Rb}$ approximates the cost of routing in Algorithm~\ref{alg:routing} to factor $O(1)$, and $R$ has $O(sTn)=O(n\log^5n(\log\logn)^{O(1)})$ nonzero entries. The algorithm succeeds w.h.p., and runs in time $O(s^2Tn\log n)=O(n\log^{10}n(\log\logn)^{O(1)})$.
\end{restatable}

Finally, with \Cref{cor:cost} and \Cref{lem:compute-basis,lem:compute-R}, we prove our main result below:

\RLone*
\BP
By \Cref{lem:compute-R}, we can compute a matrix $R$ that approximates the cost of routing in Algorithm~\ref{alg:routing} to factor $O(1)$. By \Cref{cor:cost}, this cost of routing is at most $O(kw T)\cd\opt(b)$, and it is clearly at least $\opt(b)$. Thus, $R$ approximates $\opt(b)$ by an $O(kwT)=O(\log^3n)$ factor. The requirements on $R$ are guaranteed by \Cref{lem:compute-R}.
\EP

\subsubsection{Proof of Approximation}\label{sec:apx}



We first begin with a few invariants of Algorithm~\ref{alg:routing}, whose proofs are trivial by inspection:

\begin{invariant}\label{inv:R}
At the end of iteration $t$, $R$ satisfies demand vector $b_{t+1}$.
\end{invariant}
\BP
Suppose by induction on $t$ that $R$ satisfies demand vector $b_t$ at the beginning of iteration $t$.
Recall that for $R$ to satisfy $b_{t+1}$ at the end of iteration $t$, we must have $\sum_uR(u,v)=b_{t+1}(v)$ for all $v$ by then. For each $v$, we track the change in $\sum_uR(u,v)$ and show that the total change on iteration $t$ is exactly $b_{t+1}(v)-b_t(v)$, which is sufficient for our claim.
By lines~\ref{line:17}~and~\ref{line:18}, for each $x$ satisfying $b_t(x)\ne0$ and each $j\in[s]$, the value $\sum_uR(u,h_j(x))$ increases by $b_t(x)/s$ and the value $\sum_uR(u,x)$ decreases by $b_t(x)/s$. For each $x$ with $b_t(x)\ne0$, the $s$ decreases add up to a total of $b_t(x)$. As for $b_{t+1}(v)-b_t(v)$, a demand of $b_t(v)$ is not transferred over to $b_{t+1}(v)$ if $b_t(v)\ne0$, and $b_{t+1}(v)$ increases by $b_t(x)/s$ for each $x,j$ with $h_j(x)=v$. Altogether, the differences in $\sum_u R(u,v)$ and $b_{t+1}(v)-b_t(v)$ match.
\EP

\begin{invariant}\label{inv:R*}
$R^*_{t+1}$ satisfies demand vector $b_{t+1}$.
\end{invariant}
\BP
Suppose by induction on $t$ that $R^*_{t}$ satisfies demand vector $b_t$; the base case $t=0$ is trivial. For each $v$ satisfying $b_t(v)\ne0$ and each $j\in[s]$, the value $\sum_uR^*_{t+1}(u,h_j(v))$ increases by $\sum_{x,v}R^*_t(x,v)/s$ (line~\ref{line:21}), which by induction is exactly $b_t(v)/s$. This matches the increase of $b_{t+1}(h_j(v))$ by $b_t(v)/s$ on line~\ref{line:19}.
\EP

\begin{invariant}\label{inv:mult}
For each pair $(x,y)$ with $R^*_{t+1}(x,y)\ne0$, $x-y$ has all coordinates an (integral) multiple of $w^t$.
\end{invariant}
\BP
The only changes to $R^*_{t+1}$ are the $R^*_{t+1}(h_j(x),h_j(y))$ changes on line~\ref{line:21}. By definition of $h_j$, we have that $h_j(u)-h_j(v)$ is a multiple of $W=w^t$ for all $u,v$.
\EP

\RealCost*
\BP
For each trial $j\in[s]$, by construction of $h_j(x)$ (line~\ref{line:13}), we have  $\1{(h_j(x))_i-x_i} \le kW$, which incurs a cost of at most $|b_t(x)/s| \cd kW$ in the routing $R$ (lines~\ref{line:17}~and~\ref{line:18}). Over all $s$ iterations, each $x\in\R^k$ with $b_t(x)\ne0$ is responsible for at most $|b_t(x)| \cd kW$ cost. 

We now bound $\sum_x|b_t(x)|\cd kW$ in terms of $\cost(R^*_{t})$.
By \Cref{inv:R*}, for each $x$ with $b_t(x)\ne0$,
\[ \sum_{y}|R^*_{t}(x,y)| \ge \big|\sum_{y} R^*_{t}(x,y)\big| = |b_t(x)|.\]
(Here, the summation is over the finitely many $y$ that produce a nonzero summand.) Summing over all such $x$, we get
\begin{gather}
\sum_{x:b_t(x)\ne0}|b_t(x)| \le \sum_{x:b_t(x)\ne0}\sum_{y} |R^*_{t}(x,y)| \le \sum_{x,y}|R^*_{t}(x,y)| . \label{eq:cost-by-R1}
\end{gather}
  By \Cref{inv:mult}, we have $\1{x-y}\ge w^{t-1}$ for each $(x,y)$ with $R^*_t(x,y)\ne0$. Therefore,
\begin{gather}
\cost(R^*_{t}) = \sum_{x,y}|R^*_{t}(x,y)|\cd\1{x-y} \ge \sum_{x,y}|R^*_{t}(x,y)| \cd w^{t-1} . \label{eq:cost-by-R2}
\end{gather}
Thus, the cost is at most
\begin{gather}
 \sum_x|b_t(x)| \cd kW \stackrel{(\ref{eq:cost-by-R1})}\le \sum_{x,y}|R^*_{t}(x,y)| \cd kW = kw \cd \sum_{x,y}|R^*_{t}(x,y)|\cd w^{t-1} \stackrel{(\ref{eq:cost-by-R2})}\le kw \cd \cost(R^*_{t}) . \label{eq:bound-cost-R}
\end{gather}
\EP

\BCL\label{lem:support}
For each $t\in[T+1]$, $R^*_t$ has support size $n^{O(1)}$.
\ECL
\BP
For each $t\in[0,T]$, by lines~\ref{line:20}~and~\ref{line:21}, every $(x,y)$ with $R^*_t(x,y)$ is responsible for creating at most $s\le O(\log^5n)$ nonzero values in $R^*_{t+1}$. Also, $R^*_0$ is supported in $V$, so it has support size $n^{O(1)}$. Therefore, $R^*_t$ has support size at most
\[ n^{O(1)}\cd s^{T+1} = n^{O(1)}\cd(O(\log^5n))^{O(\log n / \log\log n)} = n^{O(1)} .\]
\EP

\BL\label{lem:conc}
Fix two points $(x,y)$ with $R^*_{t+1}(x,y)\ne0$, and fix a coordinate $i\in[k]$. With probability $1-n^{-\om(1)}$, we have
\begin{gather}
\f1s\sum_{j=1}^s \left| (h_j(x))_i - (h_j(y))_i \right| \le \lp1+\f1{\logn}\rp|x_i-y_i| . \label{eq:conc}
\end{gather}
\EL
\BP
Define $\de_i:=x_i-y_i$. First, if $\de_i=0\iff x_i=y_i$, then $(h_j(x))_i=(h_j(y))_i$ with probability $1$, so both sides of (\ref{eq:conc}) are zero.

Assume now that $\de_i>0$. Throughout the proof, we recommend the reader assume $W=1$ so that $\lf x\rf_W$ is simply $\lf x\rf$, etc., since the proof is unchanged upon scaling $W$. Define $\{x\}_W:= x-\lf x\rf_W$, the ``remainder'' of $x$ when divided by $W$.

Observe that for each of the $s$ independent trials, $ (h_j(x))_i - (h_j(y))_i= \lf \de_i\rf_W$ with probability $1-\{\de_i\}_W/W$ and $(h_j(x))_i - (h_j(y))_i=\lc\de_i\rc_W$ with probability $\{\de_i\}_W/W$. In particular, $\E[(h_j(x))_i - (h_j(y))_i] = \de_i$.

For $j\in[s]$, define random variable $X_j$ as the value of $\big((h_j(x))_i - (h_j(y))_i- \lf\de_i\rf_W\big)/W$ on the $j$'th independent trial, so that $X_j \in \{0,1\}$ and $\E[X_j]=\{\de_i\}_W/W$ for all $j$. We can express the LHS of (\ref{eq:conc}) as
\begin{align}
\f1s\sum_{j=1}^s \left| (h_j(x))_i - (h_j(y))_i \right| &= \f1s\sum_{j=1}^s \big((h_j(x))_i - (h_j(y))_i \big) \nonumber
\\&= \f1s \sum_{j=1}^s \big( W\cd X_j+\lf\de_i\rf_W \big) \nonumber
\\&= \f Ws \sum_{j=1}^s X_j + \lf\de_i\rf_W  \nonumber
\\&= \f Ws \sum_{j=1}^s X_j + (\de_i - \{\de_i\}_W) . \label{eq:Xj}
\end{align}

Define $\mu:=\sum_j\E[X_j]=(s/W)\,\{\de_i\}_W$. By \Cref{inv:mult} applied to iteration $t-1$, we know that $\de_i$ is a multiple of $w^{t-1}=W/w$, so $\{\de_i\} \ge W/w$, which means $\mu \ge s/w$.
Applying Chernoff bounds on the variables $X_1,\lds,X_s \in [0,1]$ with $\e:=1/\logn$, we obtain
\[ \Pr\lb \sum_{j=1}^s X_j \ge (1+\e)\mu \rb \le \exp(-\e^2\mu/3) \le \exp\lp-\f{s}{3w\log^2n}\rp=\exp(-\om(\logn)) = n^{-\om(1)}.\]
This means that with probability $1-n^{\om(1)}$,
\begin{align*}
\f1s\sum_{j=1}^s \left| (h_j(x))_i - (h_j(y))_i \right| &\stackrel{(\ref{eq:Xj})}= \f Ws \sum_{j=1}^s X_j +(\de_i - \{\de_i\}_W) 
\\&\le \f Ws(1+\e)\mu+(\de_i - \{\de_i\}_W)
\\&= (1+\e)\{\de_i\}_W+\de_i-\{\de_i\}_W
\\&=\de_i+\e\{\de_i\}_W
\\&\le(1+\e)\de_i
\\&=\lp1+\f1{\logn}\rp|x_i-y_i| ,
\end{align*}
completing (\ref{eq:conc}).

Finally, for the case $\de_i<0$, we can simply swap $x$ and $y$ and use the $\de_i>0$ case.
\EP

\Blowup*
\BP
By lines~\ref{line:20}~and~\ref{line:21}, every $(x,y)$ with $R^*_t(x,y)\ne0$ is responsible for a total cost of
\[ \sum_{j=1}^s \f{|R^*_t(x,y)|}s\cd \1{h_j(x)-h_j(y)} = \f{|R^*_t(x,y)|}s \sum_{j=1}^s \sum_{i=1}^k | (h_j(x))_i - (h_j(y))_i | .\]
We now take a union bound over all such $(x,y)$ (at most $n^{O(1)}$ many by \Cref{lem:support}). By \Cref{lem:conc}, we have that with probability $1-n^{-\om(1)}$, the total cost is at most
\[ \f{|R^*_t(x,y)|}s \sum_{j=1}^s \sum_{i=1}^k   \lp1+\f1{\logn}\rp|x_i-y_i| = \lp1+\f1{\logn}\rp |R^*_t(x,y)| \cd \1{x-y} .\] 
Summing over all such $(x,y)$, we obtain $\cost(R^*_{t+1}) \le (1+\f1{\log n})\,\cost(R^*_t)$, as desired.
\EP

\subsubsection{Proof of Sparsity}\label{sec:sparse}

\BD[History graph; originate]
Define the \emph{history graph} $H$ to be the following digraph on vertex set $V(H):=(V_0\times \{0\})\cup (V_1\times\{1\}) \cup\cds\cup (V_{T+1}\times\{T+1\})$. For every $t\in\{0,1,\lds,T\}$ and every $x,y$ such that line~\ref{line:17} is executed at least once on $R(x,y)$, add a directed edge $((x,t),(y,t+1))$ in $H$. (By \Cref{inv:inside} below, every such $x,y$ must satisfy $x\in V_t$ and $y\in V_{t+1}$.) A vertex $(x,t)\in V(H)$ \emph{originates} from vertex $v\in V=V_0$ if there is a directed path in $H$ from $(v,0)$ to $(x,t)$.
\ED

\begin{invariant}\label{inv:inside}
For each $x$ with $b_{t+1}(x)\ne0$, we have $x\in V_{t+1}$.
\end{invariant}
\BP
Every $x\in V$ with value $b_{t+1}(x)$ modified in line~\ref{line:19} is added into $V_{t+1}$ in line~\ref{line:17}.
\EP

\begin{invariant}\label{inv:cube}
For each point $v\in V$ and point $x\in V_{t}$ where $(x,t)$ originates from $v$,
\[\forall i\in[k]:\ 0\le v_i-x_i\le \sum_{j=1}^t w^{j}.\]
\end{invariant}
\BP
We prove the statement by induction on $t$; the base case $t=0$ is trivial. For iteration $t$, for each $v,x$ where $(x,t)$ originates from $v$, we have $x_i-W < (h_j(x))_i\le x_i$ for all $i\in[k],\,j\in[s]$ by definition of $h_j$. By induction, $0\le v_i-x_i\le\sum_{j=1}^{t-1}w^j$ for all $i\in[k]$. Therefore, the points $h_j(x)\in V_{t+1}$, which also originate from $v$, satisfy
\[ v_i-(h_j(x))_i \ge v_i-x_i\ge0\qquad\text{and}\qquad v_i-(h_j(x))_i\le v_i-x_i+W\le\sum_{j=1}^{t-1}w^j+W=\sum_{j=1}^tw^j \]
for all $i\in[k]$, completing the induction.
\EP

\begin{restatable}{lemma}{ExpectedSize}\label{lem:expected-size}
For each point $v\in V$ and iteration $t\in[T+1]$, the expected number of vertices $(x,t)\in V(H)$ that originate from $v$ is $O(s)$.
\end{restatable}
\BP
Fix an iteration $t\in[T+1]$. Let $r:=\sum_{j=1}^{t-1}w^j\le2w^{t-1}$; by \Cref{inv:cube}, all points $x\in V_t$ such that $(x,t)$ originates from $v$ are within the box $B:=[v_1-r,v_1]\times[v_2-r,v_2]\times\cds\times[v_k-r,v_k]$.

For each trial $j\in[s]$, consider the set $S:=\{ h_j(x) : x\in B\}$; note that every $y$ in lines~\ref{line:17}~and~\ref{line:18} for this trial satisfies $y\in S$. We claim that this set has expected size $O(1)$. To see why, observe that for each $i\in[k]$, the value $(h_j(x))_i$ over all $x\in B$ takes two distinct values with probability $r/W$ and one value with probability $1-r/W$, and these events are independent over all $i$. Moreover, if $k'\le k$ of them take two distinct values, then $|S|\le2^{k'}$, and this happens with probability $\bn k{k'}(\f rW)^{k'}(1-\f rW)^{k-k'}$. Overall, the expected size of $|S|$ is at most
\begin{align*}
\sum_{k'=0}^k \bn k{k'}\lp\f rW\rp^{k'}\lp1-\f rW\rp^{k-k'} \cd 2^{k'} &= \lp \f rW \cd 2 + \lp1-\f rW\rp \rp^k = \lp 1+\f rW\rp^k
\\&\le \lp1+\f{2w^{t-1}}{w^t}\rp^k=\lp1+\f2{\lc\log n\rc}\rp^{O(\log n)}=O(1).
\end{align*}
Over all $s$ independent trials, the sets $S$ together capture all points $y$ such that $(y,t)$ originates from $v$. The expected number of such points $(y,t)$ is therefore at most $O(s)$.
\EP

\subsubsection{Proof of Last Routing (lines~\ref{line:24}~and~\ref{line:25})}\label{sec:last}
\LastRouting*
\BP
We can follow the proof of \Cref{lem:real-cost} to obtain (\ref{eq:bound-cost-R}), where $W:=w^{T+1}$ in this case. By \Cref{inv:cube}, for each point $v\in V$ and point $x\in V_{T+1}$ where $(x,t)$ originates from $v$, we have $\1{v-x}\le kw\sum_{j=1}^{T} w^{j} = O(kw^{T+1})$. By \Cref{as:input}, the vertices $v\in V$ are at most $n^c\le w^T$ apart from each other in $\el_1$ distance. This means that the points $x\in V_{T+1}$ are at most $O(kw^{T+1})$ apart in $\el_1$ distance. Therefore, the routing on lines~\ref{line:24}~and~\ref{line:25} has cost $C\le\sum_x|b_{T+1}(x)|\cd O(kw^{T+1})$. Combining this with (\ref{eq:bound-cost-R}) gives
\[ kw\cost(R^*) \ge  \sum_x|b_{T+1}(x)|\cd kw^{T+1} \ge \Om(C) ,\]
which means $C\le O(kw)\cd\cost(R^*_{T+1})$, as desired.
\EP

\subsubsection{Computing the Matrix $R$.}\label{sec:compute-R}

First, we can modify Algorithm~\ref{alg:routing} to construct the graph $H$ without changing the running time, since every edge added to $H$ can be charged to one execution of line~\ref{line:17}.

Now for any vector $b\in\R^V$ not necessarily satisfying $\mathbbm 1 \cd b=0$, let $R_b$ be the value of $R$ once Algorithm~\ref{alg:routing} is run on $b$. 
First, we will henceforth assume the following for simplicity:
\BA\label{as:prob1}
For each $b$, every $(x,y)$ is updated at most once in $R_{b}(x,y)$ throughout Algorithm~\ref{alg:routing}.
\EA
Intuitively, \Cref{as:prob1} is true with probability $1$ because two different randomly shifted grids in Algorithm~\ref{alg:routing} align perfectly with probability $0$. More specifically, the probability that $h_j(x)=h_{j'}(x')$ for two distinct $x,j$ and $x',j'$ (possibly not even at the same iteration) is $0$.
\BL\label{lem:conc-b}
Assuming \Cref{as:prob1}, we have that w.h.p., for each $b$ and iteration $t$,
\begin{gather}
\f14kw^t\sum_{x:b_t(x)\ne0}|b_t(x)| \le \sum_{x,y:\, b_t(x)\ne0}|R_b(x,y)|\cd\1{x-y} \le \f34kw^t\sum_{x:b_t(x)\ne0}|b_t(x)| \label{eq:relate-b}
\end{gather}
\EL
\BP
Fix some $x\in V_t$, and fix a coordinate $i\in[k]$. For each trial $j\in[s]$, the difference $x_i - (h_j(x))_i$ is a uniformly random number in $[0,W)$ (where $W:=w^t$ as before). Define random variable $X_j:=(x_i-(h_j(x))_i)/W\in[0,1]$, and $\mu:=\sum_j\E[X_j]=s/2$. Applying Chernoff bounds on the variables $X_j$ with $\e:=1/4$, we obtain
\[ \Pr\lb \bigg|\sum_{j=1}^s X_j -\mu\bigg|\ge\e\mu \rb \le \exp(-\e^2\mu/3) \le \exp\lp-\Om(s)\rp = n^{-\om(1)} .\]
Therefore, with probability $1-n^{-\om(1)}$,
\[  \sum_{j=1}^s(x_i-(h_j(x))_i)=\sum_{j=1}^s WX_i \in \lb \f s4W,\, \f{3s}4W \rb .\]
Summing over all $i\in[k]$, we obtain
\[ \sum_{j=1}^s\1{x-h_j(x)}=\sum_{j=1}^s\sum_{i=1}^k(x_i-(h_j(x))_i) \in \lb \f14ksW,\,\f34ksW \rb .\]
At this point, let us assume that every statement holds in the proof so far, which is true w.h.p. Fix a demand vector $b$;  by \Cref{as:prob1}, each term in the sum $\sum_{x,y:\, b_t(x)\ne0}|R_{b}(x,y)|\cd\1{x-y}$ appears exactly once in line~\ref{line:17}, so it must appear on iteration $t$. In particular, the terms can be exactly partitioned by $x$. Every $x$ with $b_t(x)\ne0$ contributes $\sum_{j=1}^s|b_t(x)/s|\cd\1{x-h_j(x)}$
to the sum (line~\ref{line:17}), which is within $\big[\f14kW|b_t(x)|,\f34kW|b_t(x)|\big]$. Summing over all $x$ proves (\ref{eq:relate-b}).
\EP

Therefore, by \Cref{lem:conc-b}, to estimate the final routing cost $\sum_{x,y}|R_{b}(x,y)|\cd\1{x-y}$ by an $O(1)$ factor, it suffices to compute the value
\begin{gather}
\sum_tkw^t\sum_{x:b_t(x)\ne0}|b_t(x)|. \label{eq:sum-b}
\end{gather}

\begin{remark}
The purpose of reducing to summing over the values $|b_t(x)|$ is to save a factor $s$ in the running time; if we did not care about extra $\polylog(n)$ factors, we could do without it.
\end{remark}

Assuming \Cref{as:prob1}, our goal is to construct a sparse matrix $R$ so that $\1{Rb}$ equals (\ref{eq:sum-b}). To do so, our goal is to have each coordinate in $Rb$ represent $kw^tb_t(x)$ for some $t,x$ with $b_t(x)\ne0$. This has the benefit of generalizing to general demands $b$ by the following \emph{linearity} property:

\BCL\label{clm:linear}
Every value $b_t(x)$ for $t\in\{0,1,2,\lds,T+1\}$, $x\in\R^k$ is a linear function in the entries of $b\in\R^V$.
\ECL
\BP
We show this by induction on $t$; the base case $t=0$ is trivial. For each $t>0$, the initialization $b_{t+1}$ as the zero function is linear in $b$, and by line~\ref{line:19}, each update of some $b_{t+1}(y)$ adds a scalar multiple of some $b_t(x)$ to $b_{t+1}(y)$. Since $b_{t+1}(y)$ was linear in $b$ before the operation, and since $b_t(x)$ is linear in $b$ by induction, $b_{t+1}(y)$ remains linear in $b$.
\EP

To exploit linearity, we consider the set of ``basis'' functions $R_b$ where $b=\chi_v$ for some $v\in V$. (Again, note that $\chi_v$ is not a demand vector, but we do not require that property here.)

\ComputeBasis*
\BP
We first show by induction on $t$ that if $b_t(x)\ne0$ for $x\in\R^k$, then $(x,t)$ originates from $v$; the base case $t=0$ is trivial. For each $t>0$, the only way some $b_{t+1}(y)$ is updated (line~\ref{line:19}) is if there exist $x\in\R^k$ with $b_t(x)\ne0$ and $y=h_j(x)$ for some $j\in[s]$. By induction, $x$ originates from $v$, and by definition of the history graph $H$, there is a directed edge $((x,t),(y,t+1))$ in $H$ added when line~\ref{line:17} is executed for this pair $x,y$. Therefore, there is a path from $(v,0)$ to $(y,1)$ in $H$, and $y$ also originates from $v$.

Therefore, for each $t$, the number of points $x\in\R^k$ satisfying $b_t(x)$ is at most the number of vertices $(x,t)\in V(H)$ originating from $v$, which by \Cref{lem:expected-size} is $O(s)$ in expectation. 

Finally, the functions $b_t$ can be computed by simply running Algorithm~\ref{alg:routing}. $O(s)$ time is spent for each $(x,t)$ with $b_t(x)\ne0$ (assuming the entries of $R_{\chi_v}$ are stored in a hash table), giving $O(s^2)$ expected time for each iteration $t$.
\EP

\computeR*
\BP
We run Algorithm~\ref{alg:routing} for each demand $\chi_v$ \emph{over the same randomness} (in particular, the same choices of $h_j$); define $b^{\chi_v}_t$ to be the function $b_t$ on input $\chi_v$. Let $b_t$ be the functions on input $b$. By linearity (\Cref{clm:linear}), we have that for each $t,x$,
\begin{gather}
b_t(x) = \sum_{v\in V}b(v) \cd b^{\chi_v}_t(x) . \label{eq:linearity}
\end{gather}
By \Cref{lem:compute-basis}, the functions $b_t^{\chi_v}$ for all $t,\chi_v$ can be computed in $O(s^2Tn)$ total time in expectation.

We now construct matrix $R$ as follows: for each $t,x$ with $b_t^{\chi_v}(x)\ne0$ for at least one $\chi_v$, we add a row to $R$ with value $kw^t b_t^{\chi_v}$ at each entry $v\in V$. The dot product of this row with $b$, which becomes a coordinate entry in $Rb$, is exactly
\[  \sum_{v\in V} kw^tb^{\chi_v}_t(x) \cd b(v) \stackrel{(\ref{eq:linearity})}=kw^t b_t(x) .\]
Hence, $\1{Rb}$ is exactly (\ref{eq:sum-b}), which approximates the routing cost to factor $O(1)$ by \Cref{lem:conc-b}, assuming \Cref{as:prob1} (which holds with probability $1$). Finally, by \Cref{lem:compute-basis}, $R$ has $O(sTn)$ entries in expectation.


Lastly, we address the issue that the algorithm only runs quickly in expectation, not w.h.p. Our solution is standard: run the algorithm $O(\logn)$ times, terminating it early each time if the running time exceeds twice the expectation. Over $O(\logn)$ tries, one will finish successfully w.h.p., so the final running time has an extra factor of $O(\logn)$, hence $O(s^2Tn\logn)$.
\EP

\subsection{Parallel Transshipment}\secl{trans-par}
By inspection, the entire Algorithm~\ref{alg:routing} is parallelizable in $\tO(m)$ work and $\polylog(n)$ time. The only obstacle to the entire $\el_1$-oblivious routing algorithm is the initial $\el_1$-embedding step, and the only hurdle to the final proof of \thm{main} is the final low-stretch spanning step. The latter we can handle with \Cref{rmk:sp-tree}, since minimum spanning tree can be computed in parallel with Boruvka's algorithm.   We state the following corollary below to be used in our parallel algorithms. 

\LOne*

\section*{Acknowledgments}
The author thanks Yasamin Nazari, Thatchaphol Saranurak, and Goran Zuzic for helpful discussions, and Richard Peng for useful comments on multiple drafts of the paper. The author thanks Alexandr Andoni, Cliff Stein, and Peilin Zhong for pointing out a nontrivial omission of the first version of this paper, namely an issue in rounding the transshipment flow to an ``expected'' SSSP, which was later resolved by the introduction of \sec{esssp}. This section was written with only the knowledge of their work's \emph{existence} (namely, that another group had solved parallel $s$--$t$ path), and without knowing any of their techniques or proofs. The remaining sections remain fully independent of their work.

\bibliography{ref}

\newcommand{\etalchar}[1]{$^{#1}$}
\begin{thebibliography}{CKM{\etalchar{+}}11}

\bibitem[ABP18]{hopset}
Amir Abboud, Greg Bodwin, and Seth Pettie.
\newblock A hierarchy of lower bounds for sublinear additive spanners.
\newblock {\em SIAM Journal on Computing}, 47(6):2203--2236, 2018.

\bibitem[AN12]{abraham2012using}
Ittai Abraham and Ofer Neiman.
\newblock Using petal-decompositions to build a low stretch spanning tree.
\newblock In {\em Proceedings of the forty-fourth annual ACM symposium on
  Theory of computing}, pages 395--406. ACM, 2012.

\bibitem[ASZ19]{andoni}
Alexandr Andoni, Clifford Stein, and Peilin Zhong.
\newblock Parallel approximate undirected shortest paths via low hop emulators.
\newblock {\em arXiv preprint arXiv:1911.01956}, 2019.

\bibitem[BK13]{becker2013combinatorial}
Ruben Becker and Andreas Karrenbauer.
\newblock A combinatorial o(m 3/2)-time algorithm for the min-cost flow
  problem.
\newblock {\em arXiv preprint arXiv:1312.3905}, 2013.

\bibitem[BKKL16]{becker}
Ruben Becker, Andreas Karrenbauer, Sebastian Krinninger, and Christoph Lenzen.
\newblock Near-optimal approximate shortest paths and transshipment in
  distributed and streaming models.
\newblock {\em arXiv preprint arXiv:1607.05127}, 2016.

\bibitem[CKM{\etalchar{+}}11]{christiano2011electrical}
Paul Christiano, Jonathan~A Kelner, Aleksander Madry, Daniel~A Spielman, and
  Shang-Hua Teng.
\newblock Electrical flows, laplacian systems, and faster approximation of
  maximum flow in undirected graphs.
\newblock In {\em Proceedings of the forty-third annual ACM symposium on Theory
  of computing}, pages 273--282. ACM, 2011.

\bibitem[CMSV17]{cohen2017negative}
Michael~B Cohen, Aleksander Madry, Piotr Sankowski, and Adrian Vladu.
\newblock Negative-weight shortest paths and unit capacity minimum cost flow in
  {\~o} (m 10/7 log w) time*.
\newblock In {\em Proceedings of the Twenty-Eighth Annual ACM-SIAM Symposium on
  Discrete Algorithms}, pages 752--771. SIAM, 2017.

\bibitem[Coh00]{Cohen}
Edith Cohen.
\newblock Polylog-time and near-linear work approximation scheme for undirected
  shortest paths.
\newblock {\em Journal of the ACM (JACM)}, 47(1):132--166, 2000.

\bibitem[DS08]{Daitch}
Samuel~I Daitch and Daniel~A Spielman.
\newblock Faster approximate lossy generalized flow via interior point
  algorithms.
\newblock In {\em Proceedings of the fortieth annual ACM symposium on Theory of
  computing}, pages 451--460. ACM, 2008.

\bibitem[EN18]{Elkin}
Michael Elkin and Ofer Neiman.
\newblock Efficient algorithms for constructing very sparse spanners and
  emulators.
\newblock {\em ACM Transactions on Algorithms (TALG)}, 15(1):4, 2018.

\bibitem[Fin18]{fineman2018nearly}
Jeremy~T Fineman.
\newblock Nearly work-efficient parallel algorithm for digraph reachability.
\newblock In {\em Proceedings of the 50th Annual ACM SIGACT Symposium on Theory
  of Computing}, pages 457--470. ACM, 2018.

\bibitem[FN18]{forster2018faster}
Sebastian Forster and Danupon Nanongkai.
\newblock A faster distributed single-source shortest paths algorithm.
\newblock In {\em 2018 IEEE 59th Annual Symposium on Foundations of Computer
  Science (FOCS)}, pages 686--697. IEEE, 2018.

\bibitem[HP11]{har2011geometric}
Sariel Har-Peled.
\newblock {\em Geometric approximation algorithms}.
\newblock Number 173. American Mathematical Soc., 2011.

\bibitem[JL84]{JL}
William~B Johnson and Joram Lindenstrauss.
\newblock Extensions of lipschitz mappings into a hilbert space.
\newblock {\em Contemporary mathematics}, 26(189-206):1, 1984.

\bibitem[KLOS14]{KLOS}
Jonathan~A Kelner, Yin~Tat Lee, Lorenzo Orecchia, and Aaron Sidford.
\newblock An almost-linear-time algorithm for approximate max flow in
  undirected graphs, and its multicommodity generalizations.
\newblock In {\em Proceedings of the twenty-fifth annual ACM-SIAM symposium on
  Discrete algorithms}, pages 217--226. SIAM, 2014.

\bibitem[KMP14a]{UltraSpanner}
Ioannis Koutis, Gary~L Miller, and Richard Peng.
\newblock Approaching optimality for solving sdd linear systems.
\newblock {\em SIAM Journal on Computing}, 43(1):337--354, 2014.

\bibitem[KMP14b]{KMP}
Ioannis Koutis, Gary~L Miller, and Richard Peng.
\newblock Approaching optimality for solving sdd linear systems.
\newblock {\em SIAM Journal on Computing}, 43(1):337--354, 2014.

\bibitem[KNP19]{khesin2019preconditioning}
Andrey~Boris Khesin, Aleksandar Nikolov, and Dmitry Paramonov.
\newblock Preconditioning for the geometric transportation problem.
\newblock {\em arXiv preprint arXiv:1902.08384}, 2019.

\bibitem[KS92]{klein1992parallel}
Philip~N Klein and Sairam Subramanian.
\newblock A parallel randomized approximation scheme for shortest paths.
\newblock In {\em STOC}, volume~92, pages 750--758, 1992.

\bibitem[KS97]{klein1997randomized}
Philip~N Klein and Sairam Subramanian.
\newblock A randomized parallel algorithm for single-source shortest paths.
\newblock {\em Journal of Algorithms}, 25(2):205--220, 1997.

\bibitem[LLR95]{linial}
Nathan Linial, Eran London, and Yuri Rabinovich.
\newblock The geometry of graphs and some of its algorithmic applications.
\newblock {\em Combinatorica}, 15(2):215--245, 1995.

\bibitem[LS14]{lee2014path}
Yin~Tat Lee and Aaron Sidford.
\newblock Path finding methods for linear programming: Solving linear programs
  in o (vrank) iterations and faster algorithms for maximum flow.
\newblock In {\em 2014 IEEE 55th Annual Symposium on Foundations of Computer
  Science}, pages 424--433. IEEE, 2014.

\bibitem[Mad10]{Madry10}
Aleksander Madry.
\newblock Fast approximation algorithms for cut-based problems in undirected
  graphs.
\newblock In {\em {FOCS}}, pages 245--254. {IEEE} Computer Society, 2010.

\bibitem[Mad13]{madry2013navigating}
Aleksander Madry.
\newblock Navigating central path with electrical flows: From flows to
  matchings, and back.
\newblock In {\em 2013 IEEE 54th Annual Symposium on Foundations of Computer
  Science}, pages 253--262. IEEE, 2013.

\bibitem[MPVX13]{MPV}
Gary~L Miller, Richard Peng, Adrian Vladu, and Shen~Chen Xu.
\newblock Improved parallel algorithms for spanners and hopsets.
\newblock {\em arXiv preprint arXiv:1309.3545}, 2013.

\bibitem[MPX13]{miller2013parallel}
Gary~L Miller, Richard Peng, and Shen~Chen Xu.
\newblock Parallel graph decompositions using random shifts.
\newblock In {\em Proceedings of the twenty-fifth annual ACM symposium on
  Parallelism in algorithms and architectures}, pages 196--203. ACM, 2013.

\bibitem[Nes05]{nesterov2005smooth}
Yu~Nesterov.
\newblock Smooth minimization of non-smooth functions.
\newblock {\em Mathematical programming}, 103(1):127--152, 2005.

\bibitem[Pen16]{Peng}
Richard Peng.
\newblock Approximate undirected maximum flows in o (m polylog (n)) time.
\newblock In {\em Proceedings of the twenty-seventh annual ACM-SIAM symposium
  on Discrete algorithms}, pages 1862--1867. SIAM, 2016.

\bibitem[PS14]{PS}
Richard Peng and Daniel~A Spielman.
\newblock An efficient parallel solver for sdd linear systems.
\newblock In {\em Proceedings of the forty-sixth annual ACM symposium on Theory
  of computing}, pages 333--342. ACM, 2014.

\bibitem[R{\"a}c08]{Racke}
Harald R{\"a}cke.
\newblock Optimal hierarchical decompositions for congestion minimization in
  networks.
\newblock In {\em Proceedings of the fortieth annual ACM symposium on Theory of
  computing}, pages 255--264. ACM, 2008.

\bibitem[Ren88]{Renegar}
James Renegar.
\newblock A polynomial-time algorithm, based on newton's method, for linear
  programming.
\newblock {\em Mathematical Programming}, 40(1-3):59--93, 1988.

\bibitem[RST14]{RST}
Harald R{\"a}cke, Chintan Shah, and Hanjo T{\"a}ubig.
\newblock Computing cut-based hierarchical decompositions in almost linear
  time.
\newblock In {\em Proceedings of the twenty-fifth annual ACM-SIAM symposium on
  Discrete algorithms}, pages 227--238. Society for Industrial and Applied
  Mathematics, 2014.

\bibitem[She13]{Sherman13}
Jonah Sherman.
\newblock Nearly maximum flows in nearly linear time.
\newblock In {\em 2013 IEEE 54th Annual Symposium on Foundations of Computer
  Science}, pages 263--269. IEEE, 2013.

\bibitem[She17a]{ShermanArea}
Jonah Sherman.
\newblock Area-convexity, l∞ regularization, and undirected multicommodity
  flow.
\newblock In {\em Proceedings of the 49th Annual ACM SIGACT Symposium on Theory
  of Computing}, pages 452--460. ACM, 2017.

\bibitem[She17b]{Sherman17}
Jonah Sherman.
\newblock Generalized preconditioning and undirected minimum-cost flow.
\newblock In {\em Proceedings of the Twenty-Eighth Annual ACM-SIAM Symposium on
  Discrete Algorithms}, pages 772--780. SIAM, 2017.

\bibitem[ST04]{ST}
Daniel~A Spielman and Shang-Hua Teng.
\newblock Nearly-linear time algorithms for graph partitioning, graph
  sparsification, and solving linear systems.
\newblock In {\em Proceedings of the STOC}, volume~4, 2004.

\bibitem[Ye97]{Ye}
Yinyu Ye.
\newblock {\em Interior point algorithms: theory and analysis}.
\newblock Springer, 1997.

\end{thebibliography}

\appendix

\section{Vertex Sparsification and Recursion}\secl{ultra-S}
This section is dedicated to proving the vertex-sparsification lemma, \lem{ultra-S}, restated below:

\ultraS*

\subsection{Case $S=\{s\}$ of \lem{ultra-S}}

In this section, we first prove \lem{ultra-S} for the case when $S=\{s\}$ for a single source $s\in V$ in the lemma below. We then extend our result to any set $S\s V$ in \Cref{sec:S}.

\BL\leml{ultra}
Let $G=(V,E)$ be a connected graph with $n$ vertices and $(n-1)+t$ edges, and let $\al>0$ be a parameter. Let $\m A$ be an algorithm that inputs a connected graph on at most $5t$ vertices and edges and outputs an $\al$-approximate $s$-SSSP potential of that graph. Then, for any source $s\in V$, we can compute an $\al$-approximate $s$-SSSP potential of $G$ through a single call to $\m A$, plus $\tO(m)$ additional work and $\pl(n)$ additional time.
\EL

Our approach is reminiscent of the $j$-tree construction of Madry~\cite{Madry10}, but modified to handle SSSP instead of flow/cut problems.\footnote{Essentially, according to the terminology of~\cite{Madry10}, any graph with $(n-1)+t$ edges is a $5t$-tree}

First, compute a spanning tree $T$ of $G$, and let $S_0\s V$ be the endpoints of the $t$ edges in $G-T$ together with the vertex $s$, so that $|S_0|\le2t+1$. Next, let $T_{0}$ be the (tree) subgraph in $T$ whose edges consist of the union of all paths in $T$ between some pair of vertices in $S_0$. The set $T_{0}$ can be computed in parallel as follows:
\BE
 \im Root the tree $T$ arbitrarily, and for each vertex $v\in V$, compute the number $N(v)$ of vertices in $S_0$ in the subtree rooted at $v$.
 \im Compute the vertex $v$ with maximum depth satisfying $N(v)=|S_0|$; this is the lowest common ancestor $\textsf{lca}(S_0)$ of the vertices in $S_0$.
 \im The vertices in $T_{0}$ are precisely the vertices $v$ in the subtree rooted at $\textsf{lca}(S_0)$ which satisfy $N(v)\ne0$.
\EE

Let $S_3$ be the set of vertices in $T_{0}$ whose degree in $T_{0}$ is at least $3$, and let $S := S_0\cup S_3$. Starting from $T_0$, contract every maximal path of degree-$2$ vertices disjoint from $S$ into a single edge whose weight is the sum of weights of edges on that path; let $T_1$ be the resulting tree. Since every leaf in $T_0$ is a vertex in $S_0$, and since every degree-2 vertex disjoint from $S$ is contracted, the vertex set of $T_1$ is exactly $S$. We furthermore claim the following:

\BCL\clml{4t}
$T_1$ has at most $4t$ vertices and edges.
\ECL
\BP
Let $n_1$, $n_2$, and $n_{\ge3}$ be the number of vertices in $T_1$ of degree $1$, $2$, and at least $3$, respectively. Since every leaf in $T_0$ is a vertex in $S_0$, we have $n_1\ge|S_0|$. Also, since $T_1$ is a tree, it has $n_1+n_2+n_{\ge3}-1$ edges, and since the sum of degrees is twice the number of edges, we have
\[ n_1+2n_2+3n_{\ge3} \le 2(n_1+n_2+n_{\ge3}-1) \implies n_{\ge3} \le n_1-2\le|S_0|-2 .\]
The number of vertices in $T_1$ is exactly $n_1+n_{\ge3}$, which is at most $2|S_0|-2\le4 t$. The edge bound also follows since $T_1$ is a tree.
\EP

Let $G_1$ be $T_1$ together with each edge in $G-T$ added to its same endpoints (recall that no endpoint in $G-T$ is contracted). Since $T_1$ has at most $4t$ vertices and edges by \clm{4t}, and since we add $t$ additional edges to form $G_1$, the graph $G_1$ has at most $4t$ vertices and $5t$ edges. 

Finally, let $G_0$ be $T_0$ together with each edge in $G-T$ added to its same endpoints, so that $G_0$ is exactly $G_1$ with the contracted edges expanded into their original paths. Since every edge in $G-T$ is contained in $G_0$, we have that $G-G_0$ is a forest.
We summarize our graph construction below, which will be useful in \Cref{sec:S}.
\BL\leml{G1}
Let $G=(V,E)$ be a graph with $n$ vertices and $(n-1)+t$ edges, and let $T$ be an arbitrary spanning tree of $G$. We can select a vertex set $V_0\s V$ and define the graph $G_0:=G[V_0]$ such that (i) $G-G_0$ is a forest, and (ii) we can contract degree-2 paths from $G_0$ into single edges so that the resulting graph $G_1$ has at most $4t$ vertices and $5t$ edges. The contracted edges in $G_1$ have weight equal to the total weight of the contracted path. This process takes $\tO(m)$ work and $\pl(n)$ time.
\EL

It is easy to see that the aspect ratio of $G_1$ is $O(M)$.
Now, call $\m A$ on $G_1$ with $s$ as the source (recall that $s\in S_0\s S=V(G_1)$, so it is a vertex in $G_1$), obtaining an SSSP potential $\phi_1$ for $G_1$.
It remains to extend $\phi_1$ to the entire vertex set $V$.

\subsection{Extending to Contracted Paths}

First, we extend $\phi_1$ to the vertices (of degree $2$) contracted from $T_0$ to $T_1$. More precisely, we will compute a SSSP potential $\phi_0(v)$ on the vertices in $G_0$ that agrees with $\phi_1$ on $V(G_1)$.

 Define $\phi_0(v):=\phi_1(v)$ for $v\in V(G_1)$, and for each such path $v_0,v_1,\lds,v_\el$ with $v_0,v_\el\in V(G_1)$ we extend $\phi_0$ to $v_1,\lds,v_{\el-1}$ as follows:
\[\phi_0(v_j) := \min\bigg\{ \phi_1(v_0) + \sum_{i=1}^jw(v_{i-1},v_i), \ \phi_1(v_\el)+ \sum_{i=j}^{\el-1} w(v_{i},v_{i+1}) \bigg\} ;\]
note that these values are the same if we had replaced the path by its reverse ($v_\el,v_{\el-1},\lds,v_0$) instead.

\BCL\clml{dists-S}
For all $u,v\in V(G_1)$, we have $d_{G_0}(u,v)=d_{G_1}(u,v)$.
\ECL
\BP
Observe that any simple path $P$ in $G_0$ between $u,v\in V(G_1)$ must travel entirely along any path of degree-$2$ vertices sharing an edge with $P$. Therefore, for every contracted path in $G$ that shares an edge with $P$, we can imagine contracting that path inside $P$ as well. Since paths of degree-$2$ are contracted to an edge whose weight is the sum of weights of edges along that path, the total weight of $P$ does not change. Since $P$ is now a path in $G_1$, this shows that $d_{G_1}(u,v)\le d_{G_0}(u,v)$. Conversely, any path in $G_1$ can be ``un-contracted'' into a path in $G_0$ of the same length, so we have $d_{G_0}(u,v)\le d_{G_1}(u,v)$ as well, and equality holds.
\EP

\BCL
The vector $\phi_0$ is an $\al$-approximate $s$-SSSP potential of $G_0$.
\ECL
\BP
We first prove property~(2). Since $\phi_0(v)=\phi_1(v)$ for $v\in V(G_1)$, property~(2) holds for $\phi_0$ for edges $G_1$ that were not contracted from a path in $G_0$. For an edge $(u,v)$ that was contracted, there is a contracted path $v_0,v_1,\lds,v_\el$ where $u=v_j$ and $v=v_{j+1}$ for some $j$. First, suppose that
\[ \phi_0(v_j)=\phi_1(v_0) + \sum_{i=1}^jw(v_{i-1},v_i)  \iff \phi_1(v_0) + \sum_{i=1}^jw(v_{i-1},v_i) \le \phi_1(v_\el)+ \sum_{i=j}^{\el-1} w(v_{i},v_{i+1}). \]
Then,
\[ \phi_0(v_{j+1})-\phi_0(v_j) \le \bigg(\phi_1(v_0) + \sum_{i=1}^{j+1}w(v_{i-1},v_i)\bigg) - \bigg(\phi_1(v_0) + \sum_{i=1}^jw(v_{i-1},v_i)\bigg) = w(v_j,v_{j+1}) .\]
Otherwise, if
\[ \phi_0(v_j)=\phi_1(v_\el)+ \sum_{i=j}^{\el-1} w(v_{i},v_{i+1})  \iff \phi_1(v_0) + \sum_{i=1}^jw(v_{i-1},v_i) \ge \phi_1(v_\el)+ \sum_{i=j}^{\el-1} w(v_{i},v_{i+1}), \]
then
\[ \phi_0(v_{j+1}) \le \phi_1(v_\el)+ \sum_{i=j+1}^{\el-1} w(v_{i},v_{i+1}) = \phi_1(v_\el)+ \sum_{i=j}^{\el-1} w(v_{i},v_{i+1}) - w(v_j,v_{j+1})\le0 .\]
Therefore, in both cases, $\phi_0(v)-\phi_0(u)=\phi_0(v_{j+1})-\phi_0(v_j)\le w(v_j,v_{j+1})$. For the other direction $\phi_0(v)-\phi_0(u)\le w(v_j,v_{j+1})$, we can simply swap $u$ and $v$.

We now focus on property~(1). Since $\phi_0(v)=\phi_1(v)$ for $v\in V(G_1)$, and since $d_{G_0}(u,v)=d_{G_1}(u,v)$ for $u,v\in V(G_1)$ by \clm{dists-S},  property~(1) holds for $u,v\in V(G_1)$. We now prove property~(1) for vertices $v\notin V(G_1)$.

 If $v\notin V(G_1)$, then $v=v_j$ for some path $v_0,v_1,\lds,v_\el$ contracted in $G_0$ ($v_0,v_\el\in V(G_1)$). Observe that $d_0:=d_{G_0}(s,v_0)+\sum_{i=1}^jw(v_{i-1},v_i)$ is the shortest length of any (simple) path from $s$ to $v$ that passes through $v_0$, and similarly, $d_\el:=d_{G_0}(s,v_\el)+ \sum_{i=j+1}^\el w(v_{i-1},v_i)$ is the shortest length of any (simple) path from $s$ to $v$ that passes through $v_\el$. Furthermore,  $d_{G_0}(s,v)=\min\{d_0,d_\el\}$. We have
\[ \bigg(\phi_1(v_0) + \sum_{i=1}^jw(v_{i-1},v_i)\bigg) - \phi_1(s) \ge \sum_{i=1}^jw(v_{i-1},v_i) + \f1\al \cd d_{G_0}(s,{v_0}) \ge \f1\al\bigg(\sum_{i=1}^jw(v_{i-1},v_i) + d_{G_0}(s,{v_0})\bigg) = \f{d_0}\al, \]
and similarly, 
\[ \bigg(\phi_1(v_\el)+ \sum_{i=j}^{\el-1} w(v_{i},v_{i+1})\bigg) - \phi(s) \ge \sum_{i=j}^{\el-1} w(v_{i},v_{i+1}) + \f1\al \cd d_{G_0}(s,{v_\el}) \ge \f1\al\bigg( \sum_{i=j}^{\el-1} w(v_{i},v_{i+1}) + d_{G_0}(s,{v_\el})\bigg) = \f{d_\el}\al. \]
It follows that
\begin{align*}
\phi_0(v_j) -\phi_0(s) &= \min\bigg\{ \phi_1(v_0) + \sum_{i=1}^jw(v_{i-1},v_i), \ \phi_1(v_\el)+ \sum_{i=j}^{\el-1} w(v_{i},v_{i+1}) \bigg\}
\\&\ge \min\bigg\{\f{d_0}\al,\ \f{d_\el}\al\bigg\}
\\&=\f1\al\cd d_{G_0}(s,v),
\end{align*}
proving property~(1).
\EP

\subsection{Extending to Forest Components}

It remains to extend $\phi_0$ to an SSSP potential in the original graph $G$. First, recall that all edges in $G-T$ have endpoints inside $S=V(G_1)\s V(G_0)$, which means that $G-E(G_0)$ is a forest contained in $T$. Moreover, since $G_0\cap T = T_0$ is connected, every connected component (tree) in $G-E(G_0)$ shares exactly one endpoint with $V(G_0)$ (otherwise, there would be a cycle in $T$). Therefore, any simple path between two vertices in $V(G_0)$ must be contained in $G_0$. Since $G_0$ is itself an induced subgraph of $G$, in particular with the same edge weights, we have $d_G(u,v)=d_{G_0}(u,v)$ for all $u,v\in V(G_0)$.

In addition, for each component (tree) $C$ in the forest $G-E(G_0)$ that shares vertex $r$ with $V(G_0)$ (which could possibly be $s$), any path from $s$ to a vertex in $C$ must pass through $r$. In particular, the shortest path from $s$ to a vertex $v\in C$ consists of the shortest path from $s$ to $r$ (possibly the empty path, if $r=s$) concatenated with the (unique) path in $C$ from $r$ to $v$. It follows that $d_G(s,v)=d_G(s,r)+d_C(r,v)$.

With these properties of $G$ in mind, let us extend $\phi_0$ to the potential $\phi$ on $V$ as follows: for $v\in V(G_0)$, define $\phi(v):=\phi_0(v)$, and for each connected component $C$ of $G-E(G_0)$ sharing vertex $r$ with $V(G_0)$, define $\phi(v)=\phi_0(r)+d_C(r,v)$. Since $C$ is a tree, the values $d_C(r,v)$ for each $v\in V(C)$ are easily computed in parallel.

\BCL
The vector $\phi$ is an $\al$-approximate $s$-SSSP potential of $G$.
\ECL
\BP
Since $\phi_0$ and $\phi$ agree on $V(G_0)$, and since $G[V(G_0)]$ and $G_0$ agree on their edges (including their weights), property~(1) of \defn{SSSP-dual} holds for all $v\in V(G_0)$ and property~(2) holds for all $u,v\in V(G_0)$.

Now fix a connected component  $C$ of $G-E(G_0)$ sharing vertex $r$ with $V(G_0)$. For each vertex $v\in C$, we have
\[ \phi(v)-\phi(s) = \big(\phi_0(r)+d_C(r,v) \big) - \phi_0(s) \ge \f1\al \cd d_G(s,r) + d_C(r,v) \ge \f1\al\big( d_G(s,r)+d_C(r,v)\big)=\f{d_G(s,v)}\al ,\]
proving property~(1) for vertices in $C$. For property~(2), consider an edge $(u,v)$ in $C$. Since $C$ is a tree, either $d_C(r,u)=d_C(r,v)+w(u,v)$ or $d_C(r,v)=d_C(r,u)+w(u,v)$, so in both cases,
\[ |\phi(u)-\phi(v)| = \big| \big( \phi_0(r)+d_C(r,u) \big) - \big( \phi_0(r)+d_C(r,v) \big) \big| = \big|d_C(r,u)-d_C(r,v)\big| = w(u,v) ,\]
proving property~(2).
\EP

\subsection{Generalizing to $S$-SSSP}\label{sec:S}

Of course, \lem{ultra-S} requires calls to not just $s$-SSSP, but $S$-SSSP for a vertex subset $S\s V$. In this section, we generalize the algorithm to work for $S$-SSSP for any $S\s V$.

\BL
Let $G=(V,E)$ be a connected graph with $n$ vertices and $(n-1)+t$ edges, and let $\al>0$ be a parameter. Let $\m A$ be an algorithm that inputs (i) a connected graph on at most $70t$ vertices and edges with aspect ratio $O(M)$ and (ii) a source vertex $s$, and outputs an $\al$-approximate $s$-SSSP potential of that graph. Then, for any subset $S\s V$, we can compute an $\al$-approximate $S$-SSSP potential of $G$ through a single call to $\m A$, plus $\tO(m)$ additional work and $\pl(n)$ additional time.
\EL

Let $G_0$ and $G_1$ be the graphs guaranteed from \lem{G1}, and let $V_0,\,V_1\s V$ be their respective vertex sets. Consider the set $\m C$ of connected components (trees) in $G-G_0$; for each component $C\in\m C$, let $r(C)$ be the vertex shared between $C$ and $V_0$, and let $\m C$ be the set of components $C$ with $S\cap (V(C)\sm \{r(C)\})\ne\emptyset$.  Root each component $C\in\m C$ at $r(C)$, and let $C^\up$ be the subgraph of $C$ induced by the vertices that have no ancestor in $S\cap (V(C)\sm\{r(C)\})$ (see \fig{G2}); we will first focus our attention on $C^\up$. Let $d(C):=d_C(r(C), S\cap V(C)) = d_{C^\up}(r(C),S\cap V(C^\up))$ be the distance from $r(C)$ to the closest vertex in $S\cap V(C)$, which must also be in $S\cap V(C^\up)$. 

Next, consider all paths $P$ in $G_0$ that were contracted into edges in $C_1$; for each such path $P$, let $r_1(P),\,r_2(P)\in V_1$ be the two endpoints of $P$. Let $\m P$ be the paths $P$ which satisfy $S\cap (V(P)\sm\{r_1(P),r_2(P)\})\ne\emptyset$. For $i=1,2$, let $v_i(P)$ be the vertex on $P$ closest to $r_i(P)$, and define $d_i(P):=d_P(r_i(P),v_i(P))=d_P(r_i(P),S\cap V(P))$. Note that it is possible that $v_1(P)=v_2(P)$, which happens precisely when $|S\cap V(P)|=1$. Define $P^\up$ to be the union of the path from $r_1(P)$ to $v_1(P)$ and the path from $r_2(P)$ to $v_2(P)$. Again, we first focus on $P^\up$.

\begin{figure}\figl{G2}
\centering
\includegraphics[scale=.8,valign=t]{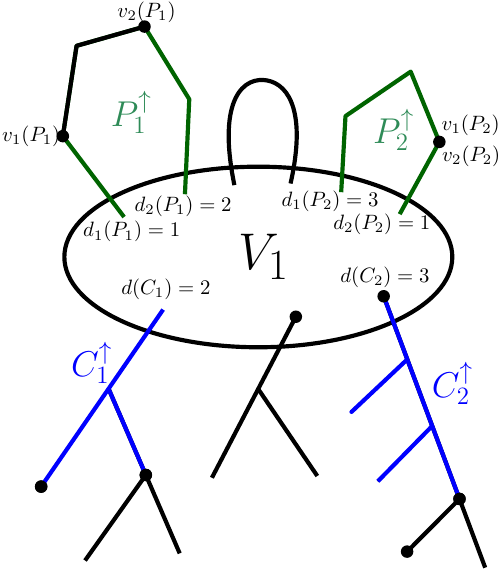}
\includegraphics[scale=.8,valign=t]{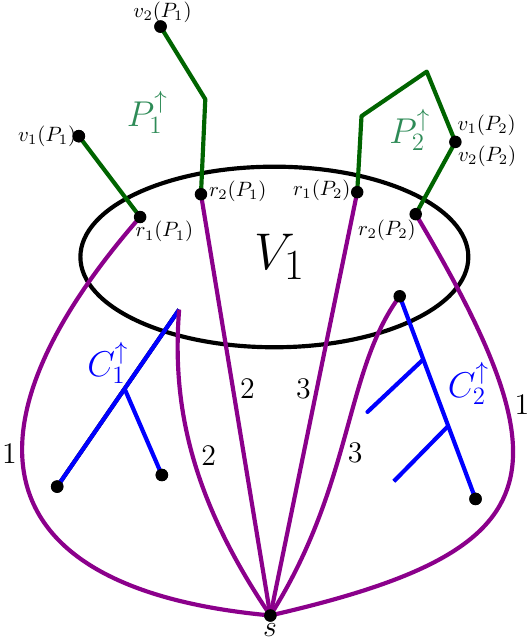}
\caption{Construction of the graph $G_2$.}
\end{figure}

We construct a graph $G_2$ as follows. The vertex set is $V_2:=V_1\cup\bigcup_{C\in\m C}V(C^\up) \cup \bigcup_{P\in\m P}V(P^\up) \cup \{s\}$ for a new vertex $s$. Add the graph $G_1$ onto the vertices $V_1$, and for each $C\in\m C$ and $P\in\m P$, add the graphs $C^\up$ and $P^\up$ into $V(C^\up)$ and $V(P^\up)$, respectively. For each vertex $v\in S\cap V_1$, add an edge of weight $0$ between $s$ and $v$, adding a total of $|S\cap V_1|\le|V_1|\le4t$ edges. Next, for each $C\in\m C$, add an edge from $s$ to $r(C)$ of weight $d(C)$, and for each $P\in\m P$, add an edge from $s$ to $r_i(P)$ of weight $d_i(P)$ for $i=1,2$. Since every component $C\in\m C$ has a distinct $r(C)\in V_1$, we have $|\m C|\le|V_1|\le4t$. Since every path $P\in\m P$ gets contracted to a (distinct) edge in $G_1$, we have $|\m P|\le|E(G_1)|\le5t$. Therefore, we add at most $13t$ edges from $s$.

\BCL\clml{dist-G2}
$G_2$ has at most $(|V_2|-1)+14t$ edges, and for every vertex $v\in V_2\sm\{s\}$, we have $d_{G_2}(s,v) \ge d_G(S,v)$.
\ECL
\BP
Since $G$ has $(n-1)+t$ edges, there exists some $t$ edges $F\s E$ such that $G-F$ is a tree. This means that $G[V_2\sm\{s\}]$ has at most $(|V_2\sm\{s\}|-1)$ edges in $G-F$, and since $|F|=t$ and we added at most $13t$ extra edges, $G_2$ has at most $(|V_2\sm\{s\}|-1)+14t$ edges.

To prove the second statement, consider a vertex $v\in V_2\sm\{s\}$, and let $P$ be the shortest path from $s$ to $v$ in $G_2$. If the first edge on the path (adjacent to $s$) its other endpoint (besides $s$) inside $V_1\cap S$, then this edge has weight $0$, and the path $P$ minus that first edge is a path in $G_1$ from $S$ to $v$ of equal weight. Then, for each edge on the path formed by contracting a path in $G$ to an edge in $G_1$, we can expand the edge back to the contracted path, obtaining a path the same weight in $G$.

Next, suppose that the first edge connects to vertex $r(C)$ for some $C\in\m C$. In this case, we replace that edge with the path in $C^\up$ from $S\cap V(C^\up)$ to $r(C^\up)$ of weight $d(C)$. The new path is a path in $G_1$ from $V$ to $v$ of the same weight, and we can expand contracted edges as in the first case.

Otherwise, the first edge must connect to a vertex $r_i(P)$ for some $P\in\m P$ and $i\in\{1,2\}$. In this case, we similarly replace that edge with the path in $P^\up$ from $S\cap T(P^\up)$ to $r_i(P^\up)$ of weight $d(P)$, and the rest of the argument is analogous.
\EP

It is clear that $G_2$ has aspect ratio $O(M)$.
We now apply \lem{ultra} on $G_2$ with source $s$ and algorithm $\m A$ (note that $70t = 5 \cd 14t$), obtaining an $s$-SSSP potential $\phi_2$ on $V_2$. W.l.o.g., we can assume that $\phi_2(s)=0$, since we can safely add any multiple of $\mathbbm1$ to $\phi_2(s)$. We now extend $\phi_2$ to $V$ by setting $\phi_2(v):=\infty$ for all $v\in V\sm V_2$.

We next define a potential $\phi_{\m C}$ on $V$ as follows. For each $C\in\m C$, let $\phi_{\m C}(r(C)):=\infty$, let $\phi_{\m C}(v):=d_C(S\cap (V(C)\sm\{r(C)\}),v)$ for $v\in V(C)\sm \{r(C)\}$ (that is, \emph{exact} distances in $C$ from $S\cap (V(C)\sm\{r(C)\}$), and assume w.l.o.g.\ that $\phi_{\m C}(v)=0$ for all $v\in S\cap V(C)$ (see \obs{shift}). For all remaining $v\in V\sm\bigcup_{C\in\m C}V(C)$, define $\phi_{\m C}(v):=\infty$. Similarly, we define potential $\phi_{\m P}$ as follows. For each $P\in\m P$, let $\phi_{\m P}(r_1(P))=\phi_{\m P}(r_2(P)):=\infty$, let $\phi_{\m P}(v):=d_P(S\cap (V(P)\sm\{r_1(P),r_2(P)\}),v)$ for $v\in V(P)$, and assume w.l.o.g.\ that $\phi_{\m P}(v)=0$ for all $v\in S\cap V(P)$; for all remaining $v\in V\sm\bigcup_{P\in\m P}V(P)$, define $\phi_{\m P}(v):=\infty$. Since each $C\in\m C$ and $P\in\m P$ is a tree, this can be done efficiently in parallel as stated below, whose proof we defer to \sec{om75}.
\begin{restatable}{lemma}{TreeSSSP}\leml{tree-SSSP}
Given a tree $T=(V,E)$ and a set of sources $S\s V$, we can compute an exact $S$-SSSP potential in $\tO(m)$ work and $\pl(n)$ time.
\end{restatable}

Finally, define $\phi(v) := \min \{ \phi_2(v), \phi_{\m C}(v), \phi_{\m P}(v) \}$. Note that for all $v\in V$, we have $\phi_i(v)\ge0$ for all $i\in\{2,\m C,\m P\}$ by \obs{shift}, so $\phi(v)\ge0$ as well.
\BCL
The vector $\phi$ is an $\al$-approximate $S$-SSSP potential of $G$.
\ECL
\BP
Since $\phi(s)=0$ for all $s\in S$, property~(0) of \defn{SSSP-dual-S} holds. We now prove property~(1). Fix a vertex $v\in V$, and suppose first that $\phi(v)=\phi_2(v)$ (i.e., the minimum is achieved at $\phi_2(v)$). Then, by \clm{dist-G2}, we have $d_{G_2}(s,v)\ge d_G(s,v)$. This, along with the guarantee $\phi_2(v)\ge \f1\al d_{G_2}(s,v)$ from $\phi_2$, implies that $\phi_2(v) \ge \f1\al d_{G_2}(s,v) \ge \f1\al d_G(s,v)$. Now suppose that $\phi(v)=\phi_{\m C}(v)$. Since $\phi_{\m C}(v)=d_C(v,S\cap V(C))$ for some $C\in\m C$, we have 
\[ \phi(v)=d_C(v,S\cap V(C)) \ge d_G(v,S\cap V(C))\ge d_G(v,S)\ge\f1\al d_G(v,S).\]
 The remaining case $\phi(v)=\phi_{\m P}(v)$ is analogous, with every instance of $C$ and $\m C$ replaced by $P$ and $\m P$, respectively.

We now focus on property~(2). Note that if $\phi_2,\phi_{\m C},\phi_{\m P}$ each satisfied property~(2), then by \Cref{obs:max}, $\phi$ would as well. In fact, a more fine-grained variant of \Cref{obs:max} states that for any edge $(u,v)\in E$, if we have $|\phi_i(u)-\phi_i(v)|\le d_G(u,v)$ for each $i\in\{2,\m C,\m P\}$, then $|\phi(u)-\phi(v)|\le d_G(u,v)$ as well. Hence, we only need to consider edges for which $|\phi_i(u)-\phi_i(v)|>d_G(u,v)$ for some $i\in\{2,\m C,\m P\}$.

We first focus on $i=2$. Observe that by property~(1) on $\phi_2$, the only edges $(u,v)\in E$ for which $|\phi_2(u)-\phi_2(v)|>d_G(u,v)$ are those where $\phi_2(u)<\infty$ and $\phi_2(v)=\infty$ or vice versa.\footnote{Let us assume for simplicity that $\infty-\infty=0$. To be more formal, we should replace each $\infty$ with some large number $M$ that exceeds the weighted diameter of the graph, so that we have $M-M=0$ instead.}
Let us assume w.l.o.g.\ that $\phi_2(u)<\infty$ and $\phi_2(v)=\infty$; by construction, we must either have $u\in S\cap V(C^\up)$ for some $C\in\m C$ or $u\in S\cap V(P^\up)$ for some $P\in\m P$. In the former case, since $\phi_{\m C}(u)=0$ and $\phi(u)\ge0$, we must have $\phi(u)=0$, and since $\phi_{\m C}(v)<\infty$ and $\phi_2(v)=\phi_{\m P}(v)=\infty$, we also have $\phi(v)=\phi_{\m C}(v)$. We thus have, for some $C\in\m C$,
\[ |\phi(u)-\phi(v)| = |\phi_{\m C}(u)-\phi_{\m C}(v)| = |0 - \phi_{\m C}(v)| = \phi_{\m C}(v) = d_C(v,S\cap V(C)) \le d_G(u,v) ,\]
so edge $(u,v)$ satisfies property~(2), as needed.

Next, consider the case $i=\m C$. By construction, the only edges $(u,r)\in E$ for which $|\phi_{\m C}(u)-\phi_{\m C}(r)|>d_G(u,v)$ are those where $\phi_{\m C}(u)<\infty$ and $\phi_{\m C}(r)=\infty$ or vice versa. Assuming again that $\phi_{\m C}(u)<\infty$ and $\phi_{\m C}(r)=\infty$, we must have $u\in V(C)$ and $r=r(C)$ for some $C\in\m C$. By construction, we must have $\phi(r)=\phi_2(r)$ and $\phi(u)=\min\{\phi_2(u),\phi_{\m C}(u)\}$. By property~(2) of $\phi_2$, we have $|\phi_2(u)-\phi_2(r)|\le w(u,r)$, so it suffices to show that $\phi_{\m C}(u)\ge\phi(r)-w(u,r)$, from which $|\phi(u)-\phi(r)|\le w(u,r)$ will follow.

By \obs{dual-upper} and the fact that $\phi(s)=0$, we have $\phi_2(r)\le d_{G_2}(s,r)$. Also, by construction of $G_2$, we have $d_{G_2}(s,r) \le w_{G_2}(s,r) = d_C(S\cap(V(C)\sm\{r(C)\})$. Therefore,
\begin{align*}
\phi_{\m C}(u) = d_C(S\cap(V(C)\sm\{r(C)\}),u) &= d_C(S\cap(V(C)\sm\{r(C)\}),r) - w(u,r)
\\&\ge d_{G_2}(u,r) - w(u,r)
\\&\ge \phi_2(r)-w(u,r) 
\\&= \phi(r)-w(u,r) ,
\end{align*}
as desired.

The case $i=\m P$ is almost identical to the case $i=\m C$, except we now have $r=r_1(P)$ or $r=r_2(P)$. Since the rest of the argument is identical, we omit the proof.
\EP

\section{Ultra-spanner Algorithm}\secl{ultra}
\newcommand{\LG}{\llfloor G\rrfloor_k}
\newcommand{\LE}{\llfloor E\rrfloor_k}
\newcommand{\LW}{\llfloor W\rrfloor_k}
\newcommand{\LS}{\llfloor S\rrfloor_k}

In this section, we present our algorithm for constructing an ultra-spanner. It is a modification of the weighted spanner algorithm of \cite{MPV}, where we sacrifice more factors in the spanner approximation for the needed ultra-sparsity.


\ultra*

Our ultra-spanner algorithm closely resembles the weighted spanner algorithm of \cite{MPV}. Their algorithm outputs an $O(k)$-spanner with $O(n^{1+1/k}\log k)$ edges, which is not ultra-sparse and therefore insufficient for our purposes. However, the $\log k$ factor of their algorithm comes from splitting the graph into $O(\log k)$ separate ones, computing a spanner for each, and taking the union of all spanners. We modify their algorithm to consider only one graph, at the cost of an extra $k$-factor in the stretch, which is okay for our application. We first introduce the subroutine \ref{EST} for \emph{unweighted} graphs from \cite{MPV} (which dates back to~\cite{miller2013parallel}) and its guarantee, whose proof we sketch for completeness. Note that while this algorithm can be adapted to the weighted setting, executing the algorithm efficiently in parallel is difficult

\begin{algorithm}[H]
\mylabel{EST}{\texttt{ESTCluster}}\caption{\ref{EST}($G=(V,E),\be\in(0,1]$), $G$ is \emph{unweighted}}
\begin{algorithmic}[1]
\State For each vertex $u$, sample $\de_u$ independently from the geometric distribution with mean $1/\be$
\State Create clusters by defining $C_u := \{v\in V:u = \arg\min_{u'\in V}d(u', v)-\de_{u'}\}$, with ties broken by a universal linear ordering of $V$. If $u\in C_u$, then $u$ is the \emph{center} of cluster $C_u$
\State Return the clusters $C_u$ along with a spanning tree on each cluster rooted at its center.
\end{algorithmic}\label{alg:EST}
\end{algorithm}

\BL\thml{EST}
For each edge in $E$, the probability that its endpoints belong to different clusters is at most $\be$.
\EL
\BP
Fix an edge $(v,v')\in E$, and let $u_1,u_2\in V$ be the vertices achieving the smallest and second-smallest values of $d(u',v)-\de_{u'}$ over all $u'\in V$, with ties broken by the linear ordering of $V$. (In particular, $v\in C_{u_1}$.) Let us condition on the choices of $u_1,u_2$ and the value of $d(u_2,v)-\de_{u_2}$. First, suppose that $u_1\le u_2$ in the linear ordering (that is, $u_1$ is preferred in the event of a tie). Then, we know that \[v'\in C_{u_1} \iff d(u_1,v')-\de_{u_1} \le d(u_2,v')-\de_{u_2} \iff \de_{u_1} \ge d(u_1,v')-d(u_2,v')+\de_{u_2} .\]
So far, we are conditioning on the event $\de_{u_1}\ge d(u_1,v)-d(u_2,v)+\de_{u_2}$. By the memoryless property of geometric variables, with probability $1-\be$, we have  $\de_{u_1}\ge (d(u_1,v)-d(u_2,v)+\de_{u_2})+1$. In that case, we also have
\[ d(u_1,v')-\de_{u_1} \le (d(u_1,v)+1)-\de_{u_1} \le(d(u_1,v)+1)-(d(u_1,v)-d(u_2,v)+\de_{u_2}+1) = d(u_2,v)+\de_{u_2}  ,\]
so $v'\in C_{u_1}$ as well and edge $(v,v')$ lies completely inside $C_{u_1}$.

 If $u_1\ge u_2$ in the linear ordering instead, then we know that
\[v\in C_{u_1} \iff d(u_1,v)-\de_{u_1} < d(u_2,v)-\de_{u_2} \iff \de_{u_1} > d(u_1,v)-d(u_2,v)+\de_{u_2} ,\]
and the proof proceeds similarly. 

Overall, for each edge in $E$, the probability that its endpoints belong to different clusters is at most $\be$.
\EP

We now proceed to our ultra-spanner algorithm.
Without loss of generality, the edge weights of $G$ range from $1$ to $W$ for some $W=\poly(n)$.
For positive real numbers $x$ and $k$, define $\llfloor x\rrfloor_k:=\max\{k^{\al}:\al\in\Z,\,k^{\al}\le x\}$ as the largest integer power of $k$ less than or equal to $x$. Let $\llfloor G\rrfloor_k=(V,\llfloor E\rrfloor_k)$ be the graph $G$ with the weight $w(u,v)$ of each edge $(u,v)\in E$ replaced by $\llfloor w(u,v)\rrfloor_k$, so that in particular, all edge weights in $G$ are now nonnegative integer powers of $k$. For each $\al\in\{0,1,2,\lds,\llfloor W\rrfloor_k\}$, define $E_\al\s \LE$ as the set of edges in $\llfloor G\rrfloor_k$ with weight $k^\al$.

\begin{algorithm}[H]
\mylabel{ultra}{\texttt{Ultraspanner}}\caption{\ref{ultra}($\LG=(V,\LE)$)}
\begin{algorithmic}[1]
\State Initialize $H_0\gets\emptyset$ \Comment{$H_i\s \LE$ will be edges contracted over the iterations}
\State Initialize $\LS\gets\emptyset$ \Comment{$\LS\s \LE$ will be the edges in the spanner}
\For {$\al=0,1,2,\lds,\llfloor W\rrfloor_k$}
  \State Let $V_0^{\al-1},V_1^{\al-1},\lds$ be the connected components of $H_{\al-1}$
  \State Let $\G_\al$ be the graph formed by starting with $(V,E_\al)$ and contracting each $V_i^{\al-1}$ into a single vertex
  \State Run \ref{EST} on the (unweighted version of) $\G_\al$ with $\be:=\f{C\ln n}{k}$ on $\G_\al$ for sufficiently large $C$, and let $F\s\G_\al$ be the forest returned
  \State Update $\LS\gets \LS\cup F$\linel{intra}
  \State Set $H_\al\gets H_{\al-1}\cup F$
  \State Add to $\LS$ all edges $e$ in $\G_\al$ whose endpoints lie in different connected components of $F$\linel{inter}
\EndFor
\State\Return$\LS$ as the spanner
\end{algorithmic}\label{alg:ultra}
\end{algorithm}

\BL\leml{alg-ultra}
If \ref{ultra} \emph{succeeds} (for a notion of success to be mentioned), then the output $\LS$ is a $k$-spanner with at most $n-1+O(\f{m\log n}k)$ edges. We can define our success condition so that it happens with probability at least $1/3$, and that we can detect if the algorithm fails (so that we can start over until it succeeds). Altogether, we can run the algorithm (repeatedly if necessary) so that w.h.p., the output $\LS$ is a $k$-spanner of $\LG$ with at most $n-1+O(\f{m\log n}k)$ edges, and it takes $\tO(m)$ work and $\tO(k)$ time.
\EL
\BP
We say that \ref{ultra} \emph{fails}  if on any iteration $\al$, some $\de_v$ in the computation of \ref{EST} satisfies $\de_v>k/6$. Observe that if the algorithm does not fail, then every call to \ref{EST} takes $\tO(k)$ time (and $\tO(m)$ work) and returns clusters with diameter at most $2\cd k/6=k/3$. We now bound the probability of failure: for any given vertex $v$ in some  $\G_\al$, the probability that $\de_v>k/6$ is $e^{-\be k/6}=e^{-(C/6)\ln n}=n^{-C/6}$. There are at most $n$ vertices in each $\G_\al$, and at most $\lf\log_k W\rf+1=O(1)$ many iterations since $G$ has bounded aspect ratio, so taking a union bound, the failure probability is at most $O(n^{-C/6+1})$.

Assume now that \ref{ultra} does not fail. Then, we prove by induction on $\al$ that the diameter of each connected component of $H_\al$ is at most $k^{\al+1}$. This is trivial for $\al=0$, and for $\al>0$, suppose by induction that the statement is true for $\al-1$. Since the algorithm does not fail, \ref{EST} returns clusters with (unweighted) diameter at most $k/3$. In the weighted $\G_\al$, these clusters have diameter at most $k/3 \cd k^\al=k^{\al+1}/3$. Observe that $H_\al$ is formed by starting with these clusters and ``uncontracting'' each vertex into a component of $H_{\al-1}$. By induction, each component in $H_{\al-1}$ has diameter at most $k^\al$. Therefore, between any two vertices in a common component of $H_\al$, there is a path between them consisting of at most $k/3$ edges of length $k^{\al}$ and at most $k/3+1$ subpaths each of length at most $k^\al$, each inside a component in $H_{\al-1}$. Altogether, the total distance is at most $k/3\cd k^\al + (k/3+1)\cd k^{\al}\le k^{\al+1}$ (assuming $k\ge3$).

Let us now argue that the stretch of $\LS$ is at most $k$ if the algorithm does not fail. Observe that the edges added to $\LS$ in \line{inter} on an iteration $\al$ are precisely the edges whose endpoints belong to different clusters in the corresponding \ref{EST} call. Conversely, any edge $e\in E_\al$ not added to $\LS$ have both endpoints in the same cluster. By the previous argument, this cluster has diameter at most $k^{\al+1}$ assuming that the algorithm does not fail. Therefore, the stretch of edge $e$ is at most $k$.

Finally, we bound the number of edges in the output $S$. By \thm{EST}, for the \ref{EST} call on iteration $\al$, every edge in $E_\al$ has its endpoints in different clusters with probability at most $\be$, which is when it is added to $\LS$ in \line{inter}. Over all iterations $\al$, the expected number of edges added to $\LS$ in \line{inter} is at most $\be m$. By Markov's inequality, with probability at least $1/2$, there are at most $2\be m$ edges added in \line{inter}. If this is not the case, we also declare our algorithm to fail. Note that the failure probability now becomes at most $1/2+O(n^{-C/6+1})\le 2/3$ (for $C$ large enough). Moreover, it is easy to see that the edges added to $S$ on \line{intra} form a forest, so at most $n-1$ are added there. Altogether, if the algorithm succeeds, then there are at most $(n-1)+2\be m=(n-1)+O(\f{m\log n}k)$ edges in the output, which is a $k$-spanner.

Lastly, \ref{ultra} can clearly be implemented to run in $\tO(m)$ work and $\tO(k)$ time. Moreover, we only need to repeat it $O(\logn)$ times before it succeeds w.h.p.
\EP

Using \lem{alg-ultra}, we now prove \thm{ultra} as follows. Let $S\s E$ be the corresponding spanner in $G$ sharing the same edges as $\LS\s\LE$ (with possibly different weights). Intuitively, since $\LG$ approximates the edge weights of $G$ up to factor $k$, the $k$-spanner $S$ should be a $k^2$-spanner on $G$. More formally, given an edge $(u,v)\in E$, let $u=v_0,v_1,\lds,v_\el=v$ be the shortest path in $\LS$. We have
\[ d_S(u,v) \le \sum_{i=1}^\el d_S(v_{i-1},v_i) \le \sum_{i=1}^\el k\cd d_{\LS}(v_{i-1},v_i) = k \cd d_{\LS}(u,v) \le k \cd kd_{\LG}(u,v) \le k^2d_G(u,v) .\]
Therefore, $S$ is a $k^2$-spanner of $G$.

\section{Sherman's Framework via Multiplicative Weights}\secl{mwu}

In this section, we provide a self-contained proof of \thm{sherman} using the \emph{multiplicative weights updates} framework at the loss of an additional $\log(n/\e)$, which can be disregarded in our parallel algorithms.

\ShermanPara*

We begin with the following classical result on solving linear programs approximately with multiplicative weights update framework. We state it without proof, since the result is a staple in advanced algorithms classes.

\BT[Solving LPs with Multiplicative Weights Update]\thml{mwu}
Let $\de\le1$ and $\om>0$ be parameters.
Consider a convex set $K\s \R^n$, a matrix $M\in\R^{m\times n}$, and a vector $c\in\R^m$. (We want to investigate approximate feasibility of the set $\{y\in K:My\le c\}$.) Let $\m O$ be an oracle that, given any vector $p\in\De_m$, either outputs a vector $y\in K$ satisfying $p^TMy\le p^Tc$ and $\norm{My-c}_\infty\le \om$, or determines that the set $\{y\in K:p^TMy\le p^Tc\}$ is infeasible. Then, consider the following algorithm:
  \BE
  \im Set $p_0\gets\f1m\mathbbm1\in\De_m$
  \im For $t=1,2,\lds,T$ where $T = O(\om^2\log m/\de^2)$:
    \BE
    \im Call oracle $\m O$ with $p_{t-1}\in\De_m$, obtaining vector $y^{(t)}$
    \im If $\m O$ determines that no $y\in K$ exists, then output $p_{t-1}$ and exit
    \im For each $j\in[m]$, set $w^{(t)}_j\gets w^{(t-1)}_j\cd\exp(\de \cd (My^{(t)}-c)_j) = \exp(\de\cd\sum_{i\in[t]}(My^{(i)}-c)_j)$
    \im For each $j\in[m]$, set $p^{(t)}_j\gets w^{(t)}_j/\sum_{i\in[m]}w^{(t)}_i$
    \EE
  \im Output $\f1T\sum_{i\in[T]}y^{(i)}\in K$
  \EE
If this algorithm outputs a vector $p\in\De_m$ on step 2(b), then the set $\{x\in K:Mx\le c\}$ is infeasible. Otherwise, if the algorithm outputs a vector $x\in K$ on step 3, then we have $Mx \le c + \de\mathbbm1$.
\ET

\BL\leml{sherman-mwu}
Consider a transshipment instance with demands $b$ and a parameter $t \ge \opt(b)/2$. Let $r$ be a parameter, and let $R\in\R^{[r]\times V}$ be a matrix satisfying  \eqn{kappa} for some parameter $\ka$. Then, there is an algorithm that performs $O((\ka/\e)^2\logn)$ matrix-vector multiplications with $A$, $A^T$, $R$, and $R^T$, as well as $O((\ka/\e)^2\logn)$ operations on vectors in $\R^r$, and outputs either
\BE
 \im an acyclic flow $f$ satisfying $\1{Cf}\le t$ and $\1{RAf-Rb} < \e t$, or
 \im a potential $\phi$ with value $b^T\phi=t$.
\EE
\EL

We will run Multiplicative Weights Update (\thm{mwu}) to determine feasibility of the region
\[ \{ y\in\R^r : \norm{y^TRAC\inv}_\infty + \f1ty^TRb \le -\e \text{ and } \norm{y}_\infty\le1 \} ,\]
which is modeled off the dual LP formulation for \ts; this connection will become clearer once \clm{output-2} is proved.

We set $c:=\e\mathbbm1$ and $K:=\{y\in\R^r:\norm y_\infty\le1\}$ in \thm{mwu}. As for matrix $M$, the constraint $\norm{y^TRAC\inv}_\infty+\f1ty^TRb\le-\e$ can be expanded into
\[ \pm y^TRAc_e\inv\chi_e + \f1ty^TRb \le -\e \quad \forall e\in E , \]
so the rows of matrix $M$ consist of the $2m$ vectors $(\pm RAc_e\inv\chi_e+\f1tRb)^T$ for each $e\in E$.  

We now specify the oracle $\m O$. On each iteration, the algorithm of \thm{mwu} computes values $p_e^+,p_e^-\ge0$ for each $e\in E$ satisfying $\sum_e(p_e^++p_e^-)=1$. The oracle needs to compute a vector $y$ satisfying $\norm y_\infty\le1$ and
\[ \sum_{e\in E}\lp p_e^+\big(y^TRAc_e\inv\chi_e+\f1ty^TRb\big)  + p_e^-\big({-y^TRAc_e\inv\chi_e}+\f1ty^TRb\big)  \rp \le -\e .\]
This inequality can be rewritten as
\[ y^T\bigg(RA\sum_{e\in E}c_e\inv(p_e^+\chi_e-p_e^-\chi_e)+ \f1tRb\bigg) \le -\e .\]
Observe that a solution $y$ exists iff
\begin{gather}
\1{RA\sum_{e\in E}c_e\inv(p_e^+\chi_e-p_e^-\chi_e)+\f1tRb} \ge \e , \label{eq:p}
\end{gather}
and if the inequality is true, then the vector $y:=-\text{sign}(RA\sum_{e\in E}c_e\inv(p_e^+\chi_e-p_e^-\chi_e)+\f1tRb)$ is a solution, so the oracle outputs it.

We set the error parameter $\de$ to be $\e/(2\ka)$. Then, either the algorithm of \thm{mwu} outputs $p$ that violates (\ref{eq:p}) on some iteration, or after a number of iterations (depending on $\rho$, which we have yet to bound), the average of all vectors $y$ computed over the iterations satisfies
\begin{gather}
\pm y^TRAc_e\inv\chi_e + \f1ty^TRb \le -\e + \f\e{2}=-\f\e{2} \quad \forall e\in E \quad\iff\quad \norm{y^TRAC\inv}_\infty + \f1ty^TRb \le -\f\e{2} . \eqnl{yy}
\end{gather}

First, suppose that the second case holds:
\BCL\clml{output-2}
Suppose \thm{mwu} outputs a vector $y$ satisfying \eqn{yy}. Then, we can compute a potential $\phi$ satisfying condition~(2).
\ECL
\BP
Consider the vector $\phi_0:=-(y^TR)^T$. We have
\[ \norm{\phi_0^TAC\inv}_\infty - \f1t\phi_0^Tb \le -\f\e{2\ka} < 0 .\]
In particular, $\f1t\phi_0^Tb>0$. Let $\phi$ be the vector $\phi_0$ scaled up so that $\f1t\phi^Tb=1$. Then,
\[ \norm{\phi^TAC\inv}_\infty - \f1t\phi^Tb  < 0 \implies \norm{\phi^TAC\inv}_\infty < \f1t\phi^Tb = 1 .\]
Since $\phi$ satisfies the transshipment dual constraints, it is a potential. Moreover, $\phi^Tb=t$, so $\phi$ satisfies condition~(2).
\EP

Let us now consider the first case:
\BCL
Suppose \thm{mwu} outputs values $p_e^+,p_e^-\ge0$ satisfying  $\sum_e(p_e^++p_e^-)=1$ and
\[ \1{RA\sum_{e\in E}c_e\inv(p_e^+\chi_e-p_e^-\chi_e)+\f1tRb} < \e. \]
Then, we can compute an acyclic flow satisfying condition~(1).
\ECL
\BP
Let us construct a flow $f\in\R^E$ defined as $f := -t\sum_ec_e\inv(p_e^+\chi_e-p_e^-\chi_e)$. We have \[\1{Cf}=\1{-t\sum_{e\in E}(p_e^+\chi_e-p_e^-\chi_e)}=-t\sum_{e\in E}|p_e^+-p_e^-|\le t\sum_{e\in E}(p_e^++p_e^-)= t\] and
\[ \1{-RAf+Rb} < \e t ;\]
it remains to show that $f$ is acyclic. Fix an edge $e=(u,v)$, so that $A\chi_e=\chi_u-\chi_v$. We have that $f$ flows from $u$ to $v$ iff
\[ f_e>0 \iff -tc_e\inv ( p^+_e - p^-_e ) > 0 \iff p^+_e < p^-_e .\]
By the construction of $p^\pm_e$ from \thm{mwu},
\begin{align*}
p^+_e < p^-_e &\iff \sum_{i\in[t]}\big((RAc_e\inv\chi_e + \f1tRb)^Ty^{(i)}-\e\big) < \sum_{i\in[t]}\big((-RAc_e\inv\chi_e + \f1tRb)^Ty^{(i)}-\e\big)
\\&\iff (RAc_e\inv\chi_e)^T\sum_{i\in[t]}y^{(i)} < (-RAc_e\inv\chi_e)^T\sum_{i\in[t]}y^{(i)}
\end{align*}
Let $\overline y:=\sum_{i\in[t]}y^{(i)}$, so that this becomes
\begin{align*}
(RAc_e\inv \chi_e)^T\overline y < (-RAc_e\inv\chi_e)^T\overline y &\iff \overline y^TRc_e\inv(\chi_u-\chi_v) < -\overline y^TRc_e\inv(\chi_u-\chi_v) \\&\iff \overline y^TRc_e\inv\chi_u < \overline y^TRc_e\inv\chi_v.
\end{align*} 
Therefore, $f$ will only flow from $u$ to $v$ if $(\overline y^TR)_u < (\overline y^TR)_v$; it follows that such a flow cannot produce any cycles.
\EP
We now bound the value $\om$ needed for \thm{mwu}, which in turn bounds the number of iterations $T$.
\BCL
In \thm{mwu}, we can set $\om:=3\ka+\e$.
\ECL
\BP
We need to show that $\norm{My-c}_\infty\le3\ka+\e$. Since $c=\e\mathbbm1$, we instead show that $\norm{My}_\infty\le3\ka$, which suffices.

Consider any values $p_e^+,p_e^-$ satisfying $\sum_e(|p_e^+|+|p_e^-|)=1$ (note that we allow negative values here), and define  $f := -t\sum_ec_e\inv(p_e^+\chi_e-p_e^-\chi_e)$ as before, so that $\1{Cf}\le t$.

Recall that the $2m$ rows of matrix $M$ are the vectors $(\pm RAc_e\inv\chi_e+\f1tRb)^T$ for each $e\in E$, which we index by $M_e^\pm$. Similarly, group the values $p_e^+,p_e^-$ into a vector $p$ indexed by $p_e^\pm$. Consider the vector
\begin{align*} 
(p^TM)^T&=\sum_{e\in E} (p_e^+M_e^++p_e^-M_e^-) ^T
\\&= RA\sum_{e\in E}c_e\inv(p_e^+\chi_e-p_e^-\chi_e) + \f1tRb \sum_{e\in E}(p_e^++p_e^-)
 \\&= -\f1tRAf+\f1tRb\sum_{e\in E}(p_e^++p_e^-) .
\end{align*}
To avoid clutter, define $z=\sum_{e\in E}(p_e^++p_e^-)\le1$. We bound the norm
\begin{align*}
\1{p^TM} = \1{-\f1tRAf+\f ztRb}
&\le \f1t\lp \1{RAf} + z\1{Rb} \rp
\\&\le \f1t(\ka\cd\opt(Af) + z\cd\ka\cd\opt(b))
\\&\le \f1t(\ka\cd\1{Cf}+z\cd\ka\cd\opt(b))
\\&\le\f1t(\ka\cd t + z\cd \ka\cd2t) \le 3\ka.
\end{align*}
It follows that $p^TMy \le \1{p^TM}\norm y_\infty \le 3\ka \cd 1 = 3\ka$. Since $p$ was an arbitrary vector satisfying $\1 p=1$, we conclude that $\norm{My}_\infty\le3\ka$, as promised.
\EP

\BR
We only used the upper bound in (\ref{eq:kappa}); the lower bound will become useful when we work with condition~(1) in the lemma. Moreover, we will not need that the flow $f$ is acyclic, but this property may be useful in other applications.
\ER

We now claim that \thm{sherman-para} follows from \lem{sherman-mwu}. We apply \lem{sherman-mwu} $O(\log(n/\e))$ times by binary-searching on the value of $t$, which we ensure is always at least $\opt(b)/2$ as required by \lem{sherman-mwu}. Begin with $t=\poly(n)$, an upper bound on $\opt(b)$. First, while \lem{sherman-mwu} outputs a flow $f$ (case~(1)), set $t\gets \f{1+\e}2t$ for the next iteration. We claim that the new $t':=\f{1+\e}2t$ still satisfies $t'\ge \opt(b)/2$:
\[ \opt(b) \le \1{f}+\opt(Af-b) \le \1{f}+\1{RAf-Rb}\le t+\e t \implies \f{\opt(b)}2 \le \lp\f{1+\e}2\rp t=t' .\]
Repeat this until \lem{sherman-mwu} outputs a potential $\phi$ instead, which signifies that $\opt(b) \ge b^T\phi=t$. At this point, we know that $\opt(b)\in[t,2t]$, and we can properly run binary search in this range, where a flow $f$ returned for parameter $t$ signifies that $\opt(b)\le t+\e t$, and a potential $\phi$ returned means that $\opt(b)\ge t$. Through binary search, we can compute two values $t_\el,t_r$ such that $t_r-t_\el\le\e\,\opt(b)$ and $\opt(b) \in (t_\el,t_r)$.

Then, run \lem{sherman-mwu} with parameter $t=t_\el/(1+\e)$. We claim that we must obtain a potential $\phi$, not a flow $f$: if we obtain a flow $f$ instead, then
\[ \opt(b) \le \1{f}+\opt(Af-b) \le \1{f}+\1{RAf-Rb} \le t+\e t=t_\el ,\]
contradicting the assumption that $\opt(b)>t_\el$. This potential $\phi$ satisfies
$b^T\phi=\f{t_\el}{1+\e} \ge \f{t_r-\e\,\opt(b)}{1+\e}\ge \f{\opt(b)-\e\,\opt(b)}{1+\e} = (1-O(\e))\opt(b)$, which is almost optimal. At this point, we can compute a $(1+\e)$-approximation of $\opt(b)$, but we still need a transshipment flow $f$.

Next, run \lem{sherman-mwu} with parameter $t=t_r$; we claim that we must obtain a flow $f$ this time: if we obtain a potential $\phi$ instead, then $\opt(b)\ge b^T\phi=t_r$, contradicting the assumption that $\opt(b)<t_r$. This flow satisfies $\1{f}\le t_r\le t_\el+\e\,\opt(b)\le\opt(b)+\e\,\opt(b)=(1+\e)\opt(b)$. However, we are not done yet, since $f$ does not satisfy $Af=b$; rather, we only know that $\opt(Af-b)\le\1{R(Af-b)}\le \e\,\opt(b)$. The key idea is to solve transshipment again with demands $b_1:=Af-b$; if we can obtain a $(1+\e)$-approximate flow $f_1$ satisfying $\1{f_1}\le\opt(b_1)=\opt(Af-b)\le\e\,\opt(b)$ and $\1{R(Af_1-b_1)} \le \e\,\opt(b_1)\le\e^2\opt(b)$, then the composed flow $f+f_1$ has cost
\[ \1{f+f_1}\le\1{f}+\1{f_1}\le(1+\e)\opt(b) + (1+\e)\opt(b_1) \le (1+\e)\opt(b)+(1+\e)\e\,\opt(b),\]
which is $(1+O(\e))\opt(b)$. We can continue this process, defining $b_i:=Af_{i-1}-b_{i-1}$ and computing  a flow $f_i$ satisfying $\1{f_i}\le\opt(b_i)=\opt(Af_{i-1}-b_{i-1})\le \e\,\opt(b_{i-1})\le\e^i\opt(b)$ and $\1{R(Af_i-b_i)}\le\e\,\opt(b_i) \le \e^{i+1}\opt(b)$ and adding it on to $f+f_1+f_2+\cds+f_{i-1}$. Assuming $\e\le1/2$, say, we can stop after $T:=O(\logn)$ iterations, so that the leftover demands $b_{T+1}$ satisfies $\opt(b_{T+1}) \le \f1{n^3}\opt(b)$. At this point, we can simply run an $n$-approximate algorithm to demands $b_{T+1}$ by routing through a minimum spanning tree (see \Cref{rmk:sp-tree}), computing a flow $f_{T+1}$ satisfying $\1{f_{T+1}}\le n\cd\f1{n^3}\opt(b)\le\e\,\opt(b)$, assuming $\e\ge1/n^2$. (If $\e=O(1/n^2)$, then a transshipment algorithm running in time $\tO(1/\e^2)\ge\tO(n^4)$ is trivial.) The final flow $f+f_1+f_2+\cds+f_{T+1}$ has cost at most $(1+O(\e))\opt(b)$.

We have thus computed $\phi$ satisfying $\1{f} \le (1+O(\e))\opt(b) \le (1+O(\e))b^T\phi$, so $(f,\phi)$ is an $(1+O(\e))$-approximate flow-potential pair. Finally, we can reset $\e$ a constant factor smaller to obtain a $(1+\e)$-approximation. This concludes the algorithm of \thm{sherman-para}; it remains to bound the running time.


In each iteration of \thm{mwu}, we perform $O(1)$ matrix-vector multiplications with $A$, $A^T$, $R$, and $R^T$, as well as an additional $O(m)$ work and $\pl(m)$ time, and the same holds for the oracle $\m O$. This requires $O(n+m+M)$ total work. By \thm{mwu}, there are $O(\om^2\log m/\de^2) = O((\ka/\e)^2\logn)$ iteration inside \lem{sherman-mwu} to compute one flow $f$ or potential $\phi$. Finally, \lem{sherman-mwu} is called $\pl(n)$ times as described above, hence the promised running time.

\section{Transshipment to Expected SSSP: Sequential}\label{sec:esssp}
\newcommand{\orE}{\overrightarrow E}
\newcommand{\orG}{\overrightarrow G}
\newcommand{\orf}{\vec f}
\newcommand{\orw}{\vec w}
\newcommand{\orA}{\overrightarrow A}
\newcommand{\In}{\textup{in}}
\newcommand{\Out}{\textup{out}}
\newcommand{\FIn}{\orf_{\textup{in}}}
\newcommand{\FOut}{\orf_{\textup{out}}}
\newcommand{\src}{\textsf{\textup{src}}}
\newcommand{\TES}{W_{\textup{ESSSP}}}
\newcommand{\freq}{\textsf{\textup{freq}}}
\newcommand{\len}{\textsf{\textup{length}}}
\newcommand{\cyc}{\textsf{\textup{cycles}}}





In this section, we devise an algorithm that solves the approximate \emph{expected} single-source shortest path problem, defined below, using multiple \emph{sequential} calls to approximate transshipment. The fact that the recursive calls are made sequentially does not immediately imply a parallel algorithm, but in \sec{parallel-esssp}, we show how to save enough computation between the recursive calls to ensure a parallel algorithm. This extra step is more technical than insightful, hence its deferral to a separate subsection.

Finally, in \sec{sample-tree}, we show how to reduce SSSP to this expected version of SSSP~\cite{becker}. Together, \Cref{sec:esssp,sec:sample-tree} form a complete proof of \thm{sample-tree}. We remark that while \sec{sample-tree} is simply a rephrasing of a similar routine in~\cite{becker} and only included for self-containment, this section is novel, albeit still inspired by \cite{becker}.

\BD[Approximate expected $s$-SSSP Tree]
Given a graph $G=(V,E)$, a source $s$, and a demand vector $b$ satisfying $b_v\ge0$ for all $v\ne s$, an \emph{$\al$-approximate expected $s$-SSSP tree} is a randomized (not necessarily spanning) tree $T$ satisfying
\[ \E\lb\sum_{v:b_v>0} b_v \cd  d_T(s,v) \rb  \le \al \sum_{v:b_v>0}b_v \cd d_G(s,v) .\]
\ED

If $b_v>0$ for all $v\ne s$, then the tree $T$ must in fact be spanning. We also remark that the term \emph{expected} has two meanings here. First, the tree $T$ is randomized, so the guarantee
\[ \sum_{v:b_v>0} b_v \cd  d_T(s,v)  \le \al \sum_{v:b_v>0}b_v \cd d_G(s,v) \]
is only satisfied in expectation. However, even if this guarantee is satisfied with probability $1$, the distances $d_T(s,v)$ are not automatically $\al$-approximate distances for \emph{every} $v$; rather, they only hold on average (weighted by $b_v$). Note that in the exact setting $\al=1$, all distances $d_T(s,v)$ are indeed exact, but this property breaks down as soon as $\al>1$.

Define $\TES(n,m,\al)$ as the work to compute an $\al$-approximate expected SSSP with arbitrary demand vector $b$ satisfying $b_v\ge0$ for all $v\ne s$. Our algorithm \ref{esssp} is itself recursive and satisfies the following recursion:

\BL
$\TES(n,m,(1+3\e)\al) \le \WTS(m,\e) + \TES(n/2,m,\al)$.
\EL

Of course, we can solve expected SSSP exactly ($\al=1$) in constant time on constant-sized graphs, so this recursion has depth at most $\log_2n$.
Unraveling this recursion, the algorithm calls transshipment at most $\log_2n$ times, and the error $(1+3\e)$ blows up multiplicatively over each recursion level, obtaining
\[ \TES(n,m,(1+3\e)^{\log_2n}) \le \log_2n\cd \WTS(m,\e) + \tO(m) .\]
Resetting the value $\e$, we can rewrite it as
\begin{gather} \TES(n,m,1+\e) \le O(\logn)\cd \WTS(m,\Th(\e/\logn)) + \tO(m) ,\eqnl{esssp-rec}\end{gather}
which is our targeted recursion for our algorithm \ref{esssp} below.


\begin{algorithm}[H]
\mylabel{esssp}{\texttt{ESSSP}}\caption{\ref{esssp}($G=(V,E),s,b,(1+3\e)\al$)}
Assumption: demand vector $b$ satisfies $b_s>0$ and $b_t\le0$ for all $t\in V\sm s$
\begin{algorithmic}[1]
\State Compute a $(1+\e$)-approximate transshipment $f$ on $G$ with demand vector $b$
\State Initialize the digraph $\orA\gets\emptyset$
\State Every vertex $u\in V\sm s$ with $\In(u)\ne\emptyset$ independently samples a random neighbor $v\in\Out(u)$ with probability $\orf(u,v)/\FOut(u)$ and adds arc $(u,v)$ to $\orA$ \linel{sample-edge}
\State Add a self-loop $(s,s)$ of zero weight to $\orA$
\State Let $A$ be the undirected version of $\orA$
\State Initialize $G'\gets(\emptyset,\emptyset)$ as an empty undirected graph \Comment{Graph to be recursed on, with $\le n/2$ vertices}
\State Initialize $b'$ as an empty vector \Comment{Demands to be recursed on}
\For{each connected component $C$ of $A$}
 \State $c(C)\gets$ total weight of edges in the (unique) cycle in $C$ (possibly the self-loop $(s,s)$)
 \State Let $T_C$ be the graph $C$ with its (unique) cycle contracted into a single vertex $r_C$ \Comment{$T_C$ is a tree}
 \State Add a vertex $v_C$ to $G'$, and set demand $b'_{v_C}\gets\sum_{v\in V(C)}b_v$
\EndFor
\For{each edge $(u,u')$ in $E$}
 \State Let $C$ and $C'$ be the connected components of $A$ containing $u$ and $u'$, respectively
 \If{$C\ne C'$}
  \State Add an edge between $v_C$ and $v_{C'}$ with weight\linel{addedge} $w(u,u')+d_{T_C}(u,r_C)+d_{T_{C'}}(u',r_{C'})+c(C)+c(C')$ 
 \EndIf
\EndFor
\State Let $s'\gets v_{C_s}$, where $C_s$ is the component of $A$ containing $s$
\State Collapse parallel edges of $G'$ by only keeping the parallel edge with the smallest weight
\State Recursively call $\ref{esssp}(G',s',b',\al)$, obtaining an $\al$-approximate expected SSSP tree $T'$ of $G'$ \linel{recurse}
\State Initialize $T\gets\emptyset$ \Comment{The expected SSSP tree}
\For{each edge $(v,v')$ in $T'$}\linel{start-T}
 \State Let $(u,u')\in E$ be the edge responsible for adding edge $(v,v')$ to $G'$
 \State Add edge $(u,u')$ to $T$
\EndFor
\For{each connected component $C$ of $A$}
 \State Remove an arbitrary edge from the (unique) cycle inside $C$, and add the resulting tree to $T$\linel{end-T}
\EndFor
\State\Return $T$
\end{algorithmic}\label{alg:ESSSP}
\end{algorithm}

Throughout the section, fix a $(1+\e)$-approximate transshipment flow $f$ satisfying the given demand vector $b$ (with $b_v\ge0$ for all $v\ne s$). The key insight in our analysis is to focus on a \emph{random walk} based on a slight modification of the transshipment flow $f$. Define a digraph $\orG=(V\cup\{\bot\},\orE,\orw)$ as follows: start from $G$ by bidirecting each edge of $E$, keeping the same weight in both directions. Then, add a new vertex $\bot$ and a single arc $(s,\bot)$ of weight $0$. Let $\overrightarrow C$ denote the diagonal matrix indexed by $\orE$, where diagonal entry $\overrightarrow C_{(u,v),(u,v)}$ is the cost of arc $(u,v)\in\orE$ under the weights $\orw$.

Next, define a flow $\orf$ on digraph $\orG$ as follows: for each edge $(u,v)\in E$, if $f_{(u,v)}>0$, then add $f_{(u,v)}$ flow to $\orf$ along the arc $(u,v)$, and if $f_{(u,v)}<0$, then add $-f_{(u,v)}$ flow to $\orf$ along the arc $(v,u)$. Lastly, add $-b_s$ flow to $\orf$ along arc $(s,\bot)$. Observe that this flow satisfies the demands $b_v$ for each $v\in V\sm s$, demand $0$ for $s$, and demand $b_s$ for $\bot$. Moreover, the cost $\1{\overrightarrow C\orf}$ of the flow $\orf$ equals $\1{Cf}$.

For each vertex $v\in V$, define $\In(v):=\{u\in V:\orf{(u,v)}>0\}$ as the neighbors of $v$ that send flow to $v$, and define $\Out(v):=\{u\in V:\orf{(v,u)}>0\}$ as the neighbors of $v$ that receive flow from $v$. For convenience, define $\FIn(v):=\sum_{u\in\In(v)}\orf(u,v)$ and $\FOut(v):=\sum_{u\in\Out(v)}\orf(v,u)$.

Define $b^+(v):=\max\{b_v,0\}$, so that $b^+(s)=0$ and $b^+(v)=b(v)$ for all $v\ne s$. Define $V^*\s V$ as the vertices $t$ for which there exists a $t\to\bot$ path using only arcs supported by the flow $\orf$ (that is, arcs $(u,v)$ with $\orf(u,v)>0$). We will only be considering vertices in $V^*$ for the rest of this section.

\BCL\clml{vstar}
For all vertices $t\in V$, if $b(t)=b^+(t)>0$, then $t\in V^*$. Moreover, for all $v\in V^*\cup\{\bot\}$, we have $\In(v)\s V^*\cup\{\bot\}$ and $\Out(v)\in V^*\cup\{\bot\}$. In other words, the vertices in $V^*\cup\{\bot\}$ are separated from the vertices in $V\sm V^*$ by arcs supported by $\orf$.
\ECL
\BP
For the first claim, let $R\s V\cup\{\bot\}$ be all vertices reachable from $t$ along arcs supported by $\orf$; we need to show that $\bot\in R$. Since there are no arcs in $\orf$ going out of $R$, by conservation of flow, we must have $\sum_{v\in R}b(v)\le0$. Since $t\in R$ and $b^+(t)=b(t)>0$, we have $\sum_{v\in R\sm t}b(v)<0$. But the only vertex in $V\cup\{\bot\}$ with negative demand is $\bot$, so it must hold that $\bot\in R$, as desired.

We now prove the second statement. If suffices to show that there are no arcs between $V^*\cup\{\bot\}$ and $V\sm V^*$. There cannot be an arc $(u,v)$ supported by $\orf$ with $u\notin V^*\cup\{\bot\}$ and $v\in V^*\cup\{\bot\}$, since that would mean $u$ can reach $\bot$ by first traveling to $v$. Suppose for contradiction that there is an arc from $V^*\cup\{\bot\}$ to $V\sm V^*$. Since there is no arc the other way, by conservation of flow, it must follow that $\sum_{v\notin V^*\cup\{\bot\}}b(v)<0$. But the only vertex $\bot$ with negative demand is $\bot$, so it cannot be that $\sum_{v\notin V^*\cup\{\bot\}}b(v)<0$, a contradiction.
\EP

For each vertex $t\in V^*$, consider the natural random walk from $t$ to $\bot$ in $\orG$, weighted by the flow $\orf$: start from $t$, and if the walk is currently at vertex $u\in V$, then travel to vertex $v\in\Out(v)$ with probability $\orf(u,v)/\FOut(u)$ (independent of all previous steps); stop when $\bot$ is reached. Let this random walk be the random variable $W_t$, which is guaranteed to stay within $V^*\cup\{\bot\}$ by \clm{vstar}.  

Given a walk $W$ in $\orG$, define $\len(W)$ to be the length of the walk under the weights $\orw$. In particular, $\len(W_t)$ is the length of the random walk $W_t$ from $t$ to $\bot$. 

 The claim below relates the transshipment cost to the expected lengths of the random walks $W_t$ for $t\in V^*$. This allows us to later charge our expected SSSP distances to the lengths of these concrete random walks, which are easier to work with than the transshipment flow itself.

\BCL\clml{len}
We have
\[ \sum_{t\in V^*} b^+(t)\cd\E[\len(W_t)] \le \1{\overrightarrow C\orf} .  \]
In other words, if, for each $t\in V^*$, we sample a random walk from $t$ to $\bot$ and multiply its length by $ b^+(t)$, then the sum of the multiplied lengths over all $t$ is at most $\1{\overrightarrow C\orf}$ in expectation.
\ECL
\BP
For each vertex $v\in V^*\cup\{\bot\}$, let $\freq_t(v)$ be the expected number of times $v$ appears in $W_t$. We first prove the following:
\BSCL\clml{out}
For all vertices $v\in V^*$, $\sum_{t\in V^*}b^+(t)\cd\E[\freq_t(v)]=\FOut(v)$.
\ESCL
\BSP
For each $t\in V^*$, treat $\E[\freq_t(v)]$ as a function from $V^*\cup\{\bot\}$ to $\R_{\ge0}$, which satisfies the following equations:
\begin{align}
\E[\freq_t(t)] &= 1+\sum_{u\in \In(t)}\E[\freq_t(u)]\cd\f{\orf(u,t)}{\FOut(u)} \eqnl{e1} \\
 \E[\freq_t(v)] &= \sum_{u\in \In(v)}\E[\freq_t(u)]\cd\f{\orf(u,v)}{\FOut(u)} \qquad \forall v\in (V^*\cup\{\bot\})\setminus t \eqnl{e2} \\
\E[\freq_t(\bot)] &= 1 \nonumber
\end{align}
Note that the equations are well-defined, since by \clm{vstar}, all vertices $u\in\In(t)$ are in $V^*$ if $t\in V^*$.
Define the function $f(v):=\sum_{t\in V^*}b^+(v)\cd\E[\freq_t(v)]$, so our goal is to show that $f(v)=\FOut(v)$ for each $v\in V^*$. We sum \Cref{eq:e1,eq:e2} as follows:
for each $t$, multiply \Cref{eq:e1,eq:e2} by $b^+(t)$, and then sum over all $t$, obtaining
\begin{align}
f(v)&= b^+(v)+\sum_{u\in \In(t)}f(u)\cd\f{\orf(u,t)}{\FOut(u)} \qquad \forall v\in V^*\cup\{\bot\} \eqnl{e3} \\
f(\bot) &= \sum_{t\in V^*}b^+(t) \eqnl{e4}
\end{align}
We first show that the solution $f(v)=\FOut(v)$ for $v\in V^*$ and $f(\bot)= \sum_{t\in V^*}b^+(t)$ satisfies the above system of equations. \Cref{eq:e4} is clearly satisfied, and for \eqn{e3}, we have
\[ \FOut(v) = b^+(v)+\FIn(v)= b^+(v)+\sum_{u\in\In(v)}\orf(u,v)= b^+(v)+\sum_{u\in \In(v)}\FOut(u)\cd\f{\orf(u,v)}{\FOut(u)} .\]
We now claim that there is a unique solution to $f$, which is enough to prove the claim. Suppose there are two solutions $f$ and $f'$ that satisfy \eqn{e3}~and~\eqn{e4}. Let $g$ be their difference: $g(v):=f(v)-f'(v)$ for all $v\in V^*$, which satisfies
\begin{align}
g(v)&= \sum_{u\in \In(t)}g(u)\cd\f{\orf(u,t)}{\FOut(u)}  \qquad \forall v\in V^*\cup\{\bot\} \eqnl{e5} \\
g(\bot) &= 0 \eqnl{e6}
\end{align}
We want to show that $g(v)=0$ for all $v\in V^*$. It suffices to show that $g(v)\le0$ for all $v\in V^*$, since \Cref{eq:e5,eq:e6} are satisfied with $g(v)$ replaced by $-g(v)$, so we would prove both $g(v)\le0$ and $-g(v)\le0$, which would give $g(v)=0$. 

To show that $g(v)\le0$ for all $v\in V^*$, let $v^*:=\arg\max_{v\in V^*}g(v)$. By \eqn{e5}, $g(v^*)$ is a weighted average of $g(u)$ over vertices $u\in\In(v^*)$, so
\[ g(v^*) =  \sum_{u\in \In(v^*)}g(u)\cd\f{\orf(u,t)}{\FOut(u)} \le  \sum_{u\in \In(v^*)}g(v^*)\cd\f{\orf(u,t)}{\FOut(u)} = g(v^*) ,\]
so the inequality must be satisfied with equality, and $g(u)=g(v^*)$ for all $u\in\In(v^*)$. Continuing this argument, any vertex $u$ for which there exists a (possibly empty) $u\to v^*$ path in $\orf$ satisfies $g(u)=g(v^*)$. Define $R:=\{u\in V^*:\text{exists }u\to v^*\text{ path in }\orf\}$ as these vertices. Suppose for contradiction that $g(v^*)>0$, or equivalently, $g(u)>0$ for all $u\in R$.  Then, summing \eqn{e5} over all $v\in R$, we obtain
\begin{align}
\sum_{v\in R}g(v) &= \sum_{v\in R}\sum_{u\in\In(v)}g(v)\cd\f{\orf(u,v)}{\Out(u)} \nonumber\\
&= \sum_{v\in R}\sum_{u\in\In(v) \cap R}g(v)\cd\f{\orf(u,v)}{\Out(u)} \nonumber\\
&= \sum_{u\in R}\lp \f{g(u)}{\Out(u)}\sum_{v\in R:u\in\In(v)}\orf(u,v) \rp \nonumber\\
&\stackrel{\mathclap{g(u)\ge0}}\le \ \, \sum_{u\in R}\lp \f{g(u)}{\Out(u)}\sum_{v\in V^*}\orf(u,v) \rp \eqnl{e7} \\
&= \sum_{u\in R}\lp \f{g(u)}{\Out(u)}\,\Out(v) \rp \nonumber\\
&= \sum_{u\in R}g(u).\nonumber
\end{align}
Since $v^*\in V^*$, there exists a $v^*\to\bot$ path in $\orf$. Since $v^*\in R$ and $\bot\notin R$, there exists an arc $(u',v')$ on the path with $u'\in R$ and $v'\notin R$ (and $\orf(u',v')>0$). Consider the inequality at \eqn{e7}. The inequality holds for each $u\in R$ in the outer summation, but for $u=u'$ in particular, we have $\sum_{v\in R:u\in\In(v)}\orf(u',v)<\sum_{v\in V^*}\orf(u,v)$. Since $g(u)>0$ by assumption, the inequality at \eqn{e7} is actually strict, which gives the contradiction $\sum_{v\in R}g(v)<\sum_{u\in R}g(u)$. It follows that $g(v)\le g(v^*)\le0$ for all $v\in V^*$.
\ESP

We now resume the proof of \clm{len}. For all $t\in V^*$, by linearity of expectation,
\[ \E[\len(W_t)] = \sum_{u\in V^*}\freq_t(u)\cd\sum_{v\in\Out(u)}\f{\orf(u,v)}{\FOut(u)}\,\orw(u,v) . \]
For each $t\in V^*$, multiply the equation by $ b^+(t)$, and sum the equations, obtaining
\begin{align*}
 \sum_{t\in V^*}   b^+(t)\cd\E[\len(W_t)] &= \sum_{t\in V^*} b^+(t) \lp \sum_{u\in V^*}\freq_t(u)\cd\sum_{v\in\Out(u)}\f{\orf(u,v)}{\FOut(u)}\,\orw(u,v) \rp \\
&= \sum_{u\in V^*}\lp\bigg(\sum_{t\in V^*} b^+(t)\cd\freq_t(u)\bigg) \sum_{v\in\Out(u)}\f{\orf(u,v)}{\FOut(u)}\,\orw(u,v) \rp \\
&\stackrel{\mathclap{\text{Sub.\,\ref{clm:out}}}}=\ \ \sum_{u\in V^*}\lp\FOut(u) \cd\sum_{v\in\Out(u)}\f{\orf(u,v)}{\FOut(u)}\,\orw(u,v) \rp \\
&= \sum_{u\in V^*}\sum_{v\in\Out(u)}{\orf(u,v)}\,\orw(u,v)\\
&\le \sum_{u\in V}\sum_{v\in\Out(u)}{\orf(u,v)}\,\orw(u,v)\\
&= \1{\overrightarrow C\orf} .
\end{align*}
This concludes \clm{len}.
\EP

We remark that all our claims so far are in expectation, and therefore do not care about dependencies between the walks $W_t$ for different $t\in V^*$. 

For each walk $W_t$, we can imagine sampling it as follows: for each vertex $u\in V^*$, sample an infinite sequence of arcs $(u,v)$ for $v\in\Out(u)$, each independent of the others and with probability $\orf(u,v)/\FOut(u)$; let these arcs be $e_1^u,e_2^u,\lds$. Once the arcs are sampled for each $u\in V^*$, the random walk $W_t$ is determined as follows: start at $t$, and if the walk is currently at $u\in V^*$, then travel along the next \emph{unused} arc in the sequence $e_1^u,e_2^u,\lds$; in other words, if we have visited vertex $u$ on the walk $k$ times before the current visit, then travel along the arc $e^{u}_{k+1}$. It is easy to see that the distribution of this random walk is exactly $W_t$.

Let us first sample the set of sequences $e^u_1,e^u_2,\lds$ for each $u\in V^*$, and then determine the walks $W_t$ using this set of sequences for all $t$; note that the walks $W_t$ are heavily dependent on each other this way. Furthermore, observe that the arcs $\{e^u_1:u\in V^*\}$ are distributed the same way as the arcs in $\orA$ from \ref{esssp}.

\BL\leml{bound-esssp}
Suppose we execute \line{sample-edge} of \ref{esssp}, so that each vertex $u$ has (independently) sampled a neighbor $v^u$ and added arc $(u,v^u)$ is added to $\orA$. Let $E$ be the event that in the infinite sequences $e^u_1,e^u_2,\lds$, we have $e^u_1=(u,v^u)$ for all $u\in V^*$.
Let $\opt=\opt_G(b)$ be the optimum transshipment cost with demands $b$ in $G$, and let $\opt'=\opt_{G'}(b')$ be the optimum transshipment cost in the recursive call at \line{recurse}. Then, by conditioning on $E$ in the random walks $W_t$, we obtain
\begin{gather} \sum_{t\in V^*} b^+(t)\cd\E[\len(W_t)\mid E] + 2\e\cd\opt \ge \E\left[\sum_{t\in V^*} ( b^+(t) \cd d_{T_C}(u,r_C) )\bigg\lvert E \right]+ \opt' \eqnl{bound-esssp} .\end{gather}
\EL
\BP
Consider the following ``greedy'' cycle-finding algorithm for walks: 
given a walk $W$, travel along the walk in the forward direction, and whenever a cycle is found, immediately remove the cycle; output the set of all cycles removed. Let $\cyc(W_t)$ be the total length of all cycles removed, where length is measured by the weights $\orw$.
We first show that $\cyc(W_t)$ must be small compared to $\opt$, so that we can later charge to arcs in these cycles.

\BSCL\clml{cycle}
Given a walk $W$, let $\cyc(W)$ be the total length of cycles computed by the cycle-finding algorithm, where length is measured by the weights $\orw$. Recall that the transshipment flow $\orf$ is a $(1+\e)$-approximation of the optimum $\opt$ (that is, $\1{\overrightarrow C\orf}\le(1+\e)\opt$); then, we have
\[ \sum_{t\in V^*} b^+(t)\cd\E[\cyc(W_t)] \le \e\cd\opt .\] 
\ESCL
\BSP
For each walk $W_t$ sampled, remove the cycles computed by the algorithm to obtain another walk $W'_t$, still from $t$ to $\bot$.  Then, let $W^-_t$ be the walk $W'_t$ minus the last vertex $\bot$, whose new last vertex must be $s$. The flow obtained by sending $ b^+(t)$ flow along the walk $W^-_t$ for each $t\in V^*\sm s$ is a transshipment flow satisfying the demands $ b^+(t)$ for each $t\ne s$. Therefore,
\begin{align*}
\opt &\le \sum_{t\in V^*\sm s} b^+(t)\cd\E[\len(W^-_t)] \\
&= \sum_{t\in V^*\sm s} b^+(t)\cd\E[\len(W'_t)]\\
&\le \sum_{t\in V^*} b^+(t)\cd\E[\len(W'_t)] \\
&= \sum_{t\in V^*} b^+(t)\cd\E[\len(W_t) - \cyc(W_t)] \\
&\le \sum_{t\in V^*} b^+(t)\cd\E[\len(W_t)] - \sum_{t\in V^*} b^+(t)\cd\E[\cyc(W_t)] \\
&\stackrel{\mathclap{\text{Clm.\,}\ref{clm:len}}}\le \ \ \ \1{\overrightarrow C\orf} - \sum_{t\in V^*} b^+(t)\cd\E[\cyc(W_t)]\\
&\le(1+\e)\opt- \sum_{t\in V^*} b^+(t)\cd\E[\cyc(W_t)],
\end{align*}
and rearranging proves the claim.
\ESP

Let us return to proving \eqn{bound-esssp}. We actually show that for \emph{any} set of infinite sequences $\{e^u_1,e^u_2,\lds:u\in V^*\}$ satisfying event $E$ (that is, $e^u_1=(u,v^u)$ for all $u\in V^*$), we have
\begin{gather} \sum_{t\in V^*} b^+(t) \cd( \len(W_t)+ 2\cyc(W_t)) \ge  \sum_{t\in V^*} b^+(t) \cd d_{T_C}(u,r_C) + \opt'. \eqnl{toshow} \end{gather}
Assuming \eqn{toshow}, taking expectations and then applying \clm{cycle} gives
\begin{align*}
 \sum_{t\in V^*} b^+(t)\cd\E[\len(W_t)\mid E] + 2\e\cd\opt &\ge\sum_{t\in V^*} b^+(t) \cd\E[ \len(W_t)\mid E]+2\sum_{t\in V^*} b^+(t)\cd\E[\cyc(W_t)\mid E] \\&=\sum_{t\in V^*} b^+(t) \cd(\E[ \len(W_t)\mid E]+2 \E[\cyc(W_t)\mid E]) \\&\stackrel{\eqn{toshow}}\ge\E[  b^+(t) \cd d_{T_C}(u,r_C)\mid E ] + \opt' \eqnl{toshow} ,
\end{align*}
as desired.

For the remainder of the proof, we will show \eqn{toshow} given \emph{any} set of arbitrary infinite sequences $\{e^u_1,e^u_2,\lds:u\in V^*\}$ satisfying $e^u_1=(u,v^u)$ for all $u\in V^*$. In particular, now that we have fixed these infinite sequences, all randomness goes away, so there are no more probabilistic arguments for the remainder of the proof.

For each $t$, we can view the random walk $W_t$ as follows: for each vertex $u\in V^*$, the first time the walk reaches $u$, it must travel along arc $(u,v^u)$ in $\orA$, and any time after that, it can choose an arbitrary arc in $\Out(u)$. 

\BSCL\clml{charge2}
Let $C$ be a connected component of $A$ in \ref{esssp} containing a vertex $u$. If the walk $W$ contains $u$, then it contains every arc on the (unique, possibly empty) path from $u$ to the (unique) cycle in $C$ (defined as the path from $u$ to the closest vertex on that cycle).
\ESCL
\BSP
Consider the first time the walk visits vertex $u$. Then, the walk must traverse the arc $(u,v^u)=e^u_1$, which is the first arc along the path from $u$ to the cycle in $C$. We can then repeat the argument with $u$ replaced by $v^u$, considering the first time the walk visits vertex $v^u$. Continuing this argument until we arrive at a vertex on the cycle proves the claim.
\ESP

\BSCL\clml{charge}
Let $C$ be a connected component of $A$ in \ref{esssp} containing a vertex $u$. Suppose a walk $W$ contains an arc $(u,v)$ with $v\ne v^u$. Then, the cycles output by the greedy cycle-finding algorithm contains  (1) every arc along the (unique, possibly empty) path from $u$ to the (unique) cycle in $C$, and (2) every arc in the (unique) cycle in $C$ in the direction given by $\orA$.
\ESCL
\BSP
For this proof only, imagine that the walk visits one vertex per unit of time, so we say that the walk reaches a vertex $v$ \emph{at time $i$} if $v$ is the $i$'th vertex of the walk. 

We first prove  statement (2). Consider the first time $i$ that the walk reaches any vertex in $C$, which must occur since vertex $u$ in $C$ is reached eventually. Then, right after time $i$, the walk will travel along $\orA$ towards the unique cycle in $C$, and then travel along $C$ in the direction given by $\orA$. The greedy cycle-finding algorithm then removes that cycle, proving statement (2).

We now prove  statement (1). If $u$ is inside the cycle of $C$, then the path is empty and there is nothing to prove. Otherwise, let $(u',v')$ be an arbitrary arc along this (nonempty) path. We first show that $(u',v')$ is traveled at least once in the walk. For this proof only, for two vertices $u,v$ in $C$ (possibly $u=v$), let $P(u,v)$ be the (possibly empty) path in $\orA$ from $u$ to $v$ (inclusive), which will always exist and be unique when we use it. Consider the first time $i$ that the walk reaches any vertex $v$ in $P(u,u')$ (possibly $u$ or $u'$). Since the walk visits vertex $u$ at least once, this `first time $i$' must occur. Then, right after time $i$, the walk must travel along $P(v,u')$, and then along arc $(u',v')$. Let the walk reach vertex $u'$ at time $i'$ (so it reaches $v'$ at time $i'+1$).

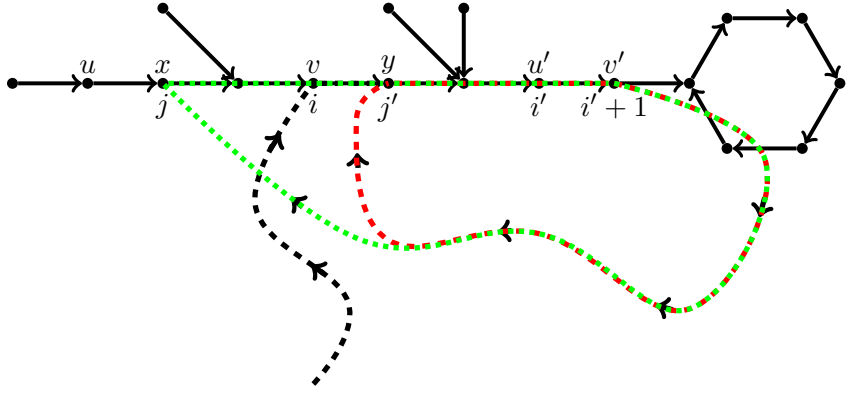
\begin{figure}\centering
\begin{tikzpicture}

\node [above] at (-5,0) {$u$};
\node [above] at (-4,0) {$x$};
\node [above] at (-2,0) {$v$};
\node [above] at (-1,0) {$y$};
\node [above] at (1,0) {$u'$};
\node [above] at (2,0) {$v'$};

\node [below] at (-4,0) {$j$};
\node [below] at (-2,0) {$i$};
\node [below] at (-1,0) {$j'$};
\node [below] at (1,0) {$i'$};
\node [below] at (2,0) {$i'+1$};

\tikzstyle{every node} = [scale=.4,fill,circle]

\node (v1) at (-6,0) {};
\node (v2) at (-5,0) {};
\node (v3) at (-4,0) {};
\node (v4) at (-3,0) {};
\node (v5) at (-2,0) {};
\node (v6) at (-1,0) {};
\node (v7) at (0,0) {};
\node (v8) at (1,0) {};
\node (v9) at (3,0) {};
\node (v15) at (2,0) {};
\draw[->,line width=1.5]  (v3) edge (v4);
\draw[->,line width=1.5]  (v1) edge (v2);
\draw[->,line width=1.5]  (v2) edge (v3);
\draw[->,line width=1.5]  (v4) edge (v5);
\draw[->,line width=1.5]  (v5) edge (v6);
\draw[->,line width=1.5]  (v6) edge (v7);
\draw[->,line width=1.5]  (v7) edge (v8);
\node (v10) at (3.5,1.732/2) {};
\node (v11) at (4.5,1.732/2) {};
\node (v12) at (5,0) {};
\node (v13) at (4.5,-1.732/2) {};
\node (v14) at (3.5,-1.732/2) {};
\draw [->,line width=1.5] (v9) edge (v10);
\draw [->,line width=1.5] (v10) edge (v11);
\draw [->,line width=1.5] (v11) edge (v12);
\draw [->,line width=1.5] (v12) edge (v13);
\draw [->,line width=1.5] (v13) edge (v14);
\draw [->,line width=1.5] (v14) edge (v9);
\draw [->,line width=1.5] (v8) edge (v15);
\draw [->,line width=1.5] (v15) edge (v9);
\draw [->,line width=2](4.0002,-1.5201) -- (3.9323,-1.7793);
\draw [->,line width=2](2.7784,-3.0134) -- (2.5562,-2.9517);
\draw [->,line width=2](0.6372,-1.9582) -- (0.4521,-1.9767);
\draw [->,line width=2](-2.1458,-1.6435) -- (-2.2939,-1.5201);
\draw [->,line width=2](-1.4053,-1.1252) -- (-1.4176,-0.9277);
\draw [dashed,line width=2] plot[smooth, tension=.7] coordinates {(-2,-4) (-1.5383,-2.9104) (-2.7706,-1.6285) (v5)};

\draw [red,dashed,line width=2] plot[smooth, tension=.7] coordinates {(v15) (4,-1) (3,-3) (1,-2) (-1.0017,-2.0856) (-1.429,-0.5154) (v6)};

\draw [green,dotted,line width=2] plot[smooth, tension=.7] coordinates {(v15) (4,-1) (3,-3) (1,-2) (-1.4091,-2.0657) (v3)};
\draw [dashed,line width=2](v5) -- (v6);
\draw [red,dashed,line width=2](v6) -- (v15);
\draw [green,dotted,line width=2](v6) -- (v15);
\draw [green,dotted,line width=2](v3) -- (v6);
\draw [->,line width=2](-1.8619,-2.5321) -- (-2.0285,-2.4148);
\draw [->,line width=2](-2.5746,-0.8815) -- (-2.4389,-0.6716);
\node (v16) at (-4,1) {};
\node (v17) at (-1,1) {};
\node at (0,1) {};
\draw [->,line width=1.5](v16) -- (v4);
\draw [->,line width=1.5](v17) -- (v7) ;
\draw [->,line width=1.5](0,1) -- (v7);
\end{tikzpicture}
\caption{The two cases of statement~(1) from \clm{charge}. The curved paths that do not align with the horizontal edges are arbitrary paths in other parts of the graph. The red dashed line marks the cycle for the first case, and the green dotted line marks the cycle for the second case. }
\label{fig:subclaim}
\end{figure}

We now show that this first traversal of arc $(u',v')$ at time $i'$ is indeed added to a cycle by the greedy algorithm. Define $v$ as before, and consider the next time $j$ the walk reaches any vertex $x$ in $P(u,v)$.  Then, right after time $j$, the walk must travel along $P(x,v)$. Since the walk contains an arc $(u,v)$ with $v\ne v^u$, vertex $u$ is reached at least twice, so this `next time $j$' must occur (we need at least \emph{twice} in case $v=u$). By time $j$, one of two things can happen (see Figure~\ref{fig:subclaim}):
\BE
\im The walk returns to some vertex $y$ in $P(v,u')$ before time $j$ (but after time $i'+1$). Let time $j'\in(i'+1,j)$ be the first such return, and let $y$ be the vertex returned to. Before time $j'$, all vertices in $P(v,u')$ were visited exactly once, so either all arcs in $P(v,v')$ were added to the same cycle before time $j$, or none of them were added to any cycle before time $j$. In the former case, the path $P(v,v')$ includes arc $(u',v')$, so we are done. In the latter case, we obtain the cycle consisting of the path $P(y,v')$, followed by the (remaining) arcs in the walk from time $i'+1$ to time $j'$. The greedy algorithm then adds this cycle, which contains arc $(u',v')$.
\im The walk does not return to any vertex in $P(v,u')$ after time $i'+1$ and before time $j$. By the same argument as the case above, either all arcs in $P(v,v')$ were added to the same cycle before time $j$, or none were added to any cycle before time $j$. Again, in the former case, we are done, so suppose the latter. Right after time $j$, the walk travels along $P(x,v)$, and no vertex on $P(x,v)$ except possibly $v$ was visited before time $j$. Therefore, the next time the walk encloses a cycle (after time $j$) is when it travels along $P(x,v)$ and reaches $v$. At this point, all arcs in $P(x,v')$ (including arc $(u',v')$) are added into the cycle that begins and ends at $v$, so we are done.
\EE
Therefore, in both cases, we are done.
\ESP

\BSCL\clml{length-bound}
For each vertex $t\in V^*$, let $C$ be the component in $A$ containing $t$. We have
\[ \len(W_t) + 2\cyc(W_t) \ge d_{T_C}(t,r_C) + d_{G'}(s,r_C) .\]
\ESCL
\BSP
Define a \emph{subwalk} of $W$ to be a contiguous subsequence of the walk when it is viewed as a sequence of vertices.
We partition the walk $W$ into subwalks as follows. Start with the first vertex $t$ of the walk, and let $C$ be the component of $A$ containing $t$. Consider the last vertex $v$ of the walk also inside $C$. If $v=s$, then we are done; otherwise, break of the subwalk from the beginning of the walk to the (last occurrence of) vertex $v$ in the walk, and recursively apply the procedure on the remaining vertices of the walk.

Let the subwalks be $W_1,W_2,\lds,W_k$ in that order, and for $i\in[k]$, let $u_i$ and $v_i$ be the first and last vertex of $W_i$, and let $C_i$ be the component of $A$ containing both $u_i$ and $v_i$. By construction, all components $C_i$ are distinct. For each $i\in[k-1]$, the arc $(v_i,u_{i+1})$ in $\orf$ is responsible for adding an arc $(v_{C_i},v_{C_{i+1}})$ in \line{addedge} of weight $w(v_i,u_{i+1})+d_{T_{C_i}}(v_i,r_{C_i})+d_{T_{C_{i+1}}}(u_{i+1},r_{C_{i+1}})+c(C_i)+c(C_{i+1})$; let this arc be $(v'_i,u'_{i+1})$. 
Consider the path from $r_C$ to $s'$ in $G'$ consisting of the edges $(v'_i,u'_{i+1})$ for $i\in[k-1]$. It suffices to show that this path has length at most $\len(W_t)+2\cyc(W_t)-d_{T_C}(t,r_C)$; the distance $d_{G'}(s,r_C)$ can only be smaller. In other words, we want to show that
\begin{align} \sum_{i=1}^{k-1} & \lp w(v_i,u_{i+1})+d_{T_{C_i}}(v_i,r_{C_i})+d_{T_{C_{i+1}}}(u_{i+1},r_{C_{i+1}})+c(C_i)+c(C_{i+1})\rp \nonumber\\
  &\le \len(W_t)+2\cyc(W_t)-d_{T_C}(t,r_C).\eqnl{bound-dist}\end{align}
For each $i\in[k-1]$, the distance $d_{T_{C_i}}(v_i,r_{C_i})$ equals the length of the (possibly empty) path $P_i$ in $C_i$ from $v_i$ to the (closest vertex on the unique) cycle in $C_i$. Since $u_{i+1}$ is not in $C_i$, we have $u_{i+1} \ne v_1^{u_{i+1}}$, so by \clm{charge}, every
arc in $P_i$ is added to some cycle by the greedy cycle-finding algorithm. Moreover, by \clm{charge}, every edge in the (unique) cycle in $C_i$ is also added to some cycle. All such arcs mentioned are distinct, so we obtain
\[ \sum_{i=1}^{k-1} \lp d_{T_{C_i}}(v_i,r_{C_i}) +c(C_i)\rp \le \cyc(W_t) .\]
For the distances $d_{T_{C_{i+1}}}(u_{i+1},r_{C_{i+1}})$ for $i\in[k-1]$, as well as the distance $d_{T_{C_{1}}}(u_{1},r_{C_{1}})=d_{T_C}(t,r_C)$, we charge them to the walk $W_t$ directly. For each $i\in[k]$, since the walk contains vertex $u_{i}$ in $C_{i}$, by \clm{charge2}, it contains all arcs on the path from $u_{i}$ to the cycle in $C_{i}$, whose weights sum to exactly  $d_{T_{C_{i}}}(u_{i},r_{C_{i}})$. Finally, we also charge the edge weights $w(v_i,u_{i+1})$ to the walk $W_t$, since the walk contains them by construction, and they are edge-disjoint from each other and all arcs in any $C_i$. Thus,
\begin{align*}
\sum_{i=1}^{k-1}&\lp w(v_i,u_{i+1})+d_{T_{C_i}}(v_i,r_{C_i})+d_{T_{C_{i+1}}}(u_{i+1},r_{C_{i+1}})+c(C_i)+c(C_{i+1})\rp \\&\le 2\sum_{i=1}^{k-1} \lp d_{T_{C_i}}(v_i,r_{C_i}) +c(C_i)\rp +\sum_{i=1}^{k-1}\lp w(v_i,u_{i+1})+d_{T_{C_{i+1}}}(u_{i+1},r_{C_{i+1}})\rp
\\&\le 2\sum_{i=1}^{k-1} \lp d_{T_{C_i}}(v_i,r_{C_i}) +c(C_i)\rp + \sum_{i=1}^{k-1}w(v_i,u_{i+1})+ \sum_{i=1}^{k}d_{T_{C_{i}}}(u_{i},r_{C_{i}})-d_{T_C}(t,r_C)
\\&\le 2\cyc(W_t) + \len(W_t)-d_{T_C}(t,r_C),
\end{align*}
proving \eqn{bound-dist}.
\ESP
We now resume the proof of \lem{bound-esssp}. Multiplying the inequality by $ b^+(t)$ for each $t$ and summing over all $t$ gives
\begin{gather} \sum_{t\in V^*} b^+(t)\cd  \len(W_t) + 2\cyc(W_t) \ge \sum_{t\in V^*} b^+(t)\cd (d_{T_C}(t,r_C) + d_{G'}(s,r_C) ) .\eqnl{sum-over-t}\end{gather}

Consider routing, for each component $C$ of $A$ and vertex $t$ in $C$, $ b^+(t)$ amount of flow along the shortest path from $s'$ to $v_C$ in $G'$. By construction of the demand vector $b'$, this is a transshipment flow that satisfies demands $b'$. Therefore, the optimum transshipment cost $\opt'$ can only be smaller (in fact, it is equal), and we obtain
\[ \opt' \le \sum_{t\in V^*} b^+(t)\cd d_{G'}(s,r_C) .\]
Together with \eqn{sum-over-t}, this concludes \eqn{toshow}, and hence \lem{bound-esssp}.
\EP

\BC\corl{bound-by-f}
Over the randomness of \ref{esssp} (in particular, the random choices on \line{sample-edge}),
\[ \E\lb \sum_{t\in V^*} b^+(t) \cd d_{T_C}(u,r_C) + \opt'  \rb \le \1{\overrightarrow C\orf} +2\e\cd\opt.\]
\EC
\BP
We apply \lem{bound-esssp} by taking the expectation of \eqn{bound-esssp} over the randomness on \line{sample-edge}, which effectively removes the conditioning by $E$, and obtain
\[ \sum_{t\in V^*} b^+(t)\cd\E[\len(W_t)] + 2\e\cd\opt \ge \E[ b^+(t) \cd d_{T_C}(u,r_C) +\opt'] .\]
This, along with $\sum_{t\in V^*} b^+(t)\cd\E[\len(W_t)] \le \1{\overrightarrow C\orf} $ from \clm{len}, finishes the proof.
\EP

\BCL\clml{T-as-good}
Define $T'$ and $T$ as in \ref{esssp}. For each component $C$ of $A$ and each vertex $v$ in $C$, we have $d_T(s,v) \le d_{T'}(s',v_C)+d_{T_C}(t,r_C)$.
\ECL
\BP
Follows easily by construction (lines~\ref{line:start-T}~to~\ref{line:end-T}), so proof is omitted.
\EP
\BCL\clml{final-approx}
Over the entire randomness of \ref{esssp} (including the recursion at \line{recurse}), we have
\[ \E\lb\sum_{t\in V} b^+(t)\cd d_T(s,t)\rb \le \al(1+3\e)\opt .\]
\ECL
\BP
By \clm{vstar}, we have $b^+(t)=0$ for all $t\in V\sm V^*$, so it suffices to prove
\[ \E\lb\sum_{t\in \textcolor{blue}{V^*}} b^+(t)\cd d_T(s,t)\rb \le \al(1+3\e)\opt .\]
By recursion, we have the guarantee 
\[ \E\lb \sum_{C:v_C\ne s'} b'_{v_C} \cd d_{T'}(s',v_C)\rb\le \al \opt .\]
For each vertex $t$, let $C_t$ be the component of $A$ containing $t$. We have
\begin{align*}
\E\lb \sum_{t\in V^*} b^+(t) \cd d_T(s,v) \rb &\stackrel{\textup{Clm.\,}\ref{clm:T-as-good}}\le \E\lb \sum_{t\in V^*} \lp b^+(t)\cd d_{T'}(s',v_{C_t}) + d_{T_{C_t}}(t,r_{C_t}) \rp \rb
\\& = \E\lb \sum_{t\in V^*}  b^+(t)\cd d_{T'}(s',v_{C_t}) \rb + \E\lb \sum_{t\in V^*} b^+(t)\cd d_{T_{C_t}}(t,r_{C_t})\rb
\\&= \E\lb \sum_{C:v_C\ne s'} b'_{v_C} \cd d_{T'}(s',v_C)\rb+ \E\lb \sum_{t\in V^*} b^+(t)\cd d_{T_{C_t}}(t,r_{C_t})\rb
\\&\le \al\E[\opt'] +  \E\lb \sum_{t\in V^*} b^+(t)\cd d_{T_{C_t}}(t,r_{C_t})\rb
\\&\le \al\lp \E[ \opt'] +  \E\lb \sum_{t\in V^*} b^+(t)\cd d_{T_{C_t}}(t,r_{C_t})\rb  \rp
\\&\stackrel{\textup{Cor.\,}\ref{cor:bound-by-f}}\le \al\lp \1{\overrightarrow C\orf}+2\e\cd\opt\rp
\\&\le \al((1+\e)\opt + 2\e\cd\opt)
\\&=\al(1+3\e)\opt,
\end{align*}
where the last inequality uses that $\orf$ is a $(1+\e)$-approximate transshipment solution.
\EP



\subsection{Parallelizing the Expected SSSP Algorithm}\secl{parallel-esssp}

\newcommand{\diam}{\textup{diam}}
\newcommand{\ol}{\overline}
\newcommand{\BLUE}{\textcolor{blue}}

In this section, we parallelize the algorithm \ref{esssp} by removing its sequential recursive calls which would, if left unchecked, blow up the parallel running time of the algorithm. This is because every time \ref{esssp} calls itself, it requires a transshipment flow of a new recursive graph, which in turn requires an $\el_1$-embedding of it.

Our solution is to avoid computing an $\el_1$-embedding (from scratch) at each step of the recursion. With an $\el_1$-embedding at hand for a given recursive graph, we can compute the transshipment flow without any recursion by invoking \cor{l1}. Naively, one might hope that a subset of the original vectors of the $\el_1$-embedding automatically produces an $\el_1$-embedding for the recursive graph instance. This is not true in general, but we can \emph{force} it to happen at a cost: we add a set of \emph{virtual} edges to the recursive graph instance (which may change its metric) that come from a \emph{spanner} of the $\el_1$-metric on a subset of the original vectors. These edges may not exist in the original graph $G$, so the shortest path tree $T$ of that instance may include edges that do not correspond to edges in the original graph. Unraveling the recursion, the final tree $T$ may also contain virtual edges not originally in $G$. However, we ensure that these virtual edges do not change the metric of the \emph{original} graph $G$. Hence, while the tree $T$ is not a \emph{spanning} SSSP tree, it still fulfills its purpose when it is called in the algorithm \ref{sssp} of \sec{sample-tree}. (If a \emph{spanning} SSSP tree is explicitly required for the final output of the original graph instance, then \ref{esssp} can be called in \ref{sssp} once on the initial recursion loop, and \ref{esssp-rec} called on every recursive instance of \ref{sssp} afterwards.)

\BL\leml{esssp-rec}
Let $G=(V,E)$ be a connected graph with $n$ vertices and $m$ edges with aspect ratio $\poly(n)$, let $\e>0$ be a parameter. 
Given graph $G$, a source $s\in V$, and an $\el_1$-embedding of it into $O(\logn)$ dimensions with distortion $\pl(n)$, we can compute, in 
$\tO(m)$ work and $\pl(n)$ time, a set $E^+$ of \emph{virtual} edges supported on $V$ and a tree $T\s E\cup E^+$ such that
 \BE
 \im For all edges $(u,v)\in E^+$, $w(u,v) \ge d_G(u,V)$, i.e., the edges $E^+$ do not change the metric of $G$
 \im $T$ is a $(1+\e)$-approximate expected $s$-SSSP tree on $G\cup E^+$
 \EE
\EL

We now present the recursive algorithm that proves \lem{esssp-rec}. Let the original graph be $G_0$ and the current recursive instance be $G$, and let $n$ and $m$ always indicate the number of vertices and edges of the original graph $G_0$. We highlight in \BLUE{blue} all modifications of \ref{esssp} that we make.

\begin{algorithm}[H]
\mylabel{esssp-rec}{\texttt{ESSSP-Rec}}\caption{\ref{esssp-rec}($G=(V,E),\BLUE{\{y_v\in\R^{k}:v\in V\}},s,b,(1+3\e)\al$)}
\BLUE{Global variables: $G_0=(V_0,E_0)$ is the original input graph; $\{x_v\in\R^{O(\logn)}:v\in V\}$ is an initial $\el_1$-embedding of $G_0$ with distortion $D=\pl(n)$ given as input}
\\ Assumption: \textcolor{blue}{$k=O(\log^2n)$ and $\{y_v\}$ satisfies $\f1{8kD}d_G(u,v)\le\1{y_u-y_v} \le (D+6)\cd d_G(u,v)$ for all $u,v\in V$;} demand vector $b$ satisfies $b_s>0$ and $b_t\le0$ for all $t\in V\sm s$
\begin{algorithmic}[1]
\State \BLUE{$\{8kD\cd y_v\}$ is an $\el_1$-embedding of $G$ with distortion $8kD(D+6)=\pl(n)$. Apply dimension reduction to obtain an $\el_1$-embedding of $G$ into $O(\logn)$ dimensions instead of $O(\log^2n$), with a slightly worse but still $\pl(n)$ distortion (see proof of \thm{embed-seq}). Using it and \cor{l1},} compute a $(1+\e$)-approximate transshipment $f$ on $G$ with demand vector $b$
\State Initialize the digraph $\orA\gets\emptyset$
\State Every vertex $u\in V\sm s$ with $\In(u)\ne\emptyset$ independently samples a random neighbor $v\in\Out(u)$ with probability $\orf(u,v)/\FOut(u)$ and adds arc $(u,v)$ to $\orA$ \linel{sample2}
\State Add a self-loop $(s,s)$ of zero weight to $\orA$
\State Let $A$ be the undirected version of $\orA$
\State Initialize $G'\gets(\emptyset,\emptyset)$ as an empty undirected graph \Comment{Graph to be recursed on, with $\le n/2$ vertices}
\State Initialize $b'$ as an empty vector \Comment{Demands to be recursed on}
\For{each connected component $C$ of $A$}
 \State $c(C)\gets$ total weight of edges in the (unique) cycle in $C$ (possibly the self-loop $(s,s)$)
 \State Let $T_C$ be the graph $C$ with its (unique) cycle contracted into a single vertex $r_C$ \Comment{$T_C$ is a tree}
 \State \textcolor{blue}{Let $v_C\in V$ be an arbitrary vertex on the cycle in $C$}
 \State Add the vertex $v_C$ to $G'$, and set demand $b'_{v_C}\gets\sum_{v\in V(C)}b_v$
\EndFor
\For{each edge $(u,u')$ in $E$}
 \State Let $C$ and $C'$ be the connected components of $A$ containing $u$ and $u'$, respectively
 \If{$C\ne C'$}
  \State Add an edge between $v_C$ and $v_{C'}$ with weight $w(u,u')+d_{T_C}(u,r_C)+d_{T_{C'}}(u',r_{C'})+c(C)+c(C')$ \linel{addedge2}
 \EndIf
\EndFor
\State \BLUE{Note that $s=v_{C_s}$}, where $C_s$ is the component of $A$ containing $s$
\State Collapse parallel edges of $G'$ by only keeping the parallel edge with the smallest weight \linel{G'}
\State \BLUE{Call \lem{l1} on the original $\el_1$-embedding vectors $\{x_v:v\in V'\}$, returning a set of vectors $\{y'_v\in\R^{O(\log^2n)}:v\in V'\}$ that satisfy \eqn{yv}}
\State \BLUE{Call \lem{Eplus} on $\{y'_v:v\in V'\}$, returning a set $E^+$ of edges supported on $V'$ that satisfy both conditions of \lem{Eplus}} \linel{Eplus}
\State Recursively call $\ref{esssp-rec}(\textcolor{blue}{G'\cup E^+,\{y'_v:v\in V'\},s},b',\al)$, obtaining an $\al$-approximate expected SSSP tree $T'$ of $G'\cup E^+$ \linel{recurse2}
\State Initialize $T\gets\emptyset$ \Comment{The expected SSSP tree}
\For{each edge $(v,v')$ in $T'\,\BLUE{\cap \,E}$}\linel{start-T2}
 \State Let $(u,u')\in E$ be the edge responsible for adding edge $(v,v')$ to $G'$
 \State Add edge $(u,u')$ to $T$
\EndFor
\BLUE{
\For{each edge $(v,v')$ in $T'\sm E$}
 \State Let $(u,u')\in E$ be the edge responsible for adding edge $(v,v')$ to $G'$
 \State Let $C$ and $C'$ be the connected components of $A$ containing $u$ and $u'$, respectively
 \State Add edge $(v,v')$ to $T$ with weight $w(v,v')-c(C)-c(C')$ \Comment{These edges are not in the original edge set $E_0$} \linel{undo}
\EndFor
}
\For{each connected component $C$ of $A$}
 \State Remove an arbitrary edge from the (unique) cycle inside $C$, and add the resulting tree to $T$\linel{end-T2}
\EndFor
\State\Return $T$
\end{algorithmic}
\end{algorithm}

One notable difference is that we always enforce the recursive vertex set $V'$ to be a subset of the current vertex set $V$. This allows us to compare the metric of the new graph $G'$ to the current graph $G$ in a more direct way. 

\BCL\clml{each-edge}
Each edge $(v_C,v_{C'})$ added on \line{addedge2} has weight at least $d_G(v_C,v_{C'})$.
\ECL
\BP
Let $(u,u')\in E$ be the edge responsible for adding the edge $(v_C,v_{C'})$. We construct a path in $G$ from $v_C$ to $v_{C'}$ whose distance is at most $w(u,u')+d_{T_C}(u,r_C)+d_{T_{C'}}(u',r_{C'})+c(C)+c(C')$. By construction of component $C$, there is a path inside $C$ from $v_C$ to $u$ of length at most $c(C)+d_{T_C}(u,r_C)$. Similarly, there is a path inside $C'$ from $u'$ to $v_{C'}$ of length at most $c(C')+d_{T_{C'}}(u',r_{C'})$. Adding the edge $(u,u')$ completes the path from $v_C$ to $v_{C'}$.
\EP

\BL\leml{l1}
Given the vectors $\{x_v\}$, we can compute, in $\tO(m)$ work and $\pl(n)$ time, a set of vectors $\{y_v\in\R^{O(\log^2n)}:v\in V'\}$ such that w.h.p., for all vertices $v_C,v_{C'}\in V'$,
\begin{align}
\1{x_{v_C}-x_{v_{C'}}} + 2(c(C)+c(C')) \le \1{y_{v_C}-y_{v_{C'}}}\le \1{x_{v_C}-x_{v_{C'}}} + 6(c(C)+c(C')) . \eqnl{yv}
\end{align}
\EL
\BP
We introduce $O(\log^2n)$ new coordinates, indexed by $(i,j)\in\Z\times[s]$. For a given component $C$ of $A$ at this instance, initialize $y_{v_C}=x_{v_C}$ without the new coordinates. Let $i\in\Z$ be the integer $i$ such that $c(C) \in [2^i,2^{i+1})$, which can take one of $O(\logn)$ values in $\Z$. For each $j\in[s]$, let the coordinate at index $(i,j)$ take value $\pm5\cd2^i/s$, each with probability $1/2$. This concludes the construction of $y_{v_C}$.

Fix vertices $v_C,v_{C'}\in V'$. Let $i,i'$ be the integers such that $c(C)\in[2^i,2^{i+1})$ and $c(C')\in[2^{i'},2^{i'+1})$. If $i\ne i'$, then by construction, the $\el_1$ distance incurred by the additional $O(\log^2n)$ coordinates is exactly $5\cd2^i+5\cd2^{i'}$. Moreover, by the input guarantee of $\{x_v:v\in V\}$,
\[ d_{G_0}(v_C,v_{C'})\le\1{x_{v_C}-x_{v_{C'}}}\le D\cd d_G(v_C,v_{C'}).\]
 Therefore,
\begin{align}
 \1{x_{v_C}-x_{v_{C'}}}+2\cd( c(C)+c(C')) &\le \1{x_{v_C}-x_{v_{C'}}}+ 2\cd2^{i+1}+2\cd2^{i'+1} \nonumber
\\&\le \1{x_{v_C}-x_{v_{C'}}}+ 5\cd2^i+5\cd2^{i'} \nonumber
\\&= \1{y_{v_C}-y_{v_{C'}}} \eqnl{dGv}
\\&\le \1{x_{v_C}-x_{v_{C'}}}+ 6\cd2^i+6\cd2^{i'} \nonumber
\\&\le \1{x_{v_C}-x_{v_{C'}}}+ 6\cd( c(C)+c(C')), \nonumber
\end{align}
as promised. Now suppose that $i=i'$. Then, for each of the $s$ coordinates $(i,j)$, there is a $1/2$ probability of contributing $0$ to $\1{y_{v_C}-y_{v_{C'}}}$, and a $1/2$ probability of contributing $6\cd2^i/s$. Since $s=O(\logn)$, by a simple Chernoff bound, the probability that the total contribution of these $s$ coordinates is in $[4\cd2^i/s,6\cd2^i/s]$ is at least $1-1/\poly(n)$. A similar bound to \eqn{dGv} proves the claim.
\EP

\BL\leml{Eplus}
Given vectors $\{y_v\in\R^{O(\log^2n)}\}$ satisfying \eqn{yv}, we can compute, in $\tO(|E|)$ work and $\pl(n)$ time, a set $E^+$ of edges supported on $V'\s V$ such that
 \BE
 \im For all edges $(v_C,v_{C'})\in E^+$, $w_{E^+}(v_C,v_{C'}) \ge d_{G_0}(v_C,v_{C'})+2(c(C)+c(C'))$,
 \im For all vertices $u,v\in V'$,
 \[ \f1{8kD}\cd d_{G'\cup E^+}(u,v) \le \1{y_u-y_v} \le (D+6) \cd d_{G'\cup E^+}(u,v) .\]
 
 \EE
\EL
\BP


Let $H$ be an $(8kD)$-spanner of the (complete) $\el_1$-metric induced by the vectors $\{y_v:v\in V'\}$. That is, for all $u,v\in V'$,
\[ \1{y_u-y_v} \le d_{H}(u,v) \le 8kD\cd\1{y_u-y_v} .\]
 We show later how to compute it efficiently, but for now, assume we have computed such a graph $H$. We set the edges $E^+$ as simply the edges of $H$ (with the same weights). 

Every edge $(v_C,v_{C'})\in E^+$ satisfies
\[ w_{E^+}(v_C,v_{C'}) \ge d_{H}(v_C,v_{C'}) \ge \1{y_{v_C}-y_{v_{C'}}} \stackrel{\eqn{dGv}}\ge \1{x_{v_C}-x_{v_{C'}}}+2(c(C)+c(C')) ,\]
and $\1{x_{v_C}-x_{v_{C'}}}\ge d_{G_0}(v_C,v_{C'})$ since $\{x_v\}$ is an $\el_1$-embedding for $G_0$, fulfilling condition~(1).
For each pair of vertices $u,v\in V'$, 
\[ d_{G'\cup E^+}(u,v)\le d_{H}(u,v) \le 8kD\cd\1{y_u-y_v} ,\]
fulfilling the lower bound in condition~(2).

For the upper bound in condition~(2), it suffices to show that, for all edges $(v_C,v_{C'})\in E'\cup E^+$,
\[ \1{y_u-y_v} \le (D+6)\cd w_{G'\cup E^+}(u,v) .\]
If $(v_C,v_{C'})\in E^+$, then
\[ \1{y_{v_C}-y_{v_{C'}}} \le d_H(v_C,v_{C'}) \le w_H(v_C,v_{C'}) = w_{G'\cup E^+}(v_C,v_{C'}) .\]
Otherwise, if $(v_C,v_{C'})\in E'$, then for the edge $(u,u')\in E$ with $u\in C$ and $u'\in C'$ responsible for this edge,
\begin{align*}
\1{y_{v_C}-y_{v_{C'}}} &\stackrel{\eqn{dGv}}\le\1{x_{v_C}-x_{v_{C'}}}+6(c(C)+c(C'))
\\&\le D\cd d_{G_0}(v_C,v_{C'}) + 6 w_{G'}(v_C,v_{C'})
\\&\stackrel{\eqn{only-inc}}\le D\cd d_{G}(v_C,v_{C'})+6 w_{G'}(v_C,v_{C'})
\\&\stackrel{\text{Clm.}\ref{clm:each-edge}}\le  D\cd w_{G'}(v_C,v_{C'}) +6 w_{G'}(v_C,v_{C'}), 
\end{align*}
assuming the following inequality holds:
\begin{gather}
d_G(u,v)\ge d_{G_0}(u,v) \qquad \forall u,v\in V.  \eqnl{only-inc}
\end{gather}
To prove \eqn{only-inc}, we assume by induction (on the recursive structure of \ref{esssp-rec}) that $d_G(u,v)\ge d_{G_0}(u,v)$ for all $u,v\in V$. By \clm{each-edge}, the edges $(u,v)$ added to $G'$ on \line{addedge2} have weight at least $d_G(u,v)\ge d_{G_0}(u,v)$. Moreover, since condition~(1) of the lemma is satisfied, we have $w(u,v)\ge d_{G_0}(u,v)$ for all edges $(u,v)\in E^+$ as well. It follows that $d_{G_0}(u,v)\le d_{G'\cup E^+}(u,v)$ for all $u,v\in V'$, fulfilling the inequality as well as completing the induction.
This concludes the upper bound of condition~(2), as well as the proof of \lem{Eplus} modulo the construction of the spanner $H$.

Finally, we show how to compute the spanner $H$ of the $\el_1$-metric induced by the vectors $\{y_v\in\R^k:v\in V\}$. We use a randomly shifted grid approach similar to the one in Algorithm~\ref{alg:routing}. Assume without loss of generality (by scaling and rounding) that all coordinates are positive integers with magnitude at most $M=\poly(n)$. We repeat the following process $O(\logn)$ times, computing a graph $H_j$ on trial $j$. For each $W$ a power of two in the range $[1,8kM]$, choose independent, uniformly random real numbers $r_1,\lds,r_k\in[0,W)$, and declare equivalence classes on the vertices where two vertices $u,v\in V$ are in the same class if $\lf (y_u)_i+r_i\rf_W=\lf(y_v)_i+r_i\rf_W$ for all coordinates $i\in[k]$. For each equivalence class $U\s V$, select an arbitrary vertex $u\in U$ as its representative, and for all $v\in U\sm\{u\}$, add an edge $(u,v)$ to $H_j$ of weight $\1{y_u-y_v}$. This concludes the construction of graph $H_j$, which has $O(|V|\log|V|)$ edges, since each value of $W$ adds at most $|V|$ edges. The spanner $H$ is the union of all graphs $H_j$ and has size $O(|V|\log|V|\logn)$.

Since all added edges $(u,v)$ have weight $\1{y_u-y_v}\ge d_G(u,v)$, distances in the final spanner $H$ are at least the distances in $G$. To show that distances in $H$ are not stretched too far, we show that for any vertices $u,v\in V$, with constant probability, each graph $H_j$ satisfies $d_{H_j}(u,v)\le 8kD\cd d_G(u,v)$. Then, since there are $\Th(\logn)$ graphs $H_j$, the probability that we do not have $d_H(u,v)\le d_{H_j}(u,v)\le 8kD\cd d_G(u,v)$ for any $j$ is $1/\poly(n)$, as promised.

Let $W$ be the smallest power of two greater than or equal to $2\1{y_u-y_v}$. Note that $\1{y_u-y_v} \le 2kM$, so such a $W$ always exists. It is not hard to show that, for each coordinate $i\in[k]$, the probability that $\lf (y_u)_i+r_i\rf_W=\lf(y_v)_i+r_i\rf_W$ is exactly $1-|(y_u - y_v)_i| / W \ge 1/2$. Therefore, the probability that $\lf (y_u)_i+r_i\rf_W=\lf(y_v)_i+r_i\rf_W$ for all $i$ is 
\begin{align*}
 \Prod_{i=1}^k\lp1-\f{|(y_u-y_v)_i|}W\rp \ge \Prod_{i=1}^k \exp\lp-2\f{|(y_u-y_v)_i|}W\rp &= \exp\lp -2\sum_{i=1}^k\f{|(y_u-y_v)_i|}W \rp \\&= \exp\lp-2\f{\1{y_u-y_v}}W\rp \ge \exp(-1) ,
\end{align*}
which is a constant. Therefore, with at least constant probability, $u$ and $v$ belong to the same equivalence class for $W$. In this case, since all vertices in this equivalence class have their vectors in a cube of side length $W$, we either added the edge $(u,v)$ of weight $\1{y_u-y_v}\le D\cd d_G(u,v)$, or we selected some vertex $v'$ and added the edges $(u,v')$ and $(v,v')$ of total weight at most
\[ \1{y_u-y_{v'}}+\1{y_v-y_{v'}} \le 2kW \le 2k \cd 4\1{y_u-y_v} \le 2k\cd 4\cd D\cd d_G(u,v) .\]
Thus, $H$ is a $8kD$-spanner w.h.p.
\EP

We now argue that, for the input graph $G_0$, source $s\in V_0$, demand vector $b_0$, and an $\el_1$-embedding $\{x_v\in\R^{O(\logn)}:v\in V_0\}$ of $G_0$ with distortion $D$, $\ref{esssp-rec}(G_0,\{x_v\},s,b_0,(1+3\e)^{\log_2n})$ returns a $(1+3\e)^{\log_2n}$-approximate expected $s$-SSSP tree. We will follow the arguments from \sec{esssp} almost line-by-line; the only changes we will highlight in \BLUE{blue}. Define $V^*,b^+,\overrightarrow C,\orf$ on the input graph $G=(V,E)$ and demands $b$ identically as in \sec{esssp}. As before, define $\opt=\opt_G(b)$.

\BCL[Restatement of \clm{T-as-good}]\clml{T-as-good2}
Define $T'$ and $T$ as in \ref{esssp-rec}. For each component $C$ of $A$ and each vertex $v$ in $C$, we have $d_T(s,v) \le d_{T'}(s',v_C)+d_{T_C}(t,r_C)$.
\ECL
\BP
Follows easily by construction (lines~\ref{line:start-T2}~to~\ref{line:end-T2}), so proof is omitted. \BLUE{Observe that the extra terms\,${}-c(C)-c(C')$ in \line{undo} are necessary here.}
\EP
\BC[Restatement of \cor{bound-by-f}]\corl{bound-by-f2}
Over the randomness of \ref{esssp-rec} (in particular, the random choices on \line{sample2}),
\[ \E\lb \sum_{t\in V^*} b^+(t) \cd d_{T_C}(u,r_C) + \opt_{G'\cup E^+}(b')  \rb \le \1{\overrightarrow C\orf} +2\e\cd\opt.\]
\EC
\noindent \BLUE{The optimum can only decrease with the addition of edges $E^+$ to $G'$, so we have}
\begin{gather} \BLUE{\opt_{G'\cup E^+}(b') \le \opt_{G'}(b') .\eqnl{cupE}}\end{gather}

\BCL[Restatement of \clm{final-approx}]\clml{final-approx2}
Over the entire randomness of \ref{esssp-rec} (including the recursion at \line{recurse2}), we have
\[ \E\lb\sum_{t\in V} b^+(t)\cd d_T(s,t)\rb \le \al(1+3\e)\opt .\]
\ECL
\BP
We follow the proof of \clm{final-approx}.
By \clm{vstar}, we have $b^+(t)=0$ for all $t\in V\sm V^*$, so it suffices to prove
\[ \E\lb\sum_{t\in \textcolor{blue}{V^*}} b^+(t)\cd d_T(s,t)\rb \le \al(1+3\e)\opt .\]
By recursion, we have the guarantee 
\[ \E\lb \sum_{C:v_C\ne s'} b'_{v_C} \cd d_{T'}(s',v_C)\rb\le \al \opt .\]
For each vertex $t$, let $C_t$ be the component of $A$ containing $t$. We have
\begin{align*}
\E\lb \sum_{t\in V^*} b^+(t) \cd d_T(s,v) \rb &\stackrel{\BLUE{\textup{Clm.\,}\ref{clm:T-as-good2}}}\le \E\lb \sum_{t\in V^*} \lp b^+(t)\cd d_{T'}(s',v_{C_t}) + d_{T_{C_t}}(t,r_{C_t}) \rp \rb
\\& = \E\lb \sum_{t\in V^*}  b^+(t)\cd d_{T'}(s',v_{C_t}) \rb + \E\lb \sum_{t\in V^*} b^+(t)\cd d_{T_{C_t}}(t,r_{C_t})\rb
\\&= \E\lb \sum_{C:v_C\ne s'} b'_{v_C} \cd d_{T'}(s',v_C)\rb+ \E\lb \sum_{t\in V^*} b^+(t)\cd d_{T_{C_t}}(t,r_{C_t})\rb
\\&\le \al\E[\opt_{G'\cup E^+}(b')] +  \E\lb \sum_{t\in V^*} b^+(t)\cd d_{T_{C_t}}(t,r_{C_t})\rb
\\& \textcolor{blue}{ \stackrel{\eqn{cupE}}\le \al\E[\opt_{G'}(b')] +  \E\lb \sum_{t\in V^*} b^+(t)\cd d_{T_{C_t}}(t,r_{C_t})\rb }
\\&\le \al\lp \E[ \opt_{G'}(b')] +  \E\lb \sum_{t\in V^*} b^+(t)\cd d_{T_{C_t}}(t,r_{C_t})\rb  \rp
\\&\stackrel{\BLUE{\textup{Cor.\,}\ref{cor:bound-by-f2}}}\le \al\lp \1{\overrightarrow C\orf}+2\e\cd\opt\rp
\\&\le \al((1+\e)\opt + 2\e\cd\opt)
\\&=\al(1+3\e)\opt,
\end{align*}
where the last inequality uses that $\orf$ is a $(1+\e)$-approximate transshipment solution.
\EP

Thus, by unraveling the recursion and repeatedly applying \clm{final-approx2}, the algorithm \ref{esssp-rec} computes a tree $T$ for $G_0$, possibly with virtual edges not in $G_0$, such that
\[ \E\lb\sum_{t\in V} b^+(t)\cd d_T(s,t)\rb \le (1+3\e)^{\log_2n}\opt .\]
It remains to show that $T$ is a ``valid'' approximate expected SSSP tree in the following sense: the virtual edges added to $T$ do not change the metric on $G_0$. This property is sufficient when the expected SSSP algorithm is used in \ref{sssp}.

For each vertex $v\in V_0$, consider the recursive instances with $v\in V$; let $C_i(v)$ be the components containing $v$ on previous levels of recursion, with $C_i$ as the value at recursion level $i$. We will need the following quick claim:

\BCL\clml{double}
Consider a recursion level $i$ with input graph $G=(V,E)$.
For all vertices $v\in V$, the lengths $c(C_j(v))$ of the cycles satisfy $c(C_{j+1}(v)) \ge 2 c(C_j(v))$ for all $j<i$.
\ECL
\BP
If $v=s$, then it holds because $c(C_j(v))=0$ for all $j$, so assume that $v\ne s$.
For a fixed recursion level $j$, by construction, every edge adjacent to $v$ has weight at least $c(C_i(v))$ in the graph $G'$ at that level. The cycle $C_{i+1}(v)$ must contain at least two such edges, so its total length is at least $2c(C_i(v))$.
\EP

For each virtual edge $(u,v)$ in $T$, there exists some recursion level $i$ and some edge $(u,v)\in E^+$ at that level of weight
\[ w_{E^+}(u,v) =  w_T(u,v) + \sum_{j=0}^i (c(C_j(u))+c(C_j(v))).\]
By \clm{double}, the sequence $c(C_0(u)),c(C_1(u)),\lds,c(C_i(u))=c(C)$ has each term at least double the previous, which means that two times the last term is at least the sum of all terms:
\[ 2c(C)=2c(C_j(u))\ge  \sum_{j=0}^ic(C_j(u)) .\]
Similarly,
\[ 2c(C')=2c(C_j(v))\ge\sum_{j=0}^ic(C_j(v)) .\]
By condition~(1) of \lem{Eplus}, 
\[ w_{E^+}(u,v) \ge d_{G_0}(u,v)+2(c(C)+c(C')) .\]
Combining these inequalities gives 
\[ w_T(u,v)= w_{E^+}(u,v) - \sum_{j=0}^i (c(C_j(u))+c(C_j(v))) \ge w_{E^+}(u,v)-2(c(C)+c(C')) \ge d_{G_0}(u,v) ,\]
as claimed.

\section{Sampling a Primal Tree}\secl{sample-tree}

In this section, we prove \thm{sample-tree}. by reducing the problems of computing a $(1+\e)$-approximate SSSP tree and potential to the approximate transshipment problem, and then using the expected SSSP tree subroutine \ref{esssp-rec} of \Cref{sec:parallel-esssp}. Most of these ideas originate from \cite{becker}, and we adapt their ideas and present them here for completeness. In particular, we  do not claim any novelty in this section.

\SampleTree*


We now briefly describe our algorithm for \thm{sample-tree}. First, we run the expected SSSP algorithm \ref{esssp-rec} with demands $\sum_{v\ne s}(\mathbbm1_v-\mathbbm1_s)$, obtaining distances that are near-optimal \emph{in expectation} in the two different ways. Of course, what we need is that \emph{all} distances are near-optimal. This is where the potential $\phi$ is useful: using it, we can approximately tell which vertices have near-optimal distances. Then, among the vertices $V'\s V\sm s$ whose distances are not near-optimal, we then compute another transshipment instance with demands $\sum_{v\in V'}(\mathbbm1_v-\mathbbm1_s)$ and repeat the process. As long as the set of remaining vertices $V'$ drops by a constant factor each round in expectation, we only require $O(\logn)$ rounds w.h.p.

To construct the potential, our strategy is simple:\ we simply take the coordinate-wise maximum of all potentials $\phi$ found over the iterations (assuming $\phi(s)=0$ always). For each vertex $v\in V\sm s$, since at least one iteration computes a near-optimal distance for $v$, the corresponding potential is also near-optimal.

Constructing the specific SSSP\ tree requires a little more care. We now describe our algorithm in pseudocode below.

\begin{algorithm}[H]
\mylabel{sssp}{\texttt{TS-to-SSSP}}\caption{\ref{sssp}($G=(V,E),\be\in(0,1],\{x_v\in\R^{O(\logn)}:v\in V\}$)}
Assumption: $\{x_v\in\R^{O(\logn)}:v\in V\}$ is an $\el_1$-embedding of $G$ with distortion $\pl(n)$
\begin{algorithmic}[1]
\State Initialize $V'\gets V\sm s$ \Comment{$V'$ is the set of vertices whose distances still need to be computed}
\State Initialize $d^*:V\sm s\to\R\cup\infty$ as $d^*(v)=\infty$ everywhere \Comment{$d^*$ tracks the best distance found for each vertex $v$}
\State Initialize $p^*:V\sm s\to V\cup\{\bot\}$ as $p^*(v)=\bot$ everywhere \Comment{$p^*$ is the ``parent'' function, used to construct the final SSSP tree}
\State Initialize $\phi^*:V\sm s\to\R$ as $\phi^*(v)=0$ everywhere \Comment{$\phi^*$ tracks the best potential found for each vertex $v$}
\While {$V'\ne\emptyset$} \linel{sssp-loop}
\State With the $\el_1$-embedding $\{x_v:v\in V\}$, call \cor{l1} on demands $\sum_v(\mathbbm1_v-\mathbbm1_s)$ to obtain $(1+\e/10)$-approximate flow-potential pair $(f,\phi)$, where $f$ is an acyclic flow and $\phi(s)=0$
 \State Set demands $b\gets \sum_{v\in V'}(\mathbbm1_v-\mathbbm1_s)$ for this iteration
 \State With the $\el_1$-embedding $\{x_v:v\in V\}$, call \ref{esssp-rec}$(G,\{x_v:v\in V\},s,b,\Th(\f\e\logn))$ to obtain a SSSP tree $T$ with $\E \lb \sum_{v\in V'} d_T(s,v)\rb \le (1+\f\e{10})\opt $\Comment{Satisfies the recursion in \eqn{esssp-rec}} 
 \State Compute distances $d_T(s,v)$ for all $v\in V'$
 \For{each vertex $v\in V'$ {in parallel}}
  \State Let $p(v)$ be the second-to-last vertex on the path from $s$ to $v$ in $T$
  \If{$d_T(s,v)<d^*(v)$} 
   \State Update $d^*(v)\gets d_T(s,v)$\linel{dT}
   \State Update $p^*(v)\gets p(v)$\linel{p*}
  \EndIf
  \State Update $\phi^*(v)\gets\max\{\phi^*(v),\phi(v)\}$
  \If{$d_T(s,v) \le (1+\e)\phi_v$} \linel{dT-small}
   \State Remove $v$ from $V'$
  \EndIf
 \EndFor
\EndWhile
\State Initialize $T^*\gets\emptyset$ \Comment{$T^*$ is the final SSSP\ tree that we compute}
\For{each vertex $v\in V\sm s$}
 \State Add edge $(v,p^*(v))$ to $T^*$
\EndFor
\State\Return SSSP tree $T^*$ and potential $\phi^*$ (augmented with $\phi^*(s):=0$)
\end{algorithmic}\label{alg:TS-to-SSSP}
\end{algorithm}

\BCL
If \ref{sssp} finishes, then the potential $\phi^*$ that it returns is a $(1+\e)$-approximate $s$-SSSP potential, and the returned $T^*$ is a $(1+\e)$-approximate SSSP tree.
\ECL
\BP
\ref{sssp} essentially takes the coordinate-wise maximum over all potentials $\phi$ that Algorithm~$\m A$ computes over the iterations. For each vertex $v\in V\sm s$, since it was removed from $V'$ at some point, some potential $\phi$ computed satisfies $(1+\e)\phi_v\ge d_T(s,v)\ge d(s,v)$ (\line{dT-small}), so the final potential $\phi^*$ also satisfies $(1+\e)\phi^*_v\ge d(s,v)$. Since each potential $\phi$ satisfies $|\phi(u)-\phi(v)|\le w(u,v)$ or each edge $(u,v)$, by \Cref{obs:max}, so does the coordinate-wise maximum $\phi^*$. Therefore, $\phi^*$ is an SSSP\ potential.

Next, we show that the graph $T^*$ returned is indeed a tree. Observe the following invariant: for each vertex $v\in V$, whenever $p^*(v)\ne\bot$, we have $d^*(p^*(v))<d^*(v)$. This is because whenever $d^*(v)$ is updated to $d_T(s,v)$ on \line{dT}, we must have $d_T(s,p^*(v))<d_T(s,v)$ for that tree $T$, so $d^*(p^*(v))$ would be updated as well if it was still at least~$d_T(s,v)$. Therefore, the edges $(v,p^*(v))$ at the end of the \textbf{while} loop must also satisfy $d^*(p^*(v))<d^*(v)$, so the edges are acyclic. Since there are $n-1$ edges total, $T^*$ is a tree. Finally, to show that $T^*$ is a $(1+\e$)-approximate SSSP tree, we show that for each vertex $v\in V$, $d_{T^*}(s,v)\le d^*(v)$. To see that this is sufficient, observe that since every vertex $v\in V\sm s$ was removed from $V'$, we have $d_T(s,v)\le(1+\e)\phi_v\le(1+\e)d(s,v)$ at some point (\line{dT-small}), so on this iteration, $d^*(v)$ would be updated to at most $(1+\e)d(s,v)$. 

We prove by induction on the ordering of $d^*(v)$ (from smallest to largest) that $d_{T^*}(s,v)\le d^*(v)$. If $p^*(v)=s$, then since $d^*(v)$ and $p^*(v)$ are updated at the same time (lines~\ref{line:dT}~and~\ref{line:p*}), the value $d^*(s)=d_T(s,v)=w(s,v)$ cannot be changed after $p^*(v)$ was set to $s$. Therefore, $d_{T^*}(s,v)=d^*(v)$. Otherwise, suppose that $p^*(v)=u\ne s$. Let $T$ be the tree computed on the iteration when $p^*(v)$ was updated to its final value. We have 
\[ d_{T^*}(s,v) = d_{T^*}(s,u)+w(u,v) \stackrel{\text{(ind.)}}\le d^*(u)+w(u,v) \le d_T(s,u)+w(u,v) ,\]
where the (ind.) stands for applying the inductive statement on vertex $u$. This completes the induction and the claim.
\EP

For the remainder of this section, we show that the \textbf{while} loop runs for only $O(\logn)$ iterations w.h.p. Consider the following potential function $\sum_{v\in V'}b_vd(s,v)$; we will show that it drops by a constant factor in expectation on each iteration of the \textup{\textbf{while}} loop. Since the graph $G$ has polynomial aspect ratio, the potential function can only decrease by a constant factor $O(\logn)$ times, so the lemma below suffices to finish \thm{sample-tree}.
\BL
On each iteration of the \textup{\textbf{while}} loop (\line{sssp-loop}), the quantity $\sum_{v\in V'}d(s,v)$ drops by a constant factor in expectation.
\EL
\BP
Define $\ol d(v):=\E[d_T(s,v)]$ over the randomness of \ref{esssp} on this iteration, which satisfies $\sum_{v\in V'}\ol d(v)\le(1+\e/10)\sum_{v\in V'}d(s,v)$ since $T$ is an $(1+\e/10)$-approximate expected SSSP tree with demands $\sum_{v\in V'}(\mathbbm1_v-\mathbbm1_s)$.
Since $(f,\phi)$ is a $(1+\e/10)$-approximate flow-potential pair, we have $\1{f}\le(1+\e/10)\sum_{v\in V}b_v\phi_v$. We have, for small enough $\e$,
\begin{align*}
 \sum_{v\in V'}\ol d(v)\le\lp1+\f\e{10}\rp\sum_{v\in V'}d(s,v)&\le\lp1+\f\e{10}\rp\opt\bigg(\sum_{v\in V'}(\mathbbm1_v-\mathbbm1_s)\bigg) \\&\le\lp1+\f\e{10}\rp\1{f} \\&\le \lp1+\f\e{10}\rp^2\sum_{v\in V}b_v\phi_v\le\lp1+\f\e4\rp\sum_{v\in V}b_v\phi_v =\lp1+\f\e4\rp\sum_{v\in V'}\phi_v ,
\end{align*}
which implies that
\[ \sum_{v\in V'} (\ol d(v)-\phi_v) \le \f\e4\sum_{v\in V'} \phi_v . \]
Observe that for all $v\in V'$, $\ol d(v)\ge d(s,v)\ge\phi_v-\phi_s=\phi_v$, so  $\ol d(v)-\phi_v\ge0$. Let $V'_{\text{good}}\s V'$ be the vertices $v\in V'$ with 
\[ \ol d(v)-\phi_v \le \f\e2\phi_v .\]
By a Markov's inequality-like argument, we have
\begin{gather} \sum_{v\in V'_{\text{good}}} \phi_v \ge \f12\sum_{v\in V'} \phi_v ;\eqnl{Vgood}\end{gather}
otherwise, we would have 
\[ \sum_{v\in V'} (d(v)-\phi_v) \ge \sum_{v\in V'\sm V'_{\text{good}}} (d(v)-\phi_v) \ge\f\e2\sum_{v\in V'\sm V'_{\text{good}}}  \phi_v > \f\e4\sum_{v\in V'} \phi_v, \] a contradiction.

For each $v\in V'_{\text{good}}$, by Markov's inequality on the nonnegative random variable $d_T(s,v)-\phi_v$ (which has expectation $\ol d(v)-\phi_v\le\f\e2\phi_v$), with probability at least $1/2$, we have
\[ d_T(s,v)-\phi_v\le\e\,\phi_v \iff d_T(s,v) \le (1+\e)\phi_v ,\]
so vertex $v$ is removed from $V'$ with probability at least $1/2$. In other words, the contribution of $v$ to the expected decrease of $\sum_{u\in V'} d(s,u)$ is at least $\f12  \phi_v$. Since
\[ d(s,v)\le d_T(s,v)\le(1+\e)\phi_v \implies \phi_v\ge(1+\e)\inv d(s,v) ,\] this expected decrease is at least $\f1{2(1+\e)} \phi_v$. Summing over all $v\in V'_{\text{good}}$, the expected decrease of $\sum_{u\in V'} d(s,u)$ is at least
\[ \sum_{v\in V'_{\text{good}}}\f1{2(1+\e)} d(s,v) \stackrel{\eqn{Vgood}}\ge \f1{4(1+\e)}\sum_{v\in V'} d(s,v) ,\]
which is a constant factor.
\EP

\section{Omitted Proofs}\secl{om}

\subsection{Proof of~\lem{aspect-ratio}}\secl{om21}
\AspectRatio*

\BP
Suppose that the demand vector $b$ satisfies $b_v\in\{0,1,2,\lds,n^C\}$ for all $v\in V$, for some constant $C$.
First, compute a minimum spanning tree $T$ of the graph $G=(V,E)$,\footnote{This can be computed work-efficiently in Parallel, e.g., with Boruvka's algorithm.} and compute the optimal transshipment cost where the input graph is $T$ instead, which is easily done efficiently since $T$ is a tree. Since $T$ is a minimum spanning tree, it is easy to see that for any vertices $u,v\in V$, we have $d_G(u,v)\le d_T(u,v)\le (n-1) \cd d_G(u,v)$, i.e., the \emph{stretch} of $T$ is at most $(n-1)$. Define $Z:=\opt_T(b)$ as the optimal \ts cost on $T$; it follows that
\begin{gather} \opt_G(b) \le Z\le (n-1)\,\opt_G(b) \eqnl{optT}.\end{gather}

 To construct $\wG$, we start with $G$ and remove all edges of weight more than $Z$ from $G$, and then add $Z/n^{C+5}$ weight to each remaining edge in the graph. Clearly, $\wG$ has aspect ratio $\poly(n)$ and satisfies $\opt_G(b)\le\opt_{\wG}(b)$. It remains to show that 
\begin{gather}\opt_{\wG}(b)\le\left(1+\f1{n^2}\right)\,\opt_G(b).\eqnl{to-show-wG}\end{gather}

The transshipment problem can be formulated as an uncapacitated minimum cost flow problem. It is well-known that if the demands of a minimum cost flow problem are integral, then there exists an optimal flow that is integral. Let $f$ be this integral flow for demands $b$. Then, $f$ cannot carry any flow along any edge with weight more than $Z$, since if it did, then it must carry at least $1$ flow along that edge, bringing its total cost to more than $Z$, contradicting the fact that $\opt_G(b)\le\opt_T(b)=Z$. It follows that removing edges with weight more than $Z$ does not affect the optimal transshipment cost.

Since $|b_v|\le n^C$ for all $v\in V$, it is also well-known that the optimal flow $f$ satisfies $|f_e|\le n^C$ for all $e\in E$. Consider the same flow $f$ on $\wG$ instead of $G$; since each edge has its weight increased by $Z/n^{C+4}$, the total increase in cost of the flow $f$ on $\wG$ is
\[ \sum_{e\in E}|f_e| \cd \f Z{n^{C+5}} \le \bn n2 \cd n^C \cd \f Z{n^{C+5}} \le \f Z{n^3} \stackrel{\eqn{optT}}\le \f{\opt_G(b)}{n^2} .\]
The cost of the optimal flow on $\wG$ can only be lower, which proves \eqn{to-show-wG}.
\EP



\subsection{Proof of \lem{embed}}\secl{om27}
\Embed*
\begin{algorithm}[H]
\caption{\texttt{L1\_embed}($G=(V,E)$)}
\begin{algorithmic}[1]
\State Let $N\gets O(\logn)$, $T\gets\lc\logn\rc$, $\e\gets1/\logn$
\For {independent iteration $i=1,2,\lds,N$}
 \For {$t=1,2,\lds,T$} \linel{for-t}
  \State Sample each vertex in $G$ independently with probability $1/2^t$; let $S$ be the sampled set \linel{sample}
  \State Compute $(1+\e)$-approximate $S$-SSSP potential $\phi_{i,t}(u)$ of $G$ through algorithm $\m A$
  \State Extend $\phi_{i,t}$ so that $\phi_{i,t}(v)=\phi_{i,t}(s)$ for all $v\in S$, so that $\phi_{i,t}(v)$ is now defined for all $v\in V$
 \EndFor
\EndFor
\State For each $v\in V$, output the vector $x(v):=\langle \f1{NT}\phi_{i,t}(v) \rangle_{i\in[N],t\in[T]} \in \R^{[N]\times[T]}$ as the $\el_1$-embedding of $v$
\end{algorithmic}\label{alg:KT}
\end{algorithm}

Fix two vertices $u,v$ throughout the proof, and define $d:=d(u,v)$; we need to show that w.h.p.,
\[ \norm{x(u)-x(v)}_1 = \f1{NT}\sum_{i\in[N],t\in[T]} \left| \phi_{i,t}(u) - \phi_{i,t}(v) \right| \in \lb \f{d}{O(\log^3n)} ,\ d \rb .\]
The upper bound is easy: by definition of approximate $s$-SSSP potential, we have $|\phi_{i,t}(u)-\phi_{i,t}(v)|\le d$ for all $i,t$, so taking the average over all $i,t$ gives $\f1{NT}\sum_{i,t} \left| \phi_{i,t}(u) - \phi_{i,t}(v) \right|\le  d $. To finish \lem{embed}, it remains to prove the lower bound, whose proof occupies the rest of this section.

\begin{restatable}{lemma}{ExistsT}\leml{exists-t}
There is a value of $t\in[T]$ such that with probability $\Om(1)$,
\[ |\phi_{i,t}(u) - \phi_{i,t}(v)| \ge \Om(\e d) .\]
\end{restatable}

For each positive $r$, define $B(u,r):=\{w\in V:d(w,u)\le r\}$ as the vertices within distance $r$ from $u$. Similarly, define $B(v,r)$ as the vertices within distance $r$ from $v$. 

\BCL\label{clm:exists-r}
There exists a value $r\in[d/6,d/3]$ and a (universal) constant $C>1$ such that
\[ |B(u,(1+2\e)r)| \le C|B(v,r)| \quad\text{or}\quad |B(v,(1+2\e)r)| \le C|B(u,r)| .\]
\ECL
\BP
First, we show that such a value $r$ must exist. If not, then we have the chain of inequalities
\[  |B(v,\f d6)| < \f1C|B(u,(1+2\e)\f d6)| < \f1{C^2}|B(v,(1+2\e)^2\f d6)|< \f1{C^3}|B(u,(1+2\e)^3\f d6)| < \cds < \f1{C^L}|B(u,(1+2\e)^L\f d6)|\]
for $L=\lf\log_{(1+2\e)}2\rf=\Th(1/\e)=\Th(\logn)$ (we assume w.l.o.g.\ that the last expression has $u$ and not $v$). For large enough $C$, this means that
\[ 1 \le |B(u,\f d6)| \le \f1{C^{\Th(\logn)}}|B(u ,(1+2\e)^L\f d6)| < \f1n |B(u ,(1+2\e)^L\f d6)| \implies |B(u ,(1+2\e)^L\f d6)|>n ,\]
which is impossible. Therefore, such a value $r$ exists.
\EP

Take the value $r$ guaranteed by \Cref{{clm:exists-r}}, and assume w.l.o.g.\ that $|B(u,(1+2\e)r)| \le C|B(v,r)|$. Pick $t\in[T]$ satisfying $2^{t-1}\le|B(v,r)|\le2^{t}$, which also means that $|B(u,(1+2\e)r)|\le O(2^t)$. Suppose we sample each vertex in $V$ with probability $1/2^t$ (\line{sample}). With probability $\Om(1)$, we sample at least one vertex in $B(v,r)$, and with probability $\Om(1)$, we sample zero vertices in $B(u,(1+2\e)r)$. Moreover, since $r + (1+2\e)r < 3r \le 3 \cd d/3=d$, the two sets $B(u,(1+2\e)r)$ and $B(v,r)$ are disjoint, so the two events are independent. Thus, with probability $\Om(1)$, we have both $S\cap B(v,r)\ne\emptyset$ and $S\cap B(u,(1+2\e)r)=\emptyset$, which implies that $d(S,v)\le r$ and $d(S,u)\ge(1+2\e)r$.

Fix an iteration $i\in[N]$, and let us condition on the previous event. Since $\phi_{i,t}$ is a $(1+\e)$-approximate $S$-SSSP potential of $G$, we have $\phi_{i,t}(v)-\phi_{i,t}(s)\le d(S,v)\le r$ by property~(1) of \defn{SSSP-dual} and $\phi_{i,t}(u)-\phi_{i,t}(s)\ge\f1{1+\e}d(S,u)\ge\f1{1+\e}(1+2\e)r=(1+\Om(\e))r$ by \obs{dual-upper}. Thus, $|\phi_{i,t}(u)-\phi_{i,t}(v)|\ge\Om(\e)\cd r \ge \Om(\e)\cd d/6= \Om(\e d)$.

Since there are $N=O(\logn)$ trials, w.h.p., one of the iterations $i\in[N]$ will satisfy $|\phi_{i,t}(u)-\phi_{i,t}(v)|\ge\Om(\e d)$ for the value of $t$ guaranteed by \lem{exists-t}. Thus, w.h.p., we have
\[\norm{x(u)-x(v)}_1 =\f1{NT}\sum_{i,t} \left| \phi_{i,t}(u) - \phi_{i,t}(v) \right| \ge \f1{NT} \Om(\e d) =\f{d}{O(\log^3n)}  ,\]
concluding \lem{exists-t}.

\subsection{Proof of \lem{tree-SSSP}}\secl{om75}
\TreeSSSP*
\BP
It suffices to compute (exact) $S$-SSSP distances on $T$, after which we simply define $\phi(v)$ as the distance to $v$ for each vertex $v$.



Define a \emph{centroid} of the tree $T$ as a vertex $v\in V$ such that every component of $T-v$ has size at most $|V|/2$. We can compute a centroid $r$ as follows: root the tree $T$ arbitrarily, and for each vertex $v$, compute the size of the subtree rooted at $v$; then, let the centroid be a vertex whose subtree has size at least $n/2$, but whose children each have subtrees of size less than $n/2$. Next, compute the distance $d_T(r,S)$ from $r$ to the closest vertex in $S$, which can be accomplished by computing SSSP on the tree with $r$ as the source. Now root the tree $T$ at $r$, and for each child vertex $v$ with subtree $T_v$, construct the following recursive instance: the tree is $T_v$ together with the edge $(v,r)$ of weight $d_T(r,S)+w(v,r)$, and the set $S$ is $(V(T_v) \cap S)\cup \{r\}$. Solve the recursive instances, and for each vertex $u\in V\sm r$, the distance $d(u)$ is the computed distance in the (unique) recursive instance $T_v$ such that $u\in T_v$.

It is clear that the above algorithm is correct and can be implemented in $\tO(m)$ work and $\pl(n)$ time.
\EP

\end{document}